\documentclass[submission, Phys]{SciPost}
\pdfoutput=1
\usepackage{amsmath,amssymb,mathtools,xspace}
\usepackage{booktabs,multirow,graphicx,tabularx,slashed}
\usepackage{hyperref}
\usepackage{color,xcolor}
\usepackage[normalem]{ulem}
\usepackage{enumitem}
\usepackage{braket}
\usepackage{stackrel}
\usepackage{tikz}
\usetikzlibrary{arrows}
\usetikzlibrary{shapes.geometric}
\usepackage{float}
\usepackage{subcaption}

\usepackage{epstopdf}
\graphicspath{{figs/}}

\makeatletter
\def\BState{\State\hskip-\ALG@thistlm}
\makeatother

\makeatletter
\@ifundefined{pdfoutput}{}{\DeclareGraphicsRule{*}{mps}{*}{}}
\makeatother

\makeatletter
\DeclareRobustCommand*{\bfseries}{%
   \not@math@alphabet\bfseries\mathbf
   \fontseries\bfdefault\selectfont
   \boldmath
}
\makeatother

\setitemize{itemsep=2pt,topsep=2pt,parsep=0pt,partopsep=0pt,leftmargin=*}
\setenumerate{itemsep=0pt,topsep=2pt,parsep=0pt,partopsep=0pt,labelindent=3pt,leftmargin=*}
\definecolor{Gcolor}{HTML}{3b528b}
\definecolor{Dcolor}{HTML}{e41a1c}

\tikzstyle{generator} = [rectangle, rounded corners, minimum width=3cm, minimum height=1cm,text centered, draw=Gcolor]
\tikzstyle{discriminator} = [rectangle, rounded corners, minimum width=3cm, minimum height=1cm,text centered, draw=Dcolor]
\tikzstyle{io} = [circle, trapezium left angle=70, trapezium right angle=110, minimum width=1cm, minimum height=1cm, text centered, draw=black]

\tikzstyle{process} = [rectangle, minimum width=1cm, minimum height=1cm, text centered, draw=black]
\tikzstyle{decision} = [rectangle, minimum width=1cm, minimum height=1cm, text centered, draw=black]

\tikzstyle{arrow} = [thick,->,>=stealth]
\usepackage{xcolor}

\newcommand{\XLangle}{\Big\langle}
\newcommand{\XRangle}{\Big\rangle}
\newcommand{\Langle}{\big\langle}
\newcommand{\Rangle}{\big\rangle}

\newcommand\one{\leavevmode\hbox{\small1\normalsize\kern-.33em1}}

\newcommand{\loss}{\mathcal{L}}

\newcommand{\qqquad}{\qquad \qquad}

\def\slashchar#1{\setbox0=\hbox{$#1$}           % set a box for #1
   \dimen0=\wd0                                 % and get its size
   \setbox1=\hbox{/} \dimen1=\wd1               % get size of /
   \ifdim\dimen0>\dimen1                        % #1 is bigger
      \rlap{\hbox to \dimen0{\hfil/\hfil}}      % so center / in box
      #1                                        % and print #1
   \else                                        % / is bigger
      \rlap{\hbox to \dimen1{\hfil$#1$\hfil}}   % so center #1
      /                                         % and print /
   \fi}

\newcommand{\ie}{\textsl{i.e.}\;}

\begin{document}
%\begin{fmffile}{feynman}

\begin{center}{\Large \textbf{
      Better Latent Spaces for Better Autoencoders
}}\end{center}

\begin{center}
Barry M. Dillon\textsuperscript{1}, 
Tilman Plehn\textsuperscript{1}, 
Christof Sauer\textsuperscript{2}, and
Peter Sorrenson\textsuperscript{3}, 
\end{center}

\begin{center}
{\bf 1} Institut f\"ur Theoretische Physik, Universit\"at Heidelberg, Germany\\
{\bf 2} Physikalisches Institut, Universit\"at Heidelberg, Germany\\
{\bf 3} Heidelberg Collaboratory for Image Processing, Universit\"at Heidelberg, Germany
\end{center}

\begin{center}
\today
\end{center}

\section*{Abstract}
{\bf Autoencoders as tools behind anomaly searches at the LHC have the
  structural problem that they only work in one direction, extracting
  jets with higher complexity but not the other way around.  To
  address this, we derive classifiers from the latent space of
  (variational) autoencoders, specifically in Gaussian mixture and
  Dirichlet latent spaces. In particular, the Dirichlet setup solves the
  problem and improves both the performance and the interpretability
  of the networks.  }

\tikzstyle{int}=[thick,draw, minimum size=2em]
%\tikzstyle{init} = [pin edge={to-,thin,black}]

\vspace{10pt}
\noindent\rule{\textwidth}{1pt}
\tableofcontents\thispagestyle{fancy}
\noindent\rule{\textwidth}{1pt}
\vspace{10pt}

\newpage
%%%%%%%%%%%%%%%%%%%%%%%%%%%%%%%%%%%%%%%%%%%%%%%%%%%%%%%%%%%%%%%%%%%%%%%%
\section{Introduction}
\label{sec:intro}

If new physics exists at the scales currently probed at the LHC, then
it is more elusive than expected and could be difficult to uncover
with traditional means and without a significant increase in the
number of analyses.  It is here that unsupervised machine learning
techniques can play a crucial role, not fulfilled by
traditional, signal-driven search techniques.

Autoencoders (AEs) with a bottleneck are the simplest unsupervised
machine learning tool. They are built on a network which maps a
high-dimensional data representation onto itself, constructing a
typical or average object. The goal is to reconstruct the input
data as accurately as possible. If we then compress  a set of layers
in the middle of this network, we cut into its expressiveness, such
that atypical objects cannot be related to the learned output any
longer. In early high-energy physics works, AEs have been shown to
classify between anomalous jets and QCD jets using the reconstruction
error as the
classifier~\cite{Heimel:2018mkt,Farina:2018fyg,Roy:2019jae,Blance:2019ibf}.

In a more sophisticated manner, variational
autoencoders~\cite{kingma2014autoencoding} (VAEs) add structure to the
bottleneck. In a first, encoding step, the high-dimensional data
representation is mapped to a lower-dimensional latent distribution;
next, the decoder learns a model to generate the input data by
sampling from this latent distribution.  After training, the latent
space should contain information about each instance of the data-set
that is not apparent in the high-dimensional input representation.  At
the LHC, it is possible to use this latent space representation of the
jets for
classification~\cite{Cerri:2018anq,Cheng:2020dal,Bortolato:2021zic},
and even anomaly detection.  The use of machine learning methods for
anomaly detection has received a lot of attention in recent
years~\cite{Collins:2018epr,Collins:2019jip,Dillon:2019cqt,Dillon_2020,Andreassen:2020nkr,Khosa:2020qrz,knapp2020adversarially,Nachman:2020lpy,Aguilar-Saavedra:2017rzt,Mikuni:2020qds,1815227,Hajer:2018kqm,Amram:2020ykb},
including a first ATLAS analysis~\cite{Aad:2020cws}, and the
LHC~Olympics~2020 community challenge has seen many groups submit
their results on a di-jet anomaly search using black-box
data-sets~\cite{Kasieczka:2021xcg}.
Additional papers studying similar unsupervised LHC problems include Ref.~\cite{DAgnolo:2018cun,DAgnolo:2019vbw,DeSimone:2018efk,Mullin:2019mmh,Romao:2019dvs,Romao:2020ojy,Nachman:2020ccu,Thaprasop:2020mzp,Komiske:2020qhg,Komiske:2019jim,Komiske:2019fks,Knapp:2020dde,Matchev:2020wwx,Romao:2020ocr,Aguilar-Saavedra:2020uhm,vanBeekveld:2020txa,Park:2020pak}.

While (V)AE-analyses based directly on anomaly scores appear
unrealistic, they allow us to collect and analyse data in a more
targeted manner, for instance as part of the trigger, in addition to
pre-scaling, or complementing minimum bias events. This kind of
application is fundamentally different from weakly supervised
analyses, which are based on some kind of signal hypotheses or
expected signal features~\cite{Collins:2021nxn}.
As a matter of fact,
the statistical nature of LHC data will typically turn supervised
techniques into weakly supervised analyses the moment we train on data
rather than Monte Carlos.  In the spirit of anomaly searches, we are
studying unsupervised learning of QCD and top jets in this paper. The
question is how a VAE can encode something related to a class label in
the latent space, and how more structured latent space geometries help
with unsupervised classification.

We start in Section~\ref{sec:problem} by demonstrating how AEs and
VAEs perform when using the compressed/latent space representations
for unsupervised jet classification.  We then study a generalisation,
in which the Gaussian prior is replaced by a Gaussian-Mixture prior
(GMVAE)~\cite{tomczak2018vae,dilokthanakul2016deep,shao2020generalized,yang2019deep,guo2018multidimensional,shu2016stochastic,hershey2007approximating}
in Section~\ref{sec:mvae}.  In Section~\ref{sec:dvae} we introduce the
Dirichlet-VAE (DVAE), which uses a compact latent space with a
Dirichlet prior~\cite{srivastava2017autoencoding,joo2019dirichlet}.
Through a specific choice of decoder architecture we can interpret the
decoder weights as the parameters of the mixture distributions in the
probabilistic model, and can visualise these to directly interpret
what the neural network is learning. Our technique shares similarities
with topic-modelling
approaches~\cite{Metodiev:2018ftz,Komiske:2018vkc,Dillon:2019cqt,Dillon_2020,Alvarez:2019knh}.
We present our conclusions in Section~\ref{sec:outlook}.

%%%%%%%%%%%%%%%%%%%%%%%%%%%%%%%%%%%%%%%%%%%%%%%%%%%%%%%%%%%%%%%%%%%%%%%%
\section{The problem with autoencoding jets}
\label{sec:problem}

While autoencoders have been proposed for unsupervised jet
classification~\cite{Heimel:2018mkt,Farina:2018fyg}, they are known
to be problematic in general applications. We briefly discuss their
issues and possible improvements through VAEs, and motivate our new
approach based on appropriately chosen latent spaces.

%%%%%%%%%%%%%%%%%%%%%%%%%%%%%%%%%%%%%%%%%%%%%%%%%%%%%%%%%%%%%%%%%%%%%%%%
\subsubsection*{Jet images and data set}
\label{sec:jets}

Throughout the paper we use the QCD and top jet samples generated for
the community top-tagging challenge outlined in
Ref.~\cite{Kasieczka:2019dbj}, in the same jet image representation as
in our earlier AE study~\cite{Heimel:2018mkt}.  We simulate the jets
with Pythia8~\cite{Sjostrand:2014zea} (default tune) using a
center-of-mass energy of 14~TeV and ignoring pile-up and multi-parton
interactions. We use Delphes~\cite{deFavereau:2013fsa} for a fast
detector simulation with the default ATLAS detector card.  The jets
are defined through the anti-$k_T$
algorithm~\cite{Cacciari:2008gp,Cacciari:2005hq} in
FastJet~\cite{Cacciari:2011ma} with a radius of $R=0.8$.  In each
event we kept only the leading jet and only if $p_T = 550~...~650$~GeV
and $|\eta|<2$.  Additionally, we require top jets to be matched to a
parton-level top within the jet radius and all parton-level decay
products to lay within the jet radius.  To define the jet constituents
we use the Delphes energy-flow algorithm and keep the leading 200
constituents from each jet for the analysis, while zero-padding the
empty entries.  We do not include particle ID or tracking information.

\begin{figure}[t]
\centering
\includegraphics[width=0.32\textwidth]{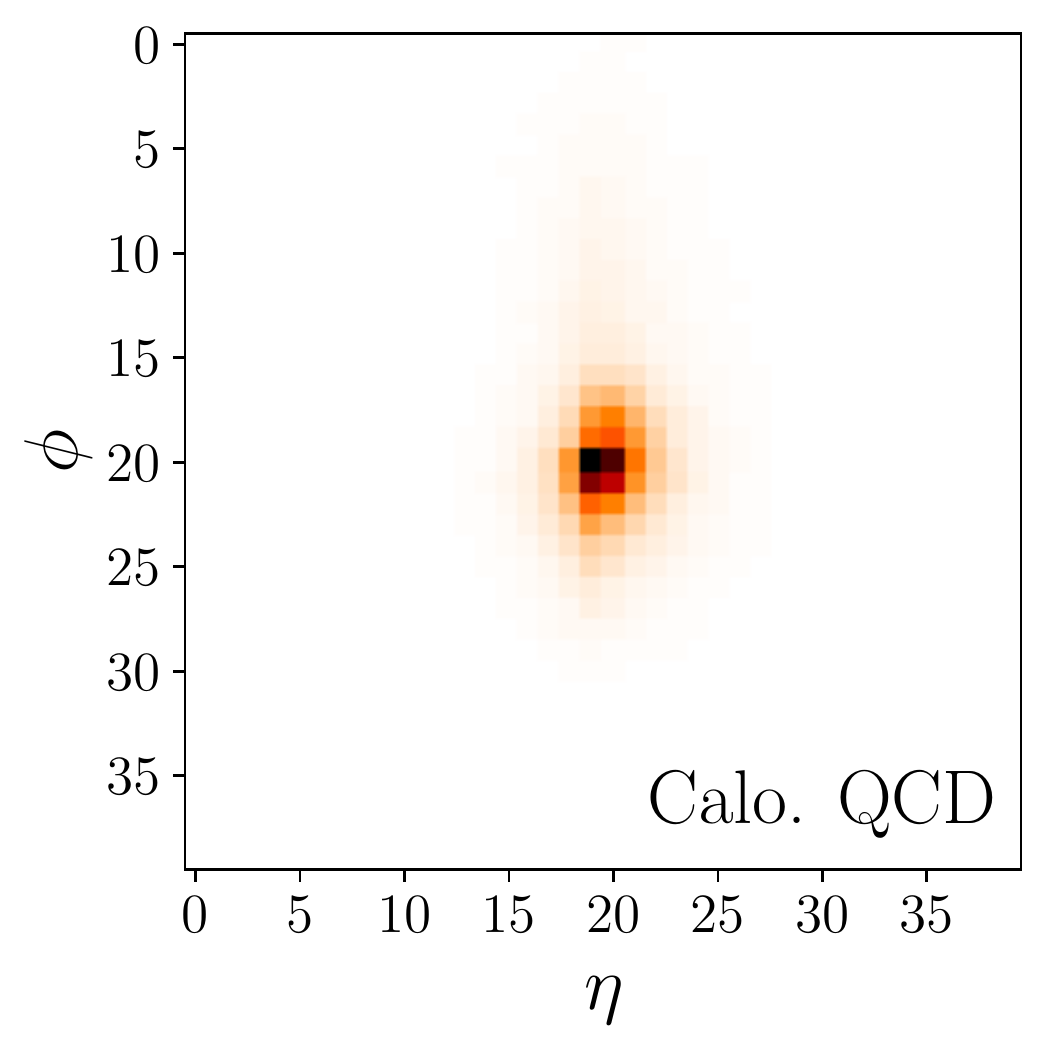}
\hspace*{0.1\textwidth}
\includegraphics[width=0.32\textwidth]{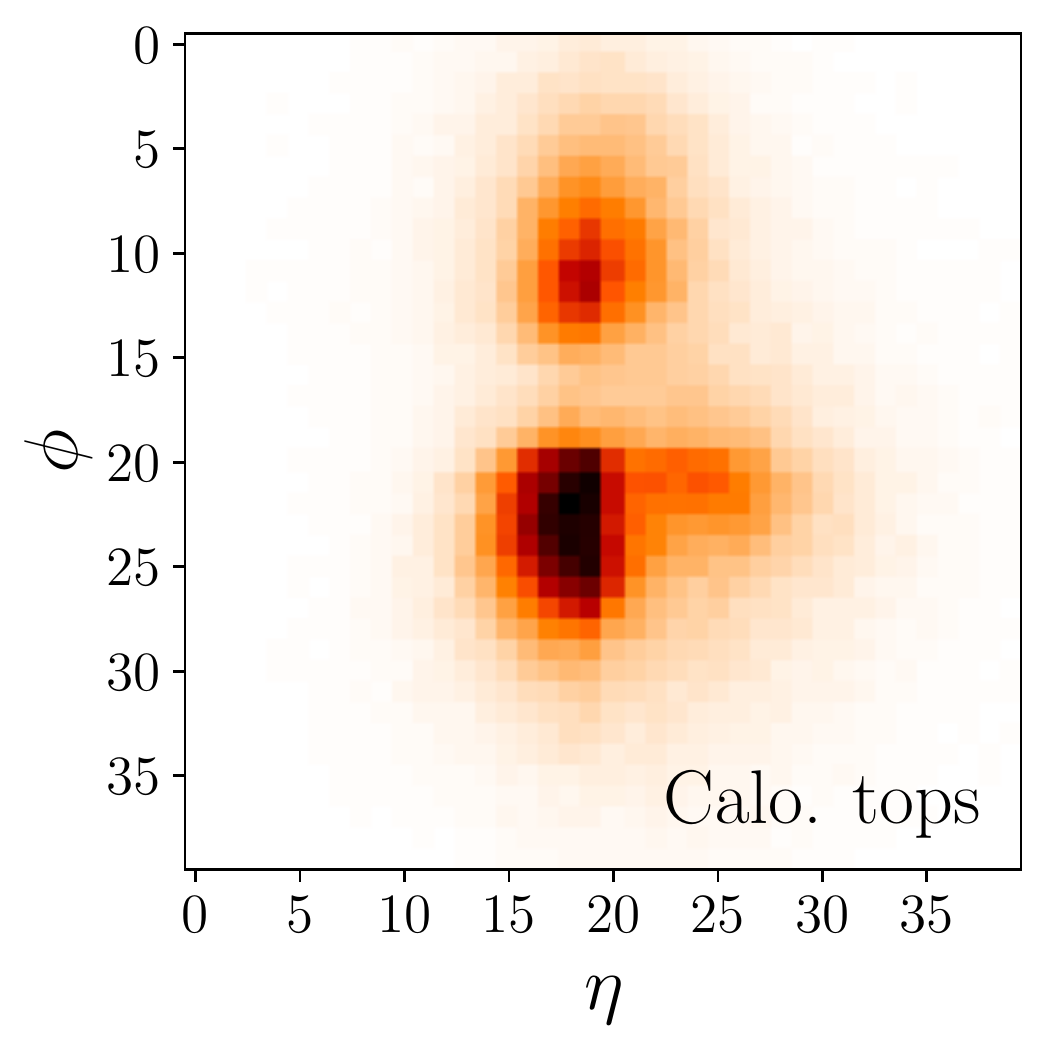}
\caption{Average of 100k QCD and top jet images after pre-processing.}
\label{fig:jetimages}
\end{figure}

For the pre-processing of the jets we follow a procedure similar to
Refs.~\cite{Heimel:2018mkt,Macaluso:2018tck}.  We do the
pre-processing at the level of the jet constituents, prior to
pixelization.  First, we center the jet using the $k_T$-weighted
centroid of the jet constituents, such that the major principle axis
points upwards.  We then flip the image in both axes so that most of
the $p_T$ is located in the lower left quadrant.  Once this is done we
pixelize the image into a 40 by 40 array, with the intensity defined
by the $p_T$ sum of the constituents per pixel.  The pixel sizes are
$[\Delta\eta,\Delta\phi]=[0.029,0.035$] and during pixelization we
crop the image to reduce the sparsity of the information within it.
In Fig.~\ref{fig:jetimages} we show the average of 10k QCD and top jet
images after the pre-processing has been applied.

Our data-set consists of 100k QCD jets and 100k top jets, and in each
analysis we use the maximum number of jets possible.  For example if
we want equal numbers of QCD and top jets we use 100k of each, and if
we want a data-set where the QCD jets are being treated as the
background and the number of top jets is varied, we keep the full 100k
QCD jets and vary the number of top jets. For all presented results we
use a 90/10 split of the data for training and testing.  The data
samples are shuffled and the testing data is selected randomly on each
run.

%----------------------------------------------------------
%----------------------------------------------------------

%%%%%%%%%%%%%%%%%%%%%%%%%%%%%%%%%%%%%%%%%%%%%%%%%%%%%%%%%%%%%%%%%%%%%%%%
\subsubsection*{The networks}
\label{sec:aevaenets}

\begin{figure}[t]
\begin{subfigure}[t]{0.44\textwidth}
\raisebox{0.9mm}{ \includegraphics[width=\textwidth]{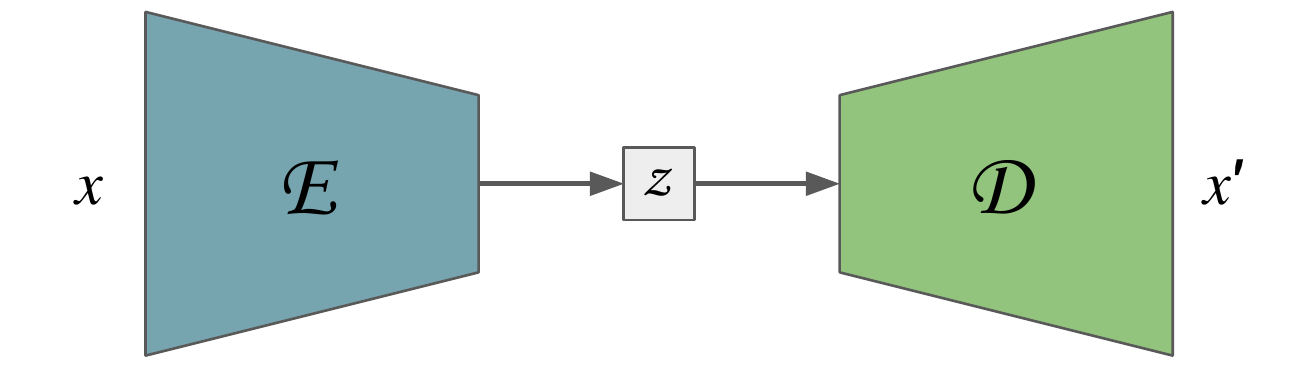} }
\caption{AE setup}
\end{subfigure}
\begin{subfigure}[t]{0.56\textwidth}
\includegraphics[width=\textwidth]{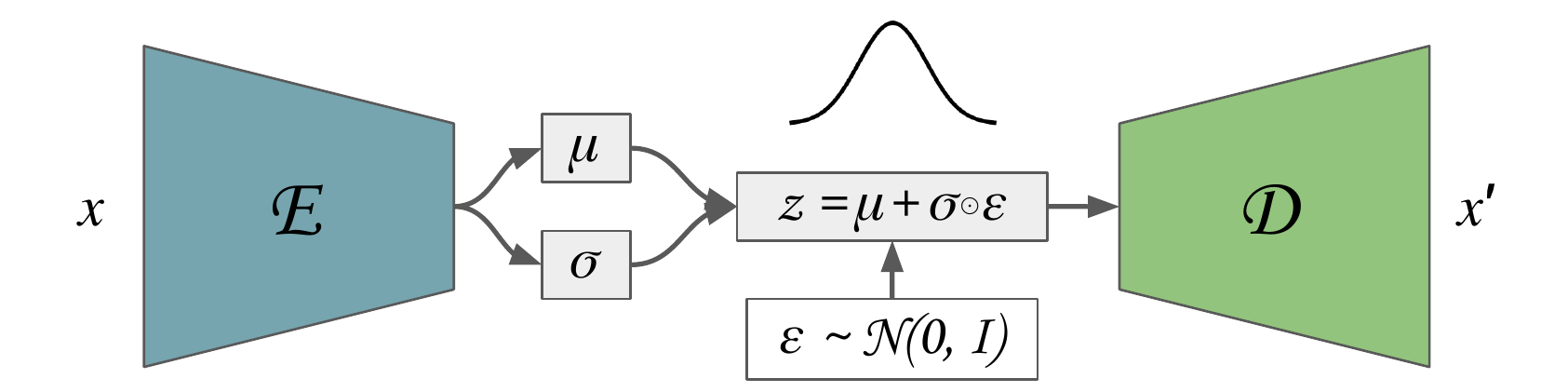} 
\caption{VAE setup}
\end{subfigure} 
\par \bigskip
\begin{subfigure}[b]{\textwidth}
\includegraphics[width=\textwidth]{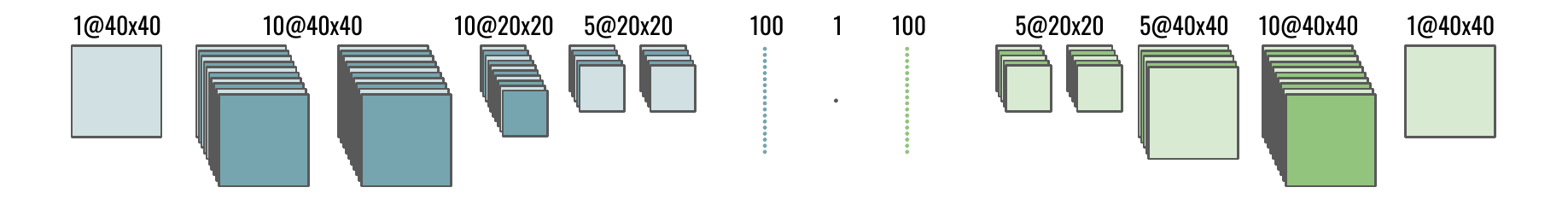}
\caption{Encoder and decoder}
\label{fig:aevae_archs_c}
\end{subfigure}
\caption{Architectures used for the AE and VAE networks in this
  study. All convolutions use a 5x5 kernel size and all layers use
  PReLU activations. Downsampling from 40x40 to 20x20 is achieved by
  average pooling, while upsampling is nearest neighbor
  interpolation.}
\label{fig:aevae_archs}
\end{figure}

Both the AE and the VAE consist of two networks illustrated in
Fig.~\ref{fig:aevae_archs}; an encoder and a decoder.  In the case of
the AE the encoder maps the data to a so-called bottleneck, for
instance an $n$-dimensional vector $z$.  In the case
of the VAE the encoder maps the data to a posterior distribution,
which in this section is taken to be a Gaussian with diagonal covariance. This structures the
encoder output and defines the latent space. An $n$-dimensional
Gaussian latent space $z$ can be described by $2n$ numbers, $n$
means $\mu_i$ and $n$ log-variances $\log\sigma_i$.  So that the
network can be optimised through back-propagation we must be able to
differentiate this sampling function with respect to the parameters of
the posterior distribution.  A re-parametrization of the sampling
step~\cite{kingma2014autoencoding} uses a vector of random numbers
$\epsilon$ from an $n$-dimensional unit Gaussian, and then defines
the sampled latent vector as $z=\mu + \sigma \odot \epsilon$.
The Hadamard product $\odot$ implies the element-wise product of the
two operands.  
The autoencoder loss used is the mean squared error
\begin{align}
\loss = \XLangle \frac{1}{2} \lVert x - x' \rVert^2 \XRangle_{p_{\text{data}}(x)} \; ,
\label{eq:aeloss}
\end{align}
where $x'$ is the reconstruction of the input jet images $x$. The VAE loss function~\cite{kingma2014autoencoding} has the form
\begin{align}
\loss = \XLangle - \Langle \log p_\theta(x|z) \Rangle_{q_\phi(z|x)} + \beta_\text{KL} D_\text{KL}(q_\phi(z|x), p(z)) \XRangle_{p_{\text{data}}(x)} \; ,
%\loss = \mathbb{E}_{p_{\text{data}}(x)}\left[ \mathbb{E}_{q_\phi(z|x)}\left[-\log p_\theta(x|z)\right] + D_\text{KL}(q_\phi(z|x) || p(z)) \right] \; ,
\label{eq:vaeloss}
\end{align}
with a learnable variational encoder $q_\phi(z|x)$ and decoder
$p_\theta(x|z)$, where $\phi$ and $\theta$ represent the parameters of
the encoder and decoder respectively.  
The $\beta_\text{KL}$ term allows us to control the relative influence
of the reconstruction (first term) and latent loss (second term) in the gradient updates. We use the mean squared error of equation \ref{eq:aeloss} as reconstruction loss, following from a choice of Gaussian $p_\theta(x|z)$. For the choice of standard normal $p(z)$, the KL divergence term has the following analytical form,
\begin{align}
D_\text{KL}(q_\phi(z|x),p(z)) = \frac{1}{2} \sum_{i=1}^n \left(\sigma_i^2 + \mu_i^2 - 1 - \log \sigma_i^2 \right),
\label{eq:klvae}
\end{align}
where $\sigma_i$ and $\mu_i$ are the outputs of the encoder for a given $x$.
We set $\beta_{KL}=10^{-4}$ as we find that this gives us optimal results.
For both the AE and the VAE the output layer of the decoder has a linear activation.
The hidden layers of the encoder and decoder architectures in both
cases use convolutional layers. All our neural
network models are constructed and optimised with
Tensorflow~\cite{abadi2016tensorflow} and
Keras~\cite{chollet2015keras}, and all use the Adam optimiser~\cite{Kingma:2014vow} with a learning rate of $0.001$.

%----------------------------------------------------------
%----------------------------------------------------------

%%%%%%%%%%%%%%%%%%%%%%%%%%%%%%%%%%%%%%%%%%%%%%%%%%%%%%%%%%%%%%%%%%%%%%%%
\subsubsection*{Reconstruction-loss tagging}
\label{sec:dvae_rec}

In the AE architecture the encoder compresses the input data into a
bottleneck, encoding each instance of the data-set in a representation
with a dimension much smaller than the dimension of the input data.
The decoder then attempts to reconstruct the input data from the
limited information encoded in the bottleneck.  The idea behind the
anomaly search is that the AE learns to compress and reconstruct the
training data very well, but different unseen data passed through the
trained AE will result in a large reconstruction error, or loss.  In
this way the AE may be used to search for data which is very different
to the training data, or even data which is part of the training data
but forms a small subclass of anomalous instances.  This strategy
fails for example in the case of QCD vs top jets, where the QCD jets
are assumed to be anomalous~\cite{Heimel:2018mkt}.  The reason appears
to be that the AE can encode simpler data with less features more
easily, so if the anomalous data is structurally simpler than the
dominant class, the AE with its reconstruction loss as the
classification metric fails.  A formal study of the limitations of AEs
has been presented recently~\cite{batson2021topological}.  In
Ref.~\cite{Heimel:2018mkt} the results appeared insensitive to
pre-processing, which can be explained by only the top jets being
sensitive to the pre-processing in a significant way.  Because the QCD
jets mostly have just a single prong, the centering of the jets is the
most relevant step.

%%%%%%%%%%%%%%%%%%%%%%%%%%%%%%%%%%%%%%%%%%%%%%%%%%%%%%%%%%%%%%%%%%%%%%%%
\subsubsection*{Latent-space tagging}
\label{sec:ae_latent}

The AE drawbacks can be avoided through using an alternative
classification metric to the reconstruction error.  In the context of
VAEs, the natural choice is to derive a metric from the latent space
embedding~\cite{Cheng:2020dal,Bortolato:2021zic}.  In this paper we
study the behaviour of such metrics systematically and provide
better-suited latent spaces.

%----------------------------------------------------------
\begin{figure}[t]
\includegraphics[width=0.32\textwidth]{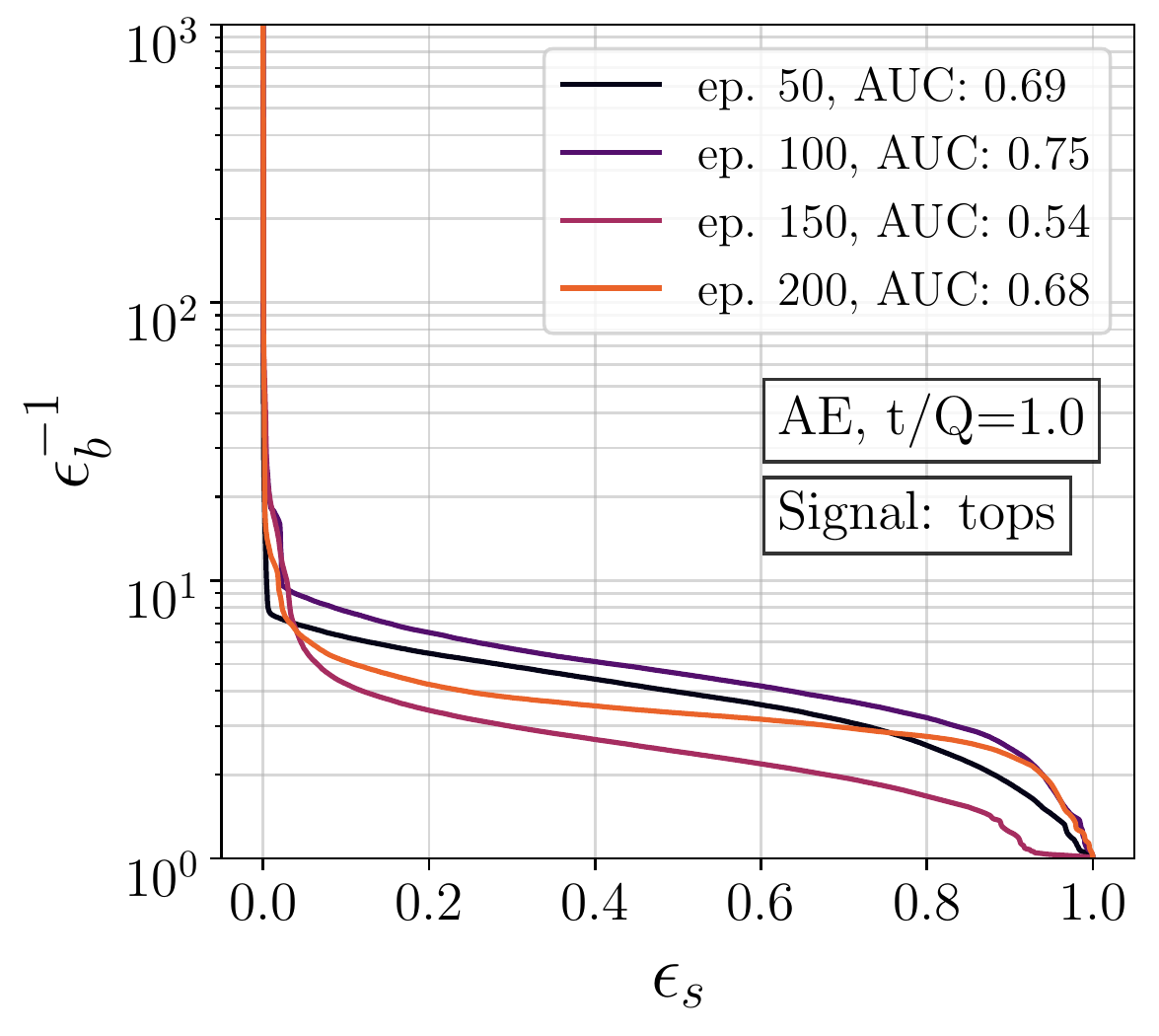}
\includegraphics[width=0.32\textwidth]{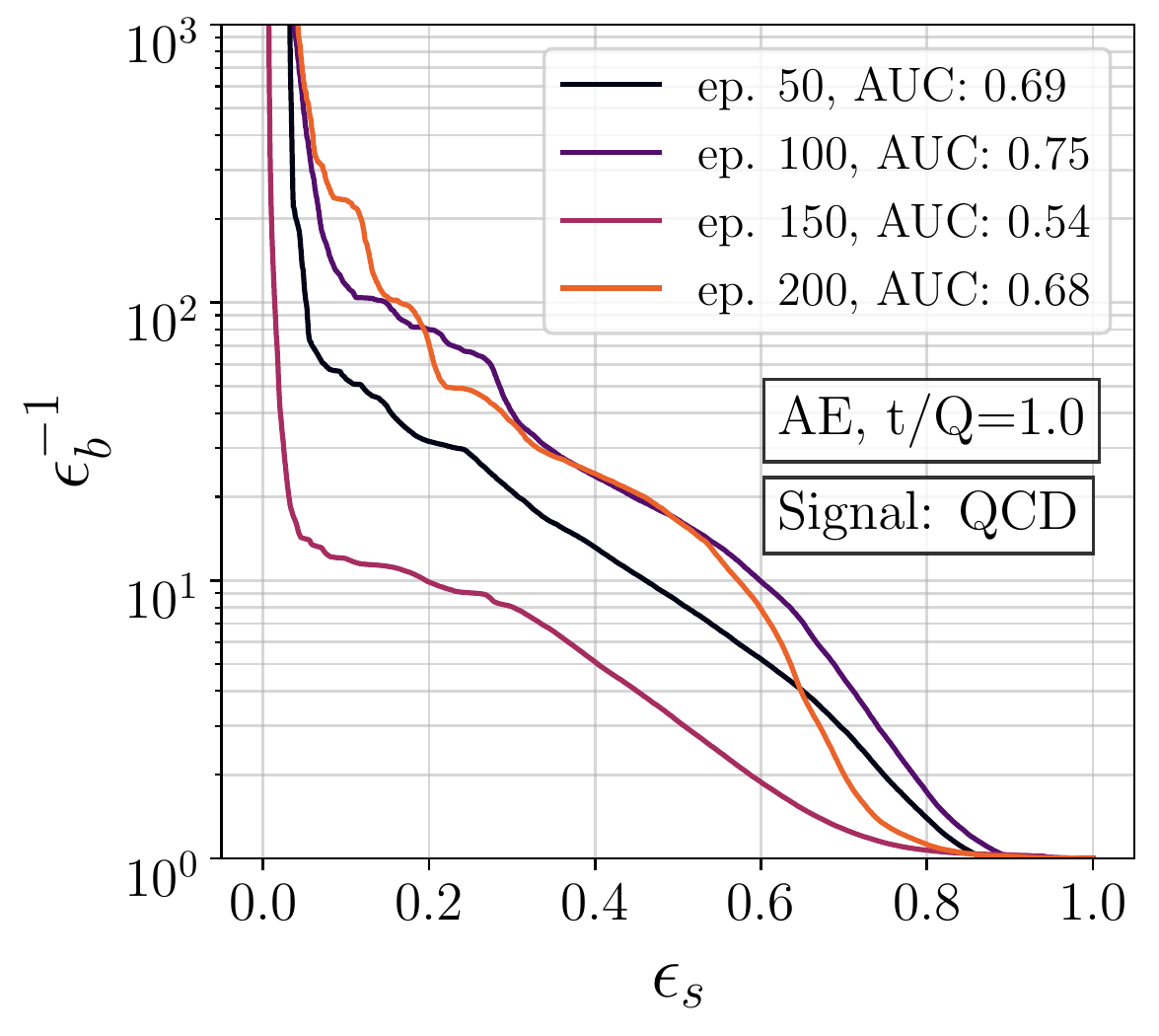}
\includegraphics[width=0.32\textwidth]{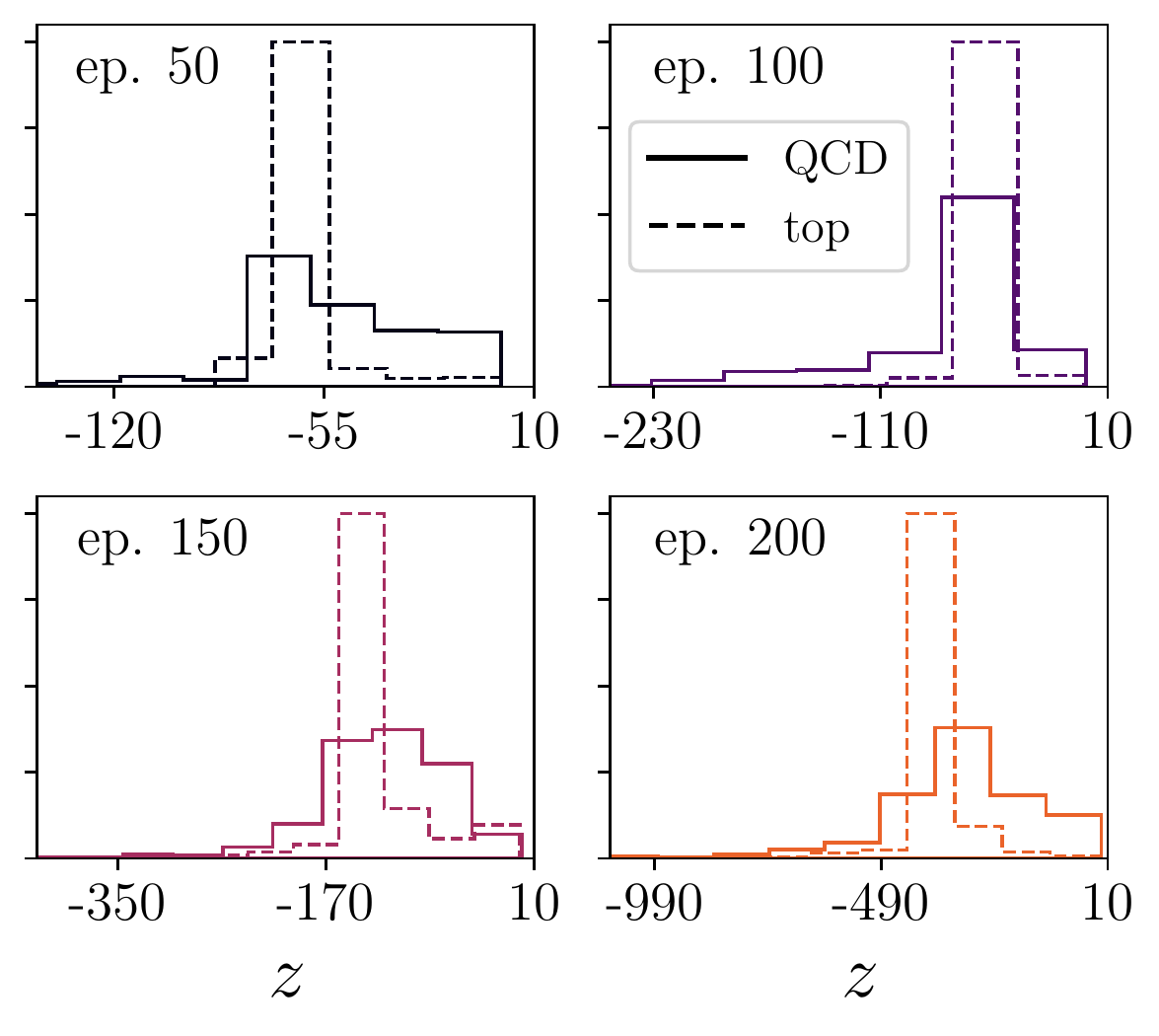} \\
\includegraphics[width=0.32\textwidth]{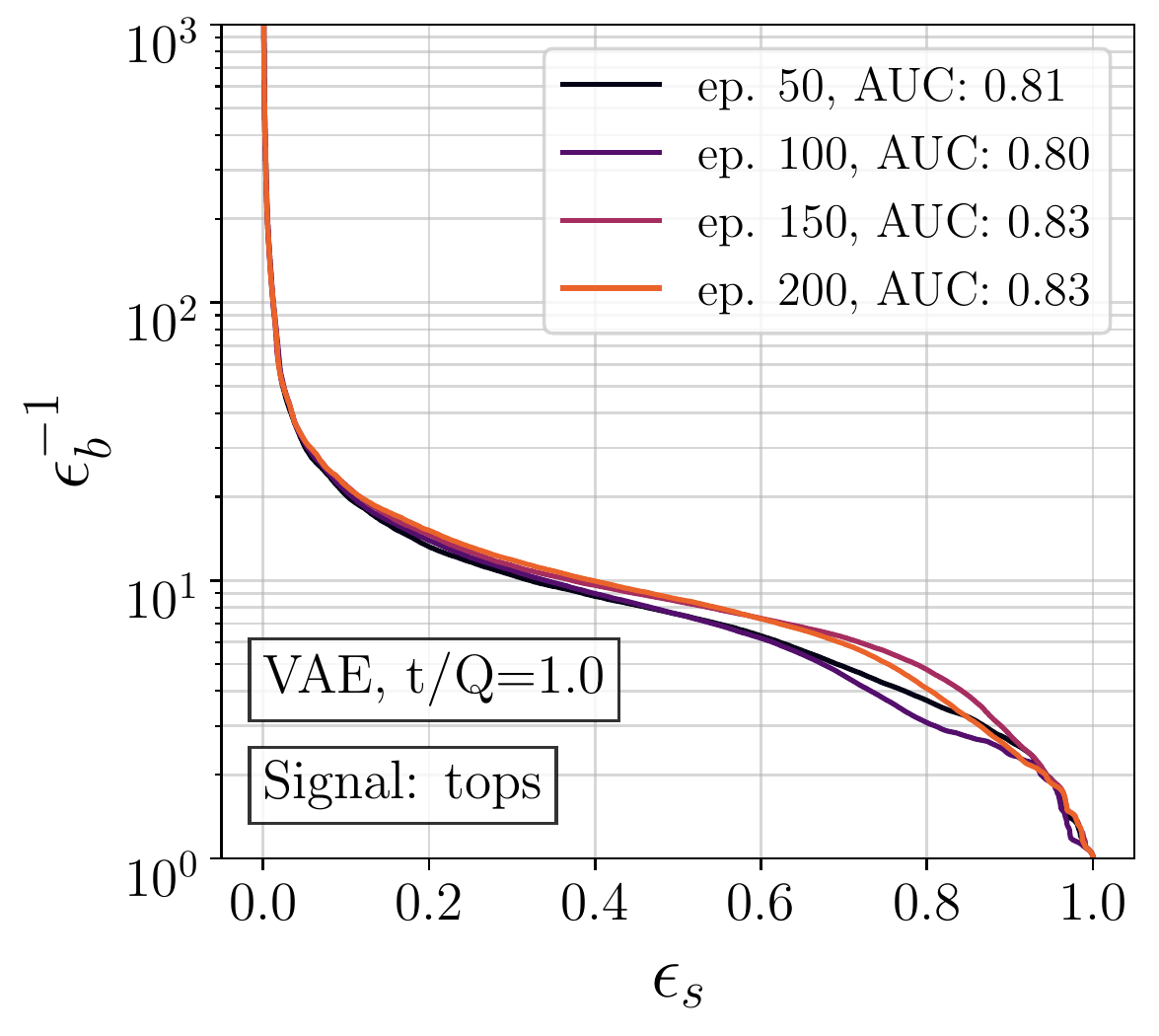}
\includegraphics[width=0.32\textwidth]{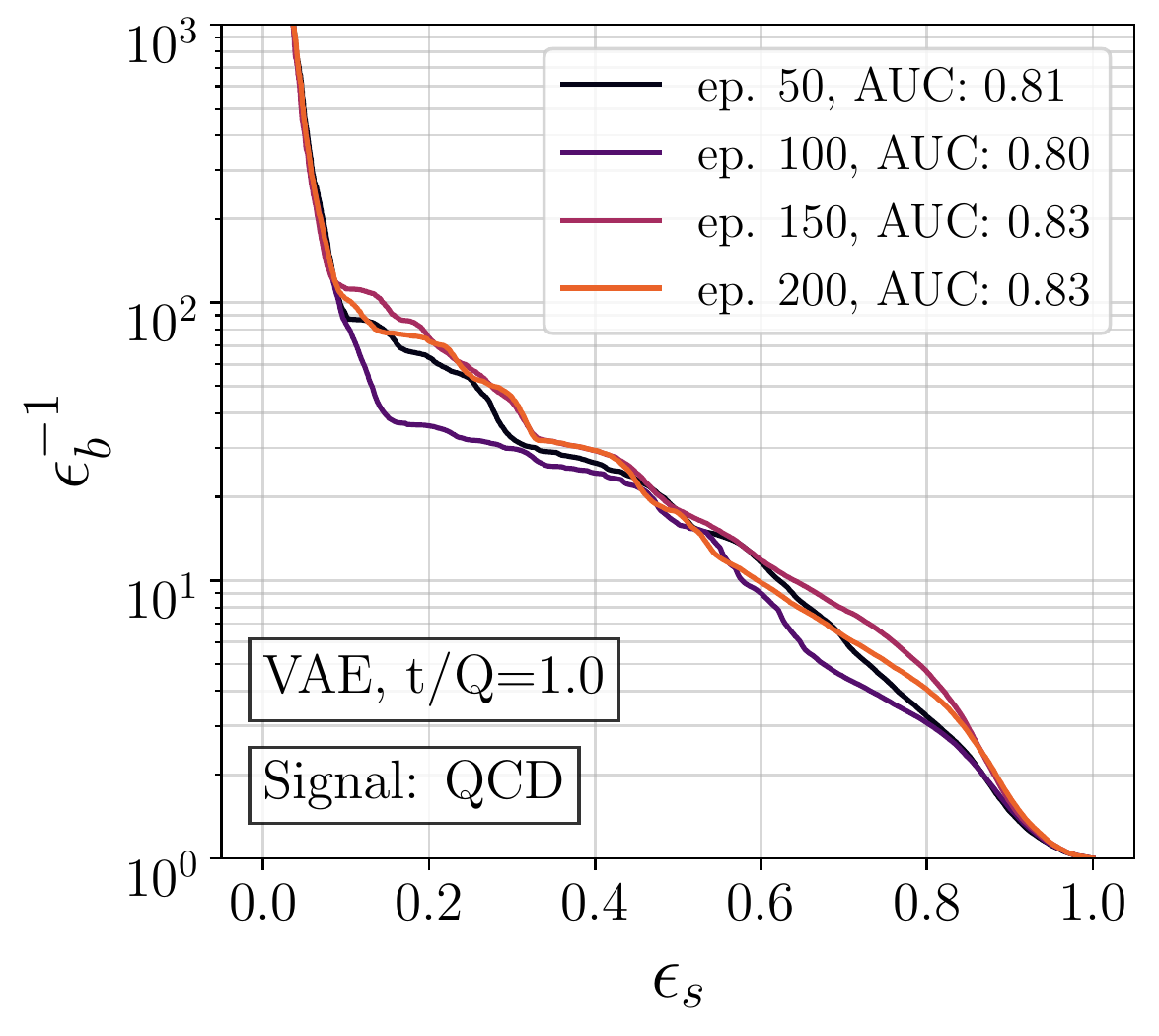}
\includegraphics[width=0.32\textwidth]{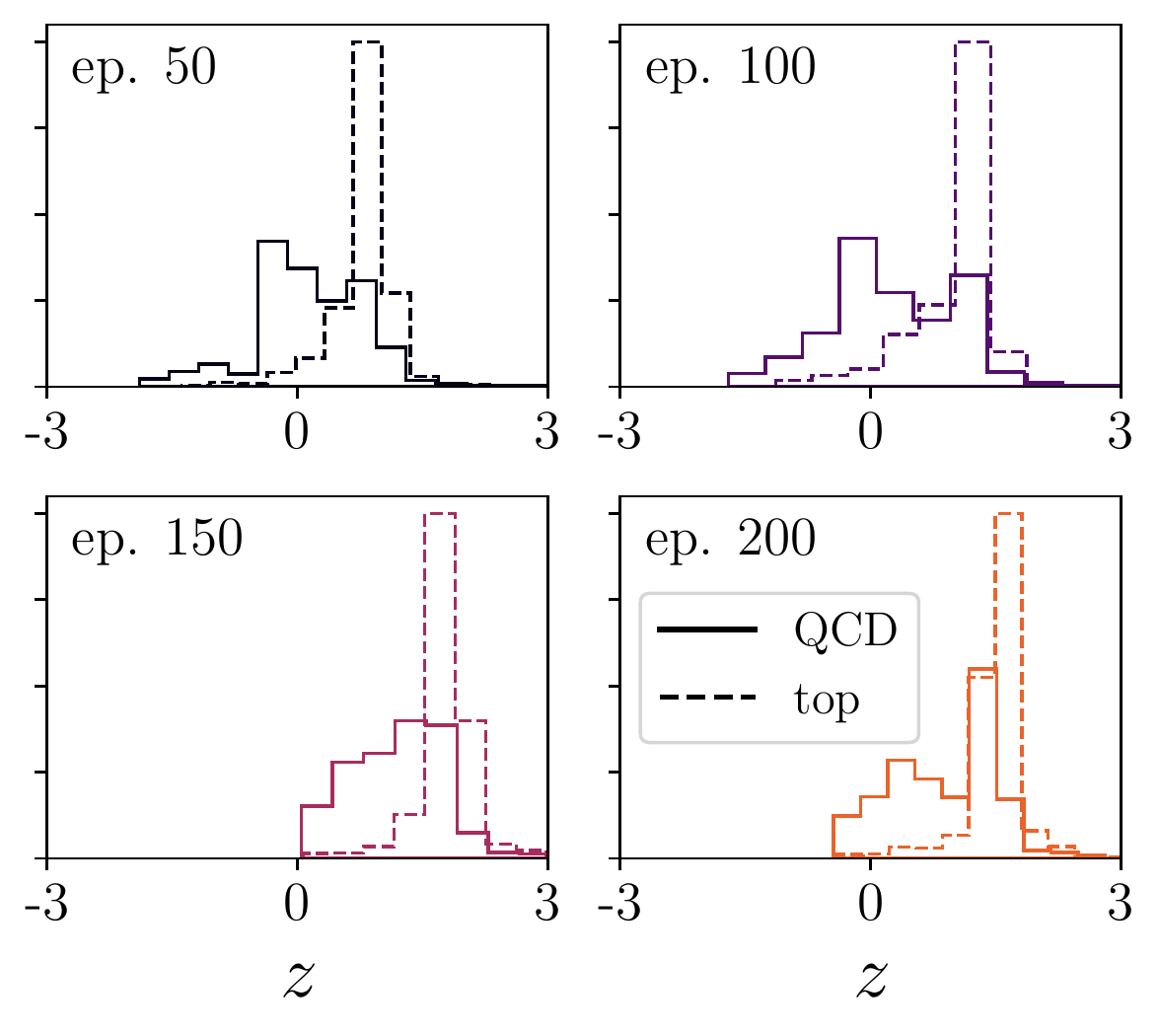}
\caption{Results for the 1D AE (top) and VAE (bottom).  In the large
  panels we show the ROC curves for tagging top and QCD as signal. In
  the small panels we show the distributions of the top and QCD jets
  in the bottleneck or latent space developed over the training.}
\label{fig:aevae}
\end{figure}
%----------------------------------------------------------

For illustration purposes, we start with a 1D bottleneck for the AE
and a 1D latent space for the VAE.  One of our main points will be
that the architecture of the network, in particular the latent space,
plays an important role in determining the type of information learned
by the network.  It has been
shown~\cite{Farina:2018fyg,Heimel:2018mkt} that with large bottlenecks
of around 30 nodes it is possible to reconstruct $40 \times 40$ jet
images with good accuracy, while at the same time achieving reasonable
top-tagging performance using the reconstruction error.  In these
cases, the latent space is unequivocally not a good classifier for top
jets because the bottleneck is too large.  With a bottleneck of this
size the network can accurately encode kinematic information about
each individual jet, allowing for an accurate per-jet reconstruction.
The trouble is that this kinematic information encoded in the
bottleneck does not tell us anything about the class the jet belongs
to, \ie top jet or QCD jet.  To learn this information the network
needs to learn how to reconstruct classes of jets, rather than
individual jets.  And to force the network to do this, we restrict the
amount of information that can be encoded in the bottleneck.  In the
case of the VAE this has a nice probabilistic interpretation, since
the mapping to the encoded space is performed by the posterior
distribution $q(z|x)$.  In 1D then the latent variable $z$ can be thought of as
a continuous generalisation of a class label.

We are interested in three aspects of the (V)AE classifiers: (i)
performance as a top-tagger, (ii) performance as a QCD-tagger, (iii)
stable encoding in the latent space.  To start, we restrict ourselves
to samples with equal numbers of 100k top and QCD jets.  We follow the
evolution of the loss and find that all models converge
before epoch 200.  The results for the AE and the VAE are shown in
Fig.~\ref{fig:aevae}. We immediately see that the AE bottleneck is not
a robust classifier; the classification is unstable and the encoding
in the bottleneck continues to evolve even after the loss function has
converged.  In the VAE plots we can see that the regularisation of the
latent space has a positive effect. The latent space of the VAE is
much more stable and structured than the AE bottleneck, even
converging to a representation in which the top jets are almost
clustered away from the QCD jets.

On the other hand, it is reasonable to assume that the most natural
latent representation for a data-set such as ours would be one which
naturally allows for a bi-modal structure.  By construction, this is
not achieved by the VAE with its unimodal Gaussian prior.

%%%%%%%%%%%%%%%%%%%%%%%%%%%%%%%%%%%%%%%%%%%%%%%%%%%%%%%%%%%%%%%%%%%%%%%%
\section{Gaussian-Mixture-VAE}
\label{sec:mvae}

While the standard VAE prior shapes a Gaussian latent space with a
single mode, a symmetric anomaly search would prefer a dual-mode
latent space to capture two classes of data. We introduce a Gaussian
mixture prior, consisting of two mixture components or modes, and show
how it performs on the benchmark from the previous section with balanced classes.  
We approximate the KL divergence between the Gaussian mixture prior and the 
latent representation of the events using a Monte Carlo estimate.
There are
papers using a Gaussian mixture as the VAE
prior~\cite{dilokthanakul2016deep,shao2020generalized,yang2019deep,guo2018multidimensional},
but most implementations are more complicated than ours, requiring
additional inference networks, as the authors wish to calculate an
analytical Kullback-Leibler (KL) divergence. In the most similar model
to ours~\cite{shu2016stochastic} the authors use an approximation to
the KL-term~\cite{hershey2007approximating} rather than a Monte Carlo
estimate.

%%%%%%%%%%%%%%%%%%%%%%%%%%%%%%%%%%%%%%%%%%%%%%%%%%%%%%%%%%%%%%%%%%%%%%%%
\subsubsection*{Gaussian mixture prior}
\label{sec:mvae_prior}

We impose a Gaussian mixture prior on the $n$-dimensional latent
space, where the Gaussian in each mixture has diagonal
covariance. Such a model has the following conditional likelihood
\begin{align}
  p(z|r) = \prod_{i=1}^n \; \frac{1}{\sqrt{2\pi} \sigma_{r,i}} \;
    \exp \left( -\frac{(z_i - \mu_{r,i})^2}{2\sigma^2_{r,i}} \right)
\end{align}
where $\mu_{r,i}$ is the mean of mixture component $r = 1~...~R$ in
dimension $i$, and $\sigma^2_{r,i}$ is the corresponding variance. The
prior is then simply calculated as $p(z) = \sum_r p(z|r) p(r)$, with
$p(r)$ being the weight of mixture component $r$.

We allow all means, variances, and mixture weights to be learned. The
means are learned as a $R \times n$ matrix $M$, the log variances as a
$R \times n$ matrix $V$, and the unnormalized log mixture weights as
a vector $a$ of length $R$. Since the mixture weights must be positive
and sum to one, we apply the softmax operation,
\begin{align}
  p(r) = \frac{e^{a_r}}{\sum_{r'} e^{a_{r'}}} \; .
\end{align}
With this notation, we can rewrite the
conditional likelihood as
\begin{align}
  p(z|r) = \prod_{i=1}^n \; \frac{1}{\sqrt{2\pi e^{V_{ri}}}} \;
  \exp \left( -\frac{(z_i - M_{ri})^2}{2e^{V_{ri}}} \right).
\end{align}
We can write this in log form,
\begin{align} 
\log \left( p(z|r) p(r) \right)
= \sum_{i=1}^n \left( -\frac{(z_i - M_{ri})^2}{2e^{V_{ri}}} - \frac{1}{2}V_{ri} \right) - \frac{n}{2}\log(2\pi) + \log \text{softmax}(a_r) \; ,
\label{eq:GMVAE-log-prob-prior}
\end{align}
from which we can apply the log-sum-exp operation to obtain the prior
log-likelihood, $\log p(z) = \log \sum_r \exp \log (p(z|r)p(r))$. In our experiments we always chose $R=2$.

%%%%%%%%%%%%%%%%%%%%%%%%%%%%%%%%%%%%%%%%%%%%%%%%%%%%%%%%%%%%%%%%%%%%%%%%
\subsubsection*{Mode-separation loss}
\label{sec:mvae_loss}

\begin{figure}[t]
\centering
\includegraphics[width=0.6\textwidth]{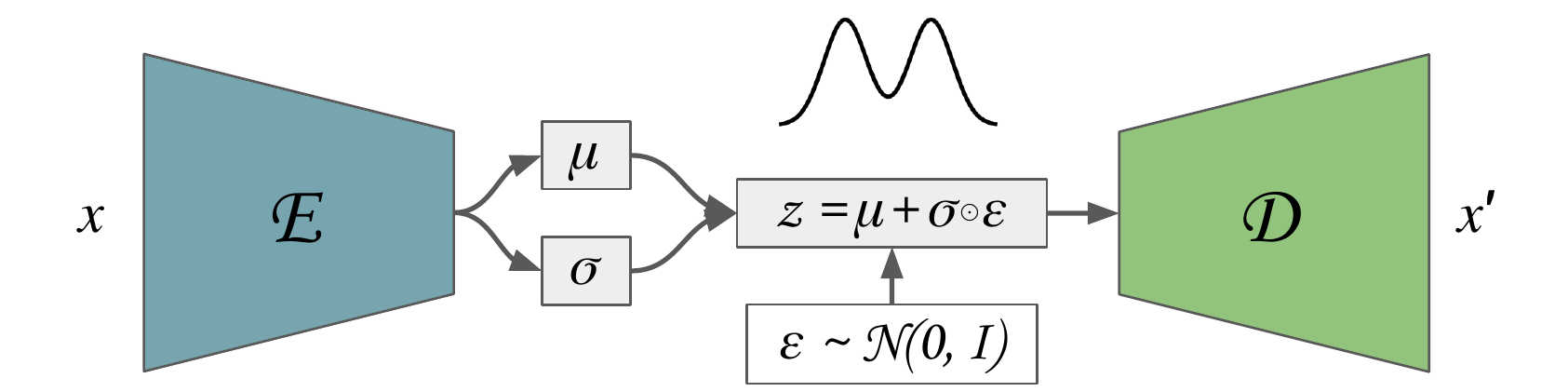} 
\caption{Architecture used for the GMVAE. The encoder and
  decoder are the same as for the (V)AE shown in
  Fig.~\ref{fig:aevae_archs_c}. In contrast to the standard VAE, the prior distribution is learnable and can be bimodal.}
\label{fig:gmvae_arch}
\end{figure}

As in a standard VAE, the GMVAE illustrated in
Fig.~\ref{fig:gmvae_arch} minimizes the negative ELBO loss,
\begin{align}
\loss = \XLangle - \Langle \log p_\theta(x|z) \Rangle_{q_\phi(z|x)} + \beta_\text{KL} D_\text{KL}(q_\phi(z|x), p(z)) \XRangle_{p_{\text{data}}(x)} \; ,
%\loss = \mathbb{E}_{p_{\text{data}}(x)}\left[ \mathbb{E}_{q_\phi(z|x)}\left[-\log p_\theta(x|z)\right] + D_\text{KL}(q_\phi(z|x) || p(z)) \right] \; ,
\label{eq:gmloss}
\end{align}
with a learnable variational encoder $q_\phi(z|x)$ and decoder
$p_\theta(x|z)$, where $\phi$ and $\theta$ represent the parameters of
the encoder and decoder, respectively.  The data-set consisting of a
mixture of QCD and top jets here is indicated by $x$.
The $\beta_\text{KL}$ term allows us to control the relative influence
of the reconstruction and latent loss terms in the gradient updates.
Exactly as was done for the VAE and AE, we use the mean squared error for the reconstruction loss and set $\beta_{KL}=10^{-4}$.

%----------------------------------------------------------
%----------------------------------------------------------

The first term in Eq.\eqref{eq:gmloss} is treated in the standard way
as a reconstruction loss, where a single sample is drawn from
$q_\phi(z|x)$ to estimate the expectation. The second term comes with
the challenge that for a Gaussian mixture we cannot calculate the
KL-divergence analytically anymore. We instead estimate it using Monte
Carlo samples~\cite{tomczak2018vae} and start by re-writing it as
\begin{align}
  D_\text{KL}(q_\phi(z|x),p(z))
%  = \mathbb{E}_{q_\phi(z|x)}\left[\log\frac{q_\phi(z|x)}{p(z)}\right]
  = \Langle \log q_\phi(z|x) \Rangle_{q_\phi(z|x)}
  - \Langle \log p(z) \Rangle_{q_\phi(z|x)} \; .
%  = \mathbb{E}_{q_\phi(z|x)}\left[\log q_\phi(z|x)\right] - \mathbb{E}_{q_\phi(z|x)}\left[\log p(z)\right].
\end{align}
The first contribution is the negative entropy
%$-H(q(z|x))$
of a Gaussian, which has the analytical form
\begin{align}
  \Langle \log q_\phi(z|x) \Rangle_{q_\phi(z|x)}
%  \mathbb{E}_{q_\phi(z|x)}\left[\log q_\phi(z|x)\right]
  %H(q(z|x))
  = - \frac{n}{2}(1 + \log(2\pi)) - \frac{1}{2} \sum_{i=1}^n \log\sigma^2_i \; .
\end{align}
The second contribution to $D_\text{KL}$ can be estimated by a single
Monte Carlo sample from $q_\phi(z|x)$, as in the reconstruction term,
using the log-likelihood of the imposed prior derived in
Eq.\eqref{eq:GMVAE-log-prob-prior}.

In testing this two-component mixture prior we observe that the mixture
components collapse onto a single mode. To prevent this, we add a
repulsive force between the modes, calculated as a function of the
Ashman distance~\cite{ashman1994detecting}. The Ashman-$D$ between two Gaussian
distributions in one dimension is
\begin{align}
D^2 = \frac{2(\mu_1 - \mu_2)^2}{\sigma^2_1 + \sigma^2_2} \; .
\end{align}
A value of $D > 2$ indicates a clear mode separation. The distance can be
extended to two $n$-dimensional Gaussians with diagonal covariance
matrices as
\begin{align}
  D^2 = \sum_{i=1}^n \frac{2(\mu_{1i} - \mu_{2i})^2}{\sigma^2_{1i} + \sigma^2_{2i}} \; ,
\end{align}
where again $D > 2$ indicates clear separation.  We encourage
bi-modality in our latent space by adding the loss term
\begin{align}
\loss_A = -\lambda_A \tanh \frac{D}{2} \; .
\end{align}
The tanh function eventually saturates and stops pushing apart the
modes.

%----------------------------------------------------------
\begin{figure}[t]
\centering
\includegraphics[width=0.32\textwidth]{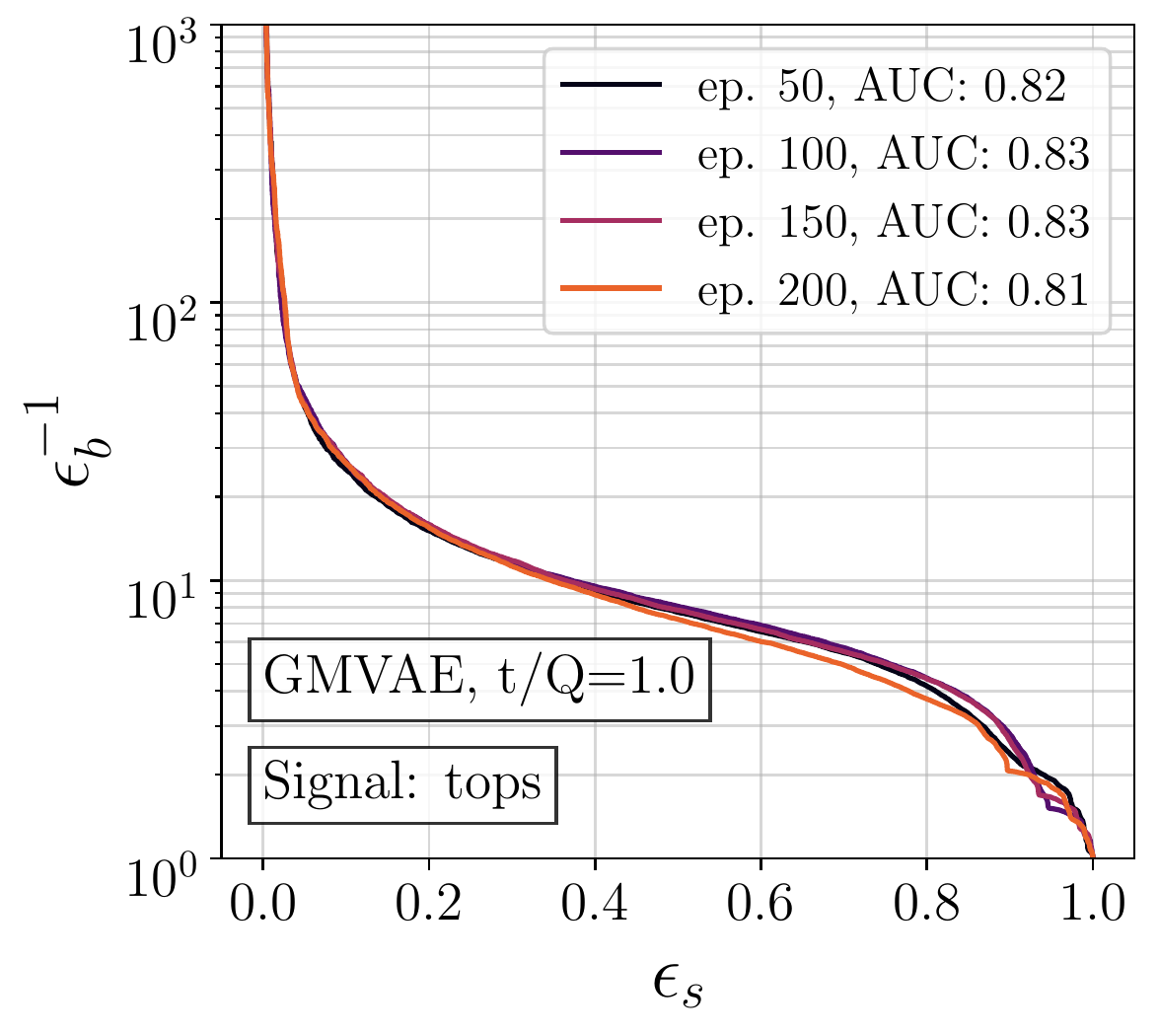}
\includegraphics[width=0.32\textwidth]{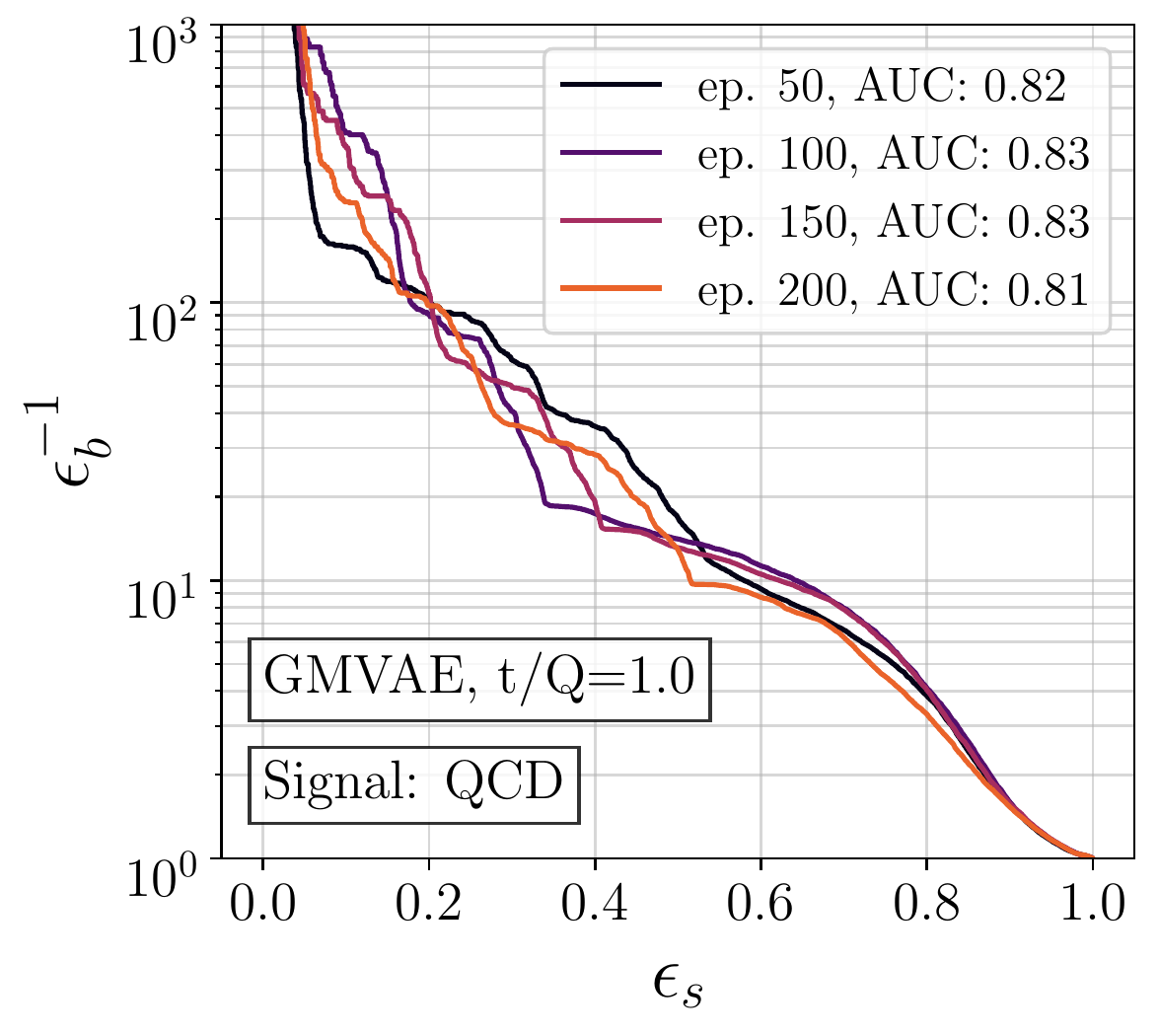}
\includegraphics[width=0.32\textwidth]{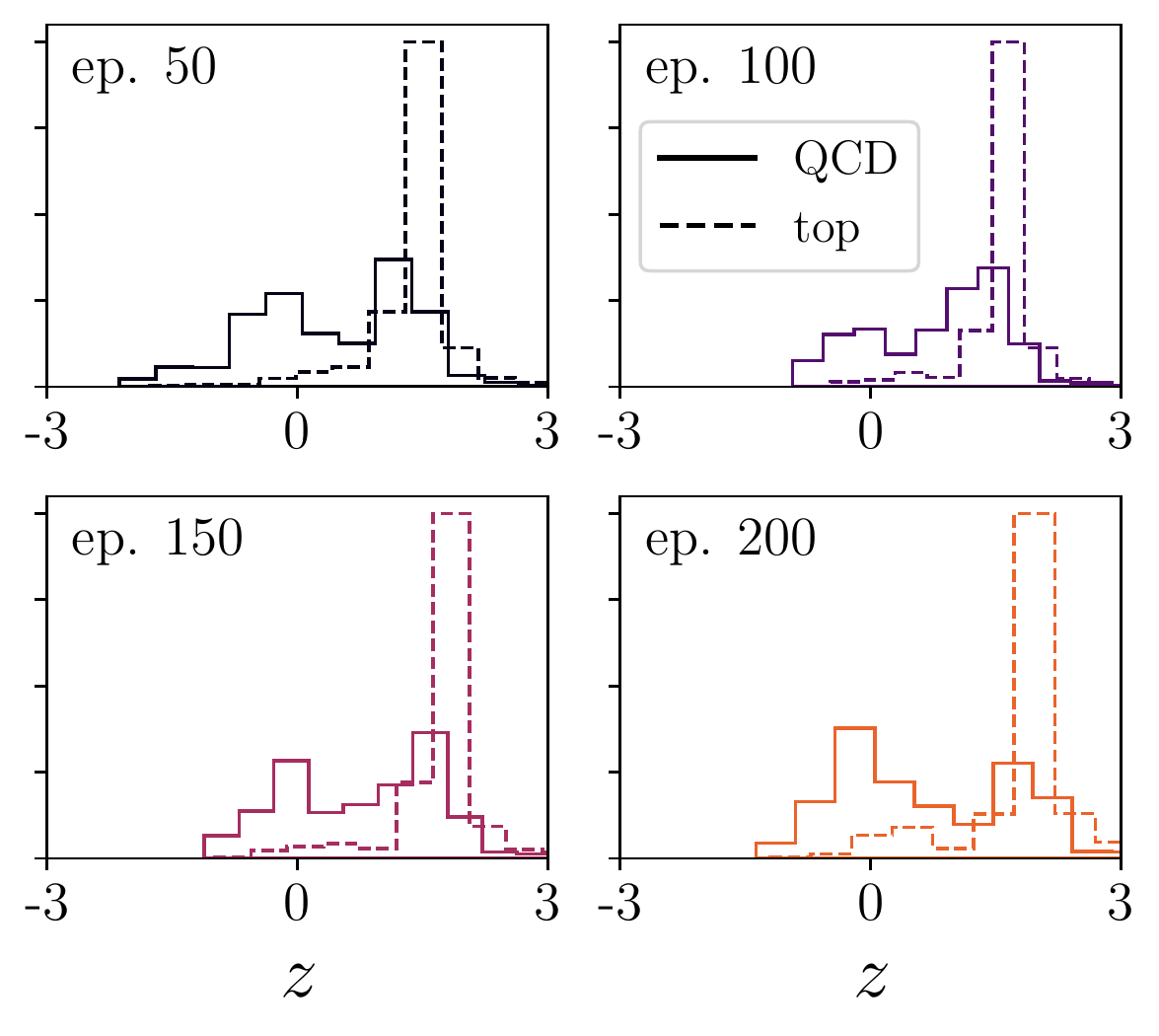}
\caption{Results for the 1D GMVAE.  In the large panels we show the
  ROC curves for tagging top and QCD as signal. In the small panels we
  show the distributions of the top and QCD jets in the latent space
  developed over the training.}
\label{fig:gmvae}
\end{figure}
%----------------------------------------------------------

%%%%%%%%%%%%%%%%%%%%%%%%%%%%%%%%%%%%%%%%%%%%%%%%%%%%%%%%%%%%%%%%%%%%%%%%
\subsubsection*{Latent-space tagging}
\label{sec:mvae_latent}

For this study of the GMVAE we use the exact same architecture as for
the VAE except for the new latent space prior, see
Fig. \ref{fig:gmvae_arch}.  We present the results for the
classification of QCD and top jets in latent space in
Fig. \ref{fig:gmvae}.  It's clear that the Gaussian mixture prior is
shaping the latent space, and the network has the same benefit as the
VAE in that the latent representation of the jets stabilises as the
training converges.  However we do not see an increase in performance
in going from the VAE to the GMVAE.  The reason for this is that while
the top jets occupy just one mode in latent space, the QCD jets occupy
both.  Upon inspecting the jets which are assigned to each mode we
find that this assignment is based on the amount of pixel activity
within the jet, rather than on the specific features in the jet.  We
also experimented with using larger latent spaces, keeping the number
of modes at two.  This resulted in much better reconstruction accuracy
and the bi-modal latent space structure persisted, however the
performance of the latent space classification did not improve.

%%%%%%%%%%%%%%%%%%%%%%%%%%%%%%%%%%%%%%%%%%%%%%%%%%%%%%%%%%%%%%%%%%%%%%%%
\section{Dirichlet-VAE}
\label{sec:dvae}

Finally, we follow an alternative approach to provide the VAE with a
geometry that leads to a mode separation.  We do this by using the Dirichlet
distribution, a family of continuous multivariate probability
distributions, as the latent space prior.
It has the following probability density function:
\begin{align}
  \mathcal{D}_\alpha(r)
  = \frac{ \Gamma( \sum_i \alpha_i ) }{ \prod_i \Gamma( \alpha_i ) } \; 
  \prod_i r_i^{\alpha_i-1},
  \qquad \text{with} \; i=1~...~R \; .
\end{align}
This $R$-dimensional Dirichlet is parameterised by $R$
hyper-parameters $\alpha_i > 0$, while the function itself is defined
on an $R$-dimensional simplex.  The Dirichlet distribution is
conjugate to the multinomial distribution and is commonly used in
Bayesian statistics as a prior in multinomial mixture models.  The
expectation values for the sampled vector components are $\langle r_i
\rangle =\alpha_i/\sum_j \alpha_j$, so as a prior it will create a
hierarchy among different mixture components.  In our application, it
imposes a compact latent space, whose latent dimensions can be
interpreted as mixture weights in a multinomial mixture
model~\cite{srivastava2017autoencoding,joo2019dirichlet}.

%%%%%%%%%%%%%%%%%%%%%%%%%%%%%%%%%%%%%%%%%%%%%%%%%%%%%%%%%%%%%%%%%%%%%%%%
\subsubsection*{Loss and network}

For a Dirichlet structure in the latent space, the re-parametrization
trick requires some attention. We opt to use a softmax Gaussian
approximation to the Dirichlet distribution~\cite{srivastava2017autoencoding}, because the
re-parametrization of Gaussian sampling is straightforward and
stable,
\begin{align}
  r \sim \text{softmax} \, \mathcal{N}(z;\tilde{\mu},\tilde{\sigma}) &\approx \mathcal{D}_\alpha(r) \notag \\
  \text{with} \quad 
\tilde{\mu}_i & = \log\alpha_i - \frac{1}{R}\sum_{i}\log\alpha_i		\notag  \\
\tilde{\sigma}_i & = \frac{1}{\alpha_i}\left(1-\frac{2}{R}\right) + \frac{1}{R^2}\sum_i\frac{1}{\alpha_i} \; .
\label{eq:softdir}
\end{align}
With this approximation the encoder network $q_\phi(r|x)$ plays the same role
as in the VAE and the GMVAE, with the encoder outputs
corresponding to the means and variances of the Gaussians in the softmax
approximation.

The loss function of the Dirichlet-VAE (DVAE) includes the usual
reconstruction loss and latent loss.  For the reconstruction loss we
use the cross-entropy between the inputs and the
outputs~\cite{srivastava2017autoencoding}.  The latent loss is given
by the KL-divergence between the per-jet latent space representation
and the Dirichlet prior, with a pre-factor $\beta_\text{KL}$.  
It is easily
calculated for the Gaussians in the softmax approximation of the
Dirichlet distribution and the Gaussians defined by the encoder
output~\cite{srivastava2017autoencoding},
\begin{align}
\loss& = \XLangle - \Langle \log p_\theta(x|r) \Rangle_{q_\phi(r|x)} + \beta_\text{KL} D_\text{KL}(q_\phi(r|x),\mathcal{D}_\alpha(r)) \XRangle_{p_{\text{data}}(x)} \; , \nonumber \\
& D_\text{KL}(q_\phi(r|x),\mathcal{D}_\alpha(r)) = \frac{1}{2}\sum_{i=1}^{R}\left(\frac{\tilde{\sigma}_i^2}{\sigma_i^2} + \frac{(\tilde{\mu}_i-\mu_i)^2}{\sigma_i^2} - R - \log\frac{\sigma_i^2}{\tilde{\sigma}_i^2}\right).
\label{eq:loss}
\end{align}
%

%----------------------------------------------------------
\begin{figure}[t]
\centering
\includegraphics[width=0.8\textwidth]{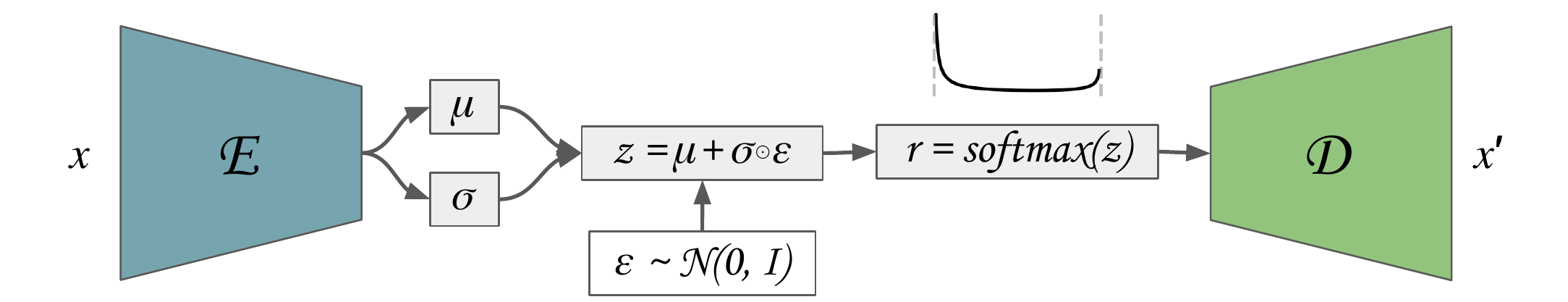}
\caption{Architecture for the Dirichlet VAE. 
The approximate Dirichlet prior is indicated by the softmax step and the bi-modal distribution shown above it.
}
\label{fig:dvae_arch}
\end{figure}
%----------------------------------------------------------

Unlike with the GMVAE, we do not make the parameters of the prior learnable.
We employ a very simple DVAE architecture, shown in
Fig.~\ref{fig:dvae_arch}.  The encoder is a fully connected network
with 1600 inputs, corresponding to the $40 \times 40$ image, and $2R$
outputs, where $R$ refers to the dimension of the Dirichlet latent
space.  We also insert a hidden layer of 100 nodes with SeLU
activations and use linear activations on the output layer.  The
decoder is also a fully connected network with $R$ nodes in the input
layer, 1600 nodes in the output layer and no hidden layers.  We do not
allow biases in the decoder, and apply a softmax activation to the
output layer\footnote{The code implementation used for the DVAE can be found at\newline \href{https://github.com/bmdillon/jet-mixture-vae}{https://github.com/bmdillon/jet-mixture-vae}.}.

The Dirichlet latent space has the advantage that the sampled $r_i$
can be interpreted as mixture weights of mixture components describing
the jets.  If we view the DVAE as an inference tool for a multinomial
mixture model, these mixture components correspond to probability
distributions over the image pixels.  These distributions enter into
the likelihood function of the model, which is parameterised by the
decoder network.  In a simple example with two mixtures, $R=2$, the
latent space is described by $r_1\in[0,1]$, with $r_0 \!=\! 1-r_1$
fixed.  The pure component distributions are given by
$p_\theta(x|r_1\!=\!0)$ and $p_\theta(x|r_1\!=\!1)$,
and any output for $r_1 \in[0,1]$ is a combination of these two
distributions. Our simple decoder architecture exactly mimics this
scenario.  In the absence of hidden layers and biases the decoder
network has exactly $R\times 1600$ parameters to describe the two
mixture component distributions.  One caveat is that we apply the
softmax activation to the linear combination of outputs on the output
layer, rather than to the outputs corresponding to each mixture
separately.  In Ref.~\cite{srivastava2017autoencoding} it is explained
how this is related to the application of an ensemble technique
called product of experts.

%%%%%%%%%%%%%%%%%%%%%%%%%%%%%%%%%%%%%%%%%%%%%%%%%%%%%%%%%%%%%%%%%%%%%%%%
\subsubsection*{Latent-space tagging}
\label{sec:dvae_latent}

%----------------------------------------------------------
\begin{figure}[t]
\includegraphics[width=0.32\textwidth]{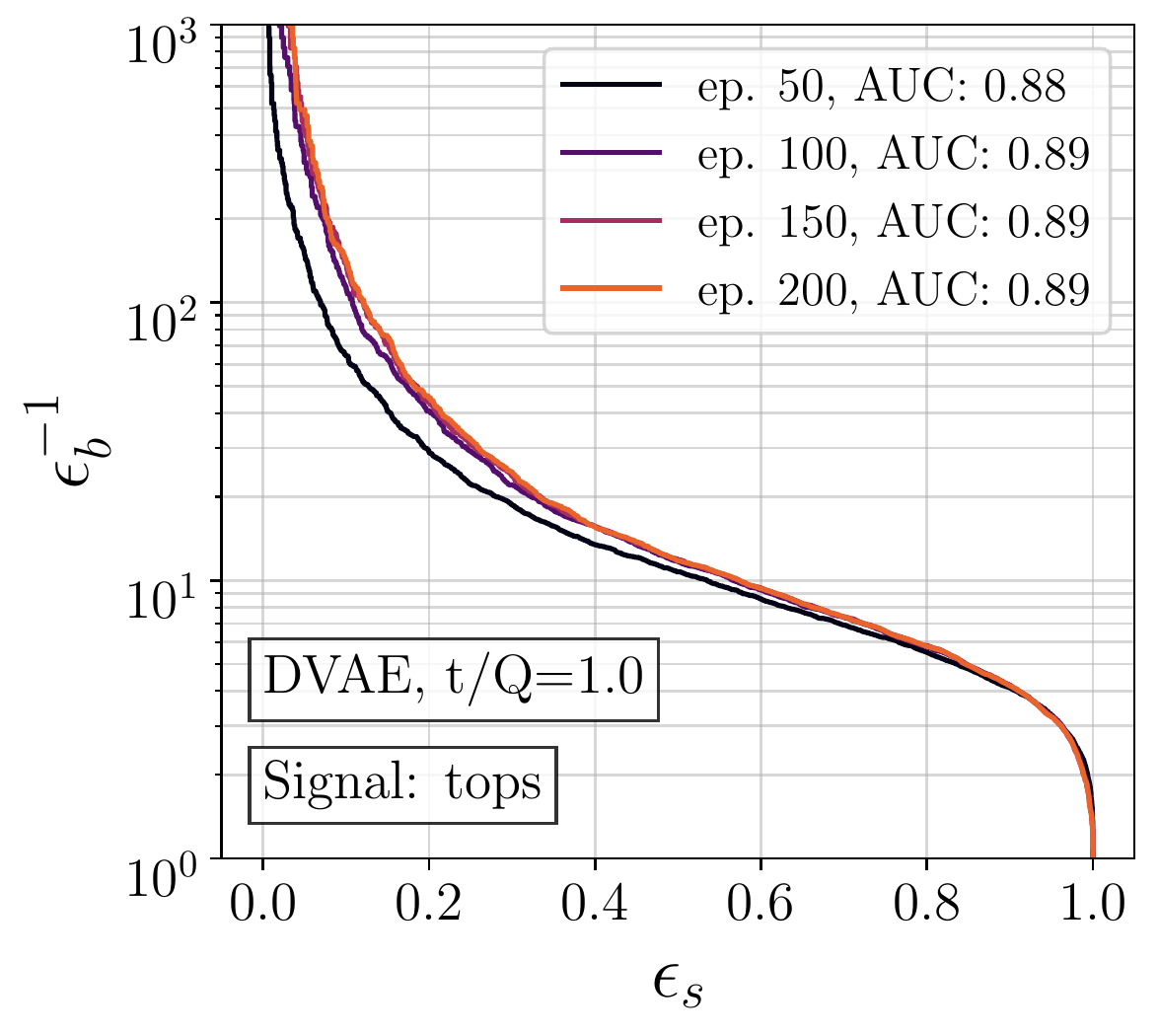}
\includegraphics[width=0.32\textwidth]{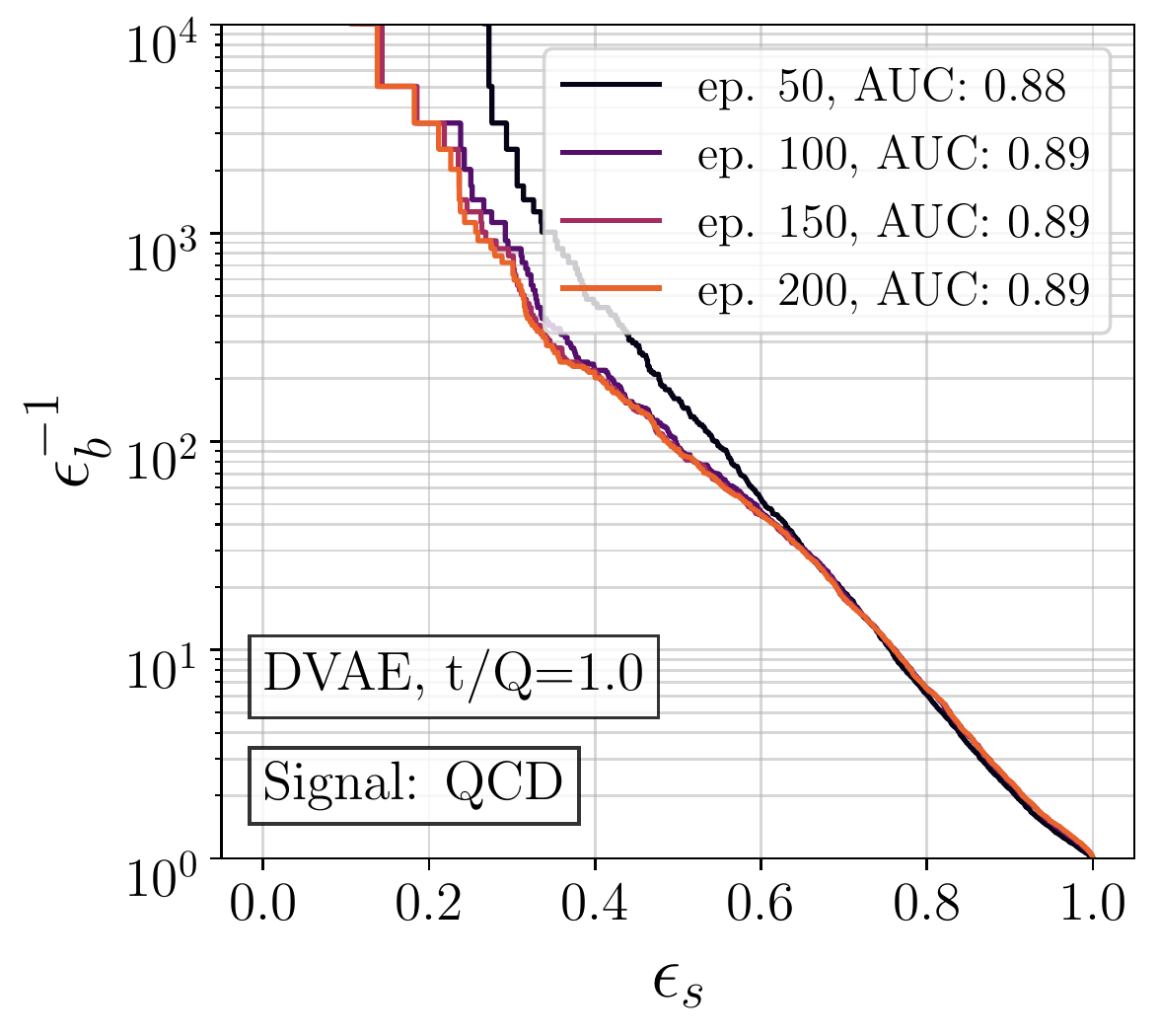}
\includegraphics[width=0.32\textwidth]{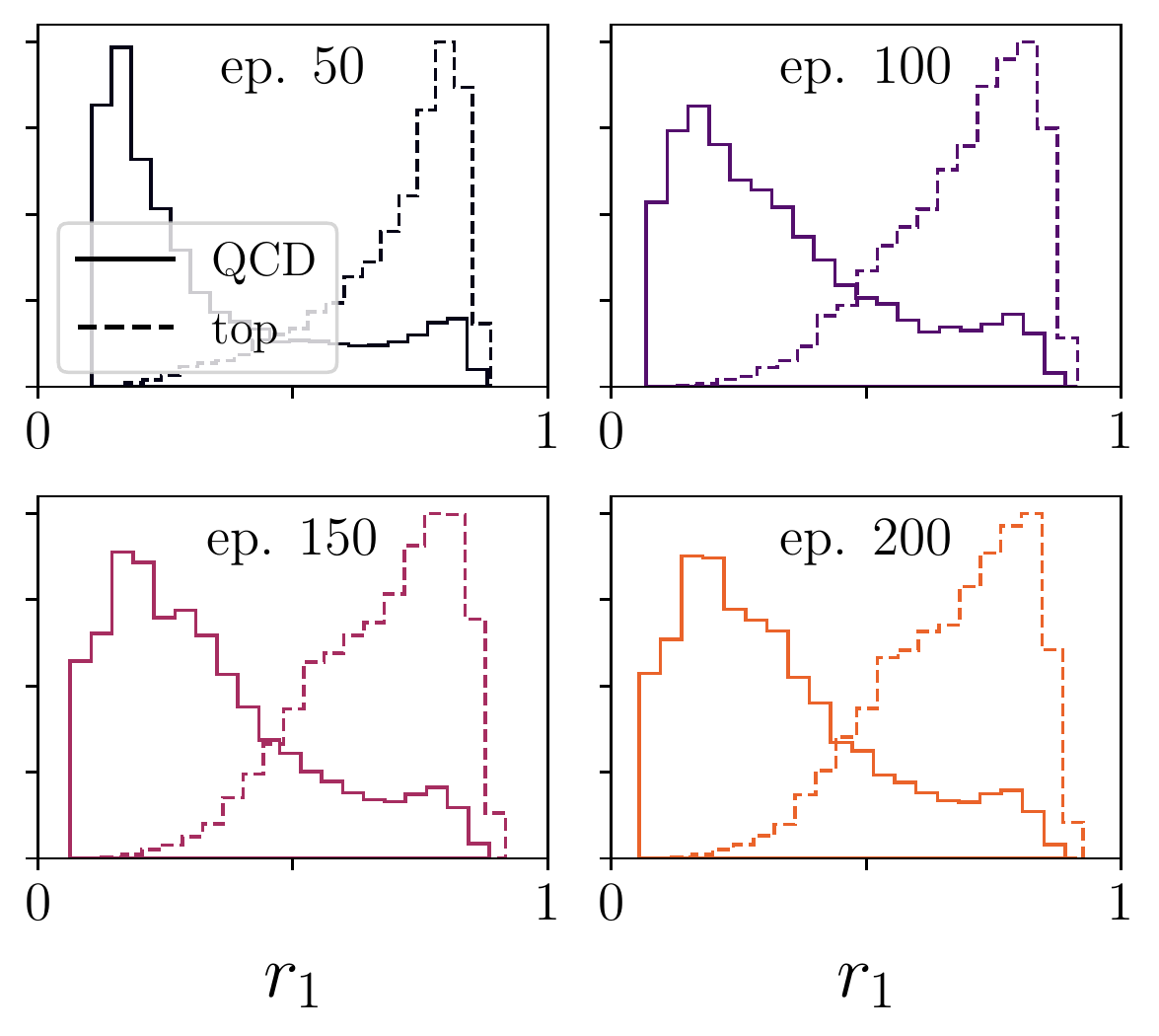}
\caption{Results for the DVAE.  In the large panels we show the
  ROC curves for tagging top and QCD as signal. In the small panels we
  show the distributions of the top and QCD jets in the latent space
  developed over the training.}
\label{fig:dvaeplot}
\end{figure}
%----------------------------------------------------------

%----------------------------------------------------------
\begin{figure}[b!]
\centering
\includegraphics[width=0.32\textwidth]{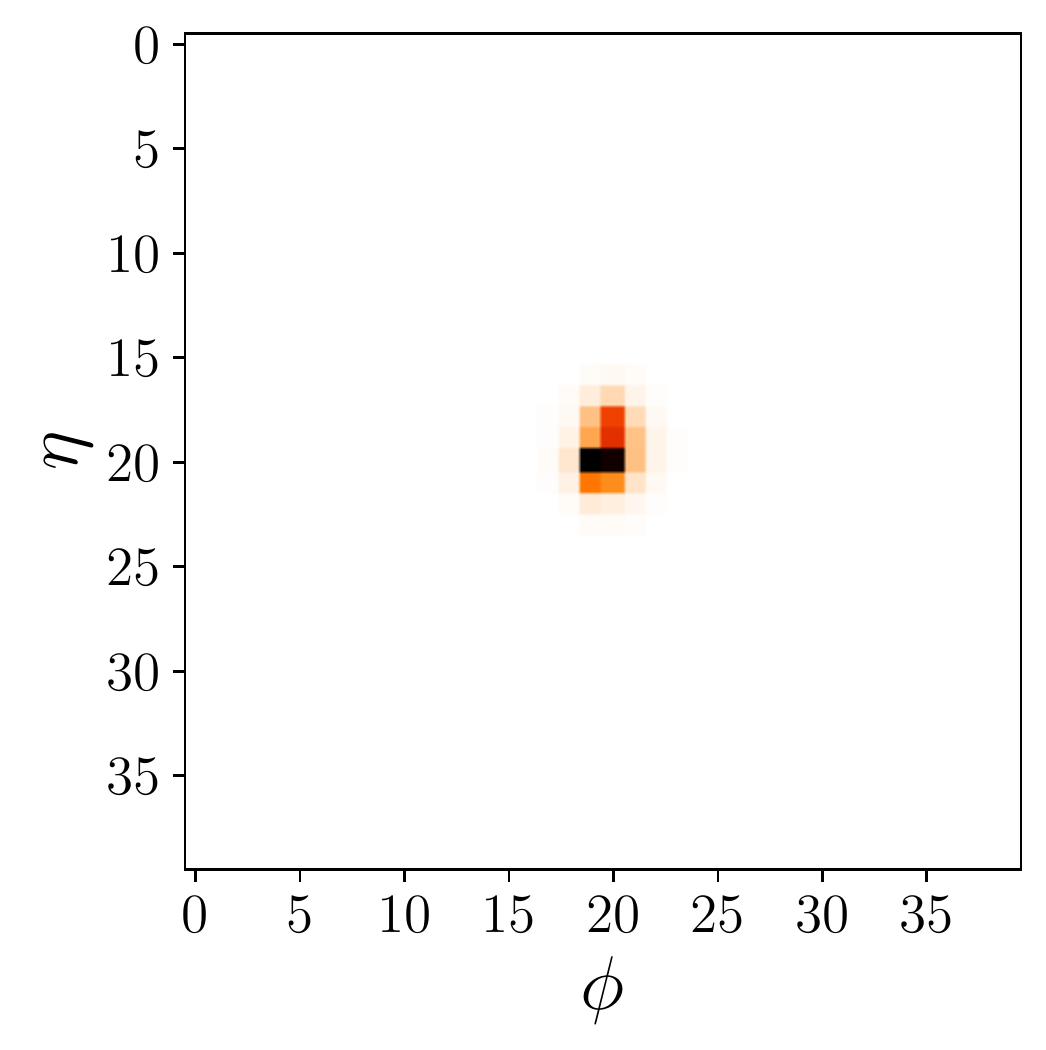}
\includegraphics[width=0.32\textwidth]{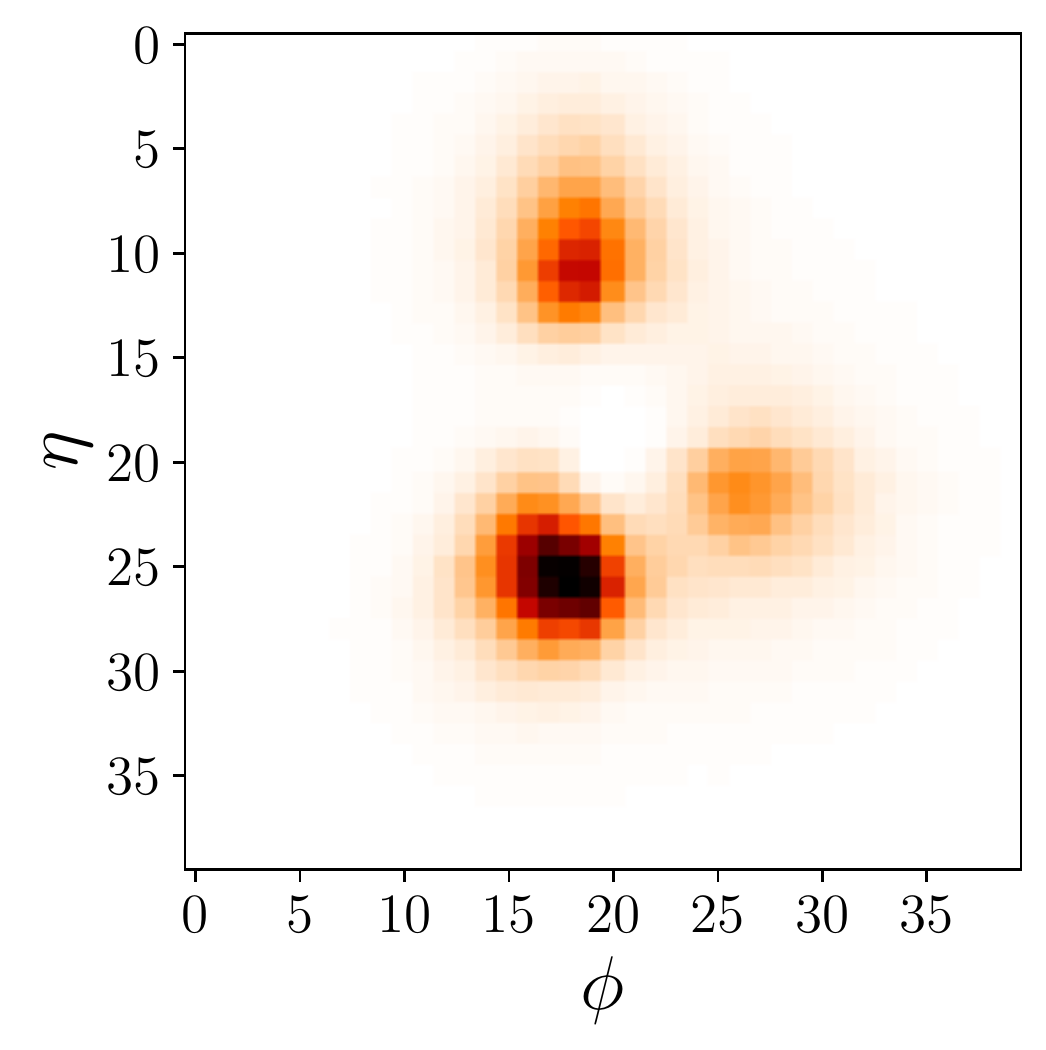}
\caption{Visualisation of the decoder weights, interpreted as the mixture distributions for $p_\theta(x|r_1\!=\!0)$ (left)
  and $p_\theta(x|r_1\!=\!1)$ (right) learned by the DVAE decoder.}
\label{fig:dvaemixtures}
\end{figure}
%----------------------------------------------------------

To compare with the AE and VAE studies in Fig.~\ref{fig:aevae} and the
GMVAE in Fig.~\ref{fig:gmvae} we start by training on 100k QCD and top
jets each.  We use a 2D Dirichlet latent space with $\alpha_{1,2} =
1$, where due to the simplex nature of the Dirichlet only one latent
space variable is independent.  We choose $\beta_\text{KL}\!=\!0.1$ so
that the prior has a large effect on the training.  
We compare QCD and top tagging and show
the latent space distributions in Fig.~\ref{fig:dvaeplot}.  The
visualisations of the decoder weights, \ie the mixture component
distributions $p_\theta(x|r_0=1)$ and
$p_\theta(x|r_0=0)$, are shown in
Fig.~\ref{fig:dvaemixtures}.

The advantages of the DVAE are immediately clear.  First, the
performance improves from an AUC of $0.83$ for the VAE to $0.89$ for
the DVAE.  Second, the latent space of the DVAE is a better and more
stable description of the multi-class jet sample. It quickly settles
into a clear bi-modal pattern with equal importance given to both
modes.  The QCD mode peaks at $r_1=0$, indicating that the
$p_\theta(x|r_1\!=\!0)$ mixture describes QCD jets, while the
top mode peaks at $r_1 = 1$, indicating that the
$p_\theta(x|r_1\!=\!1)$ mixture describes top jets.  This is
confirmed by visualizing the decoder weights in
Fig.~\ref{fig:dvaemixtures}.  Comparing these to the truth-level jet
images in Fig.~\ref{fig:jetimages} we see the clear correspondence
between the learned mixture distributions and the actual images, so
the DVAE has learned to separate the top-specific and QCD-specific
features into the different mixtures.

%%%%%%%%%%%%%%%%%%%%%%%%%%%%%%%%%%%%%%%%%%%%%%%%%%%%%%%%%%%%%%%%%%%%%%%%
\subsubsection*{Anomaly detection}
\label{sec:dvae_anom}

%----------------------------------------------------------
\begin{figure}[t]
\includegraphics[width=0.32\textwidth]{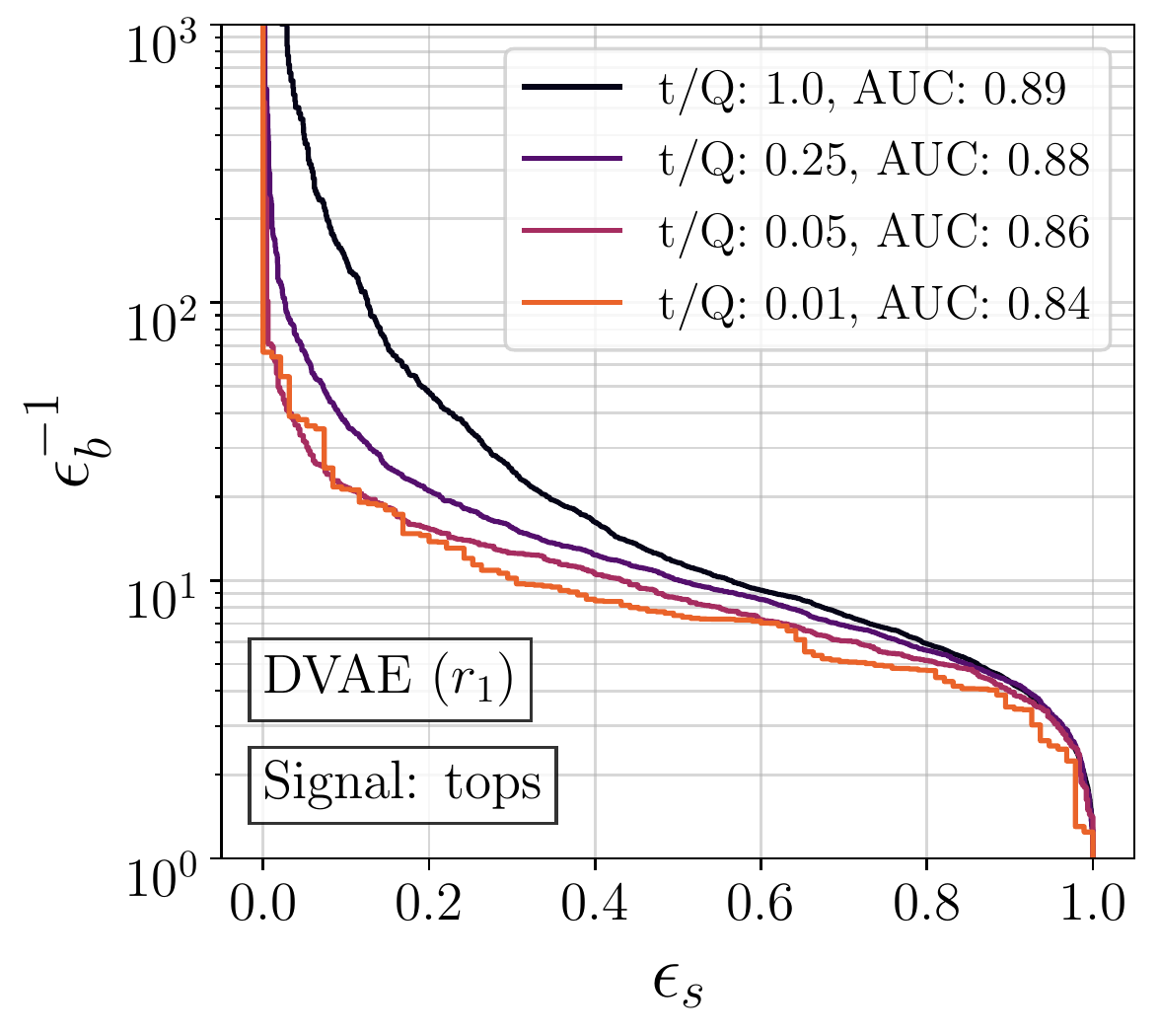}
\includegraphics[width=0.32\textwidth]{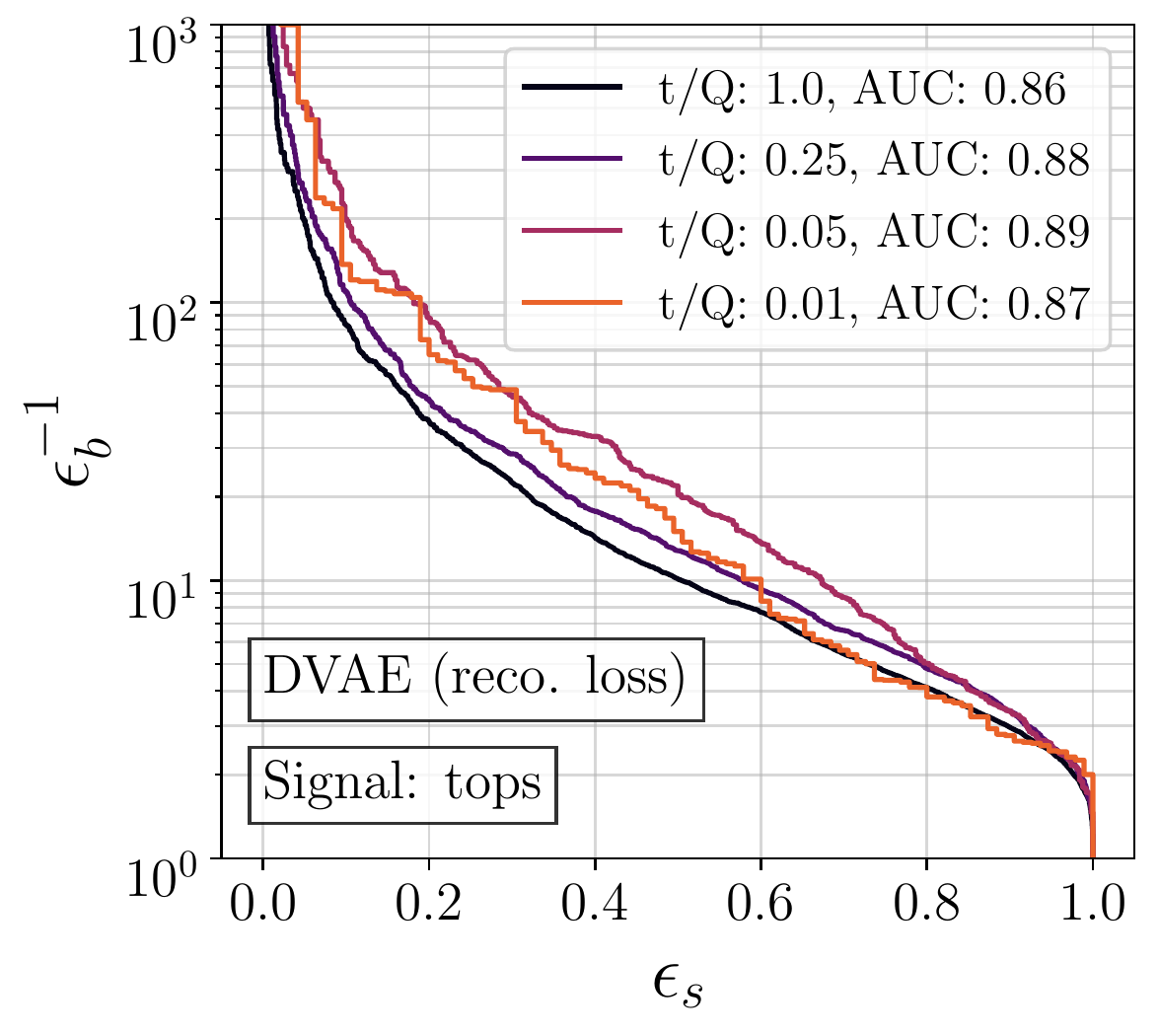} 
\includegraphics[width=0.32\textwidth]{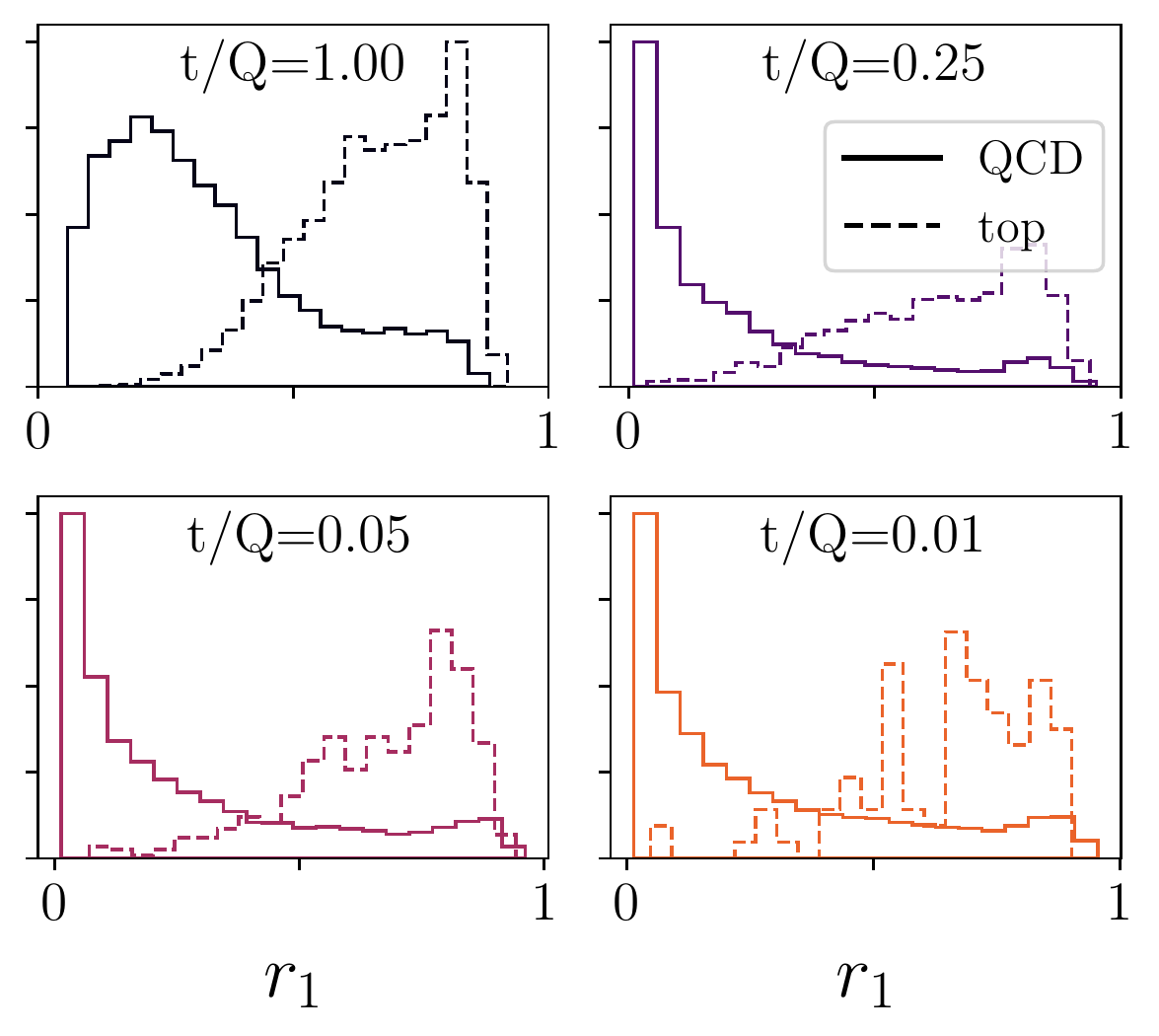} \\
\includegraphics[width=0.32\textwidth]{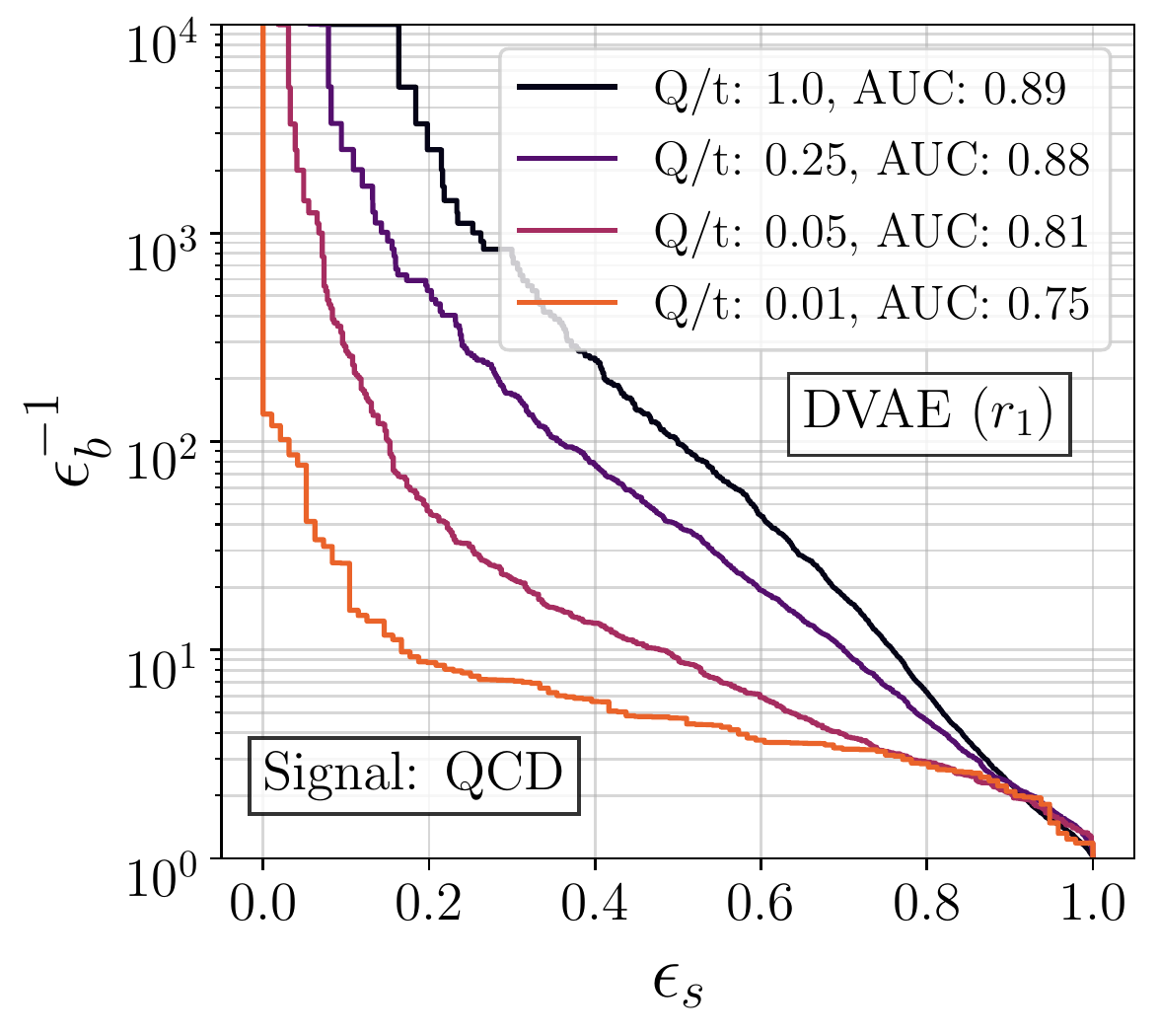}
\includegraphics[width=0.32\textwidth]{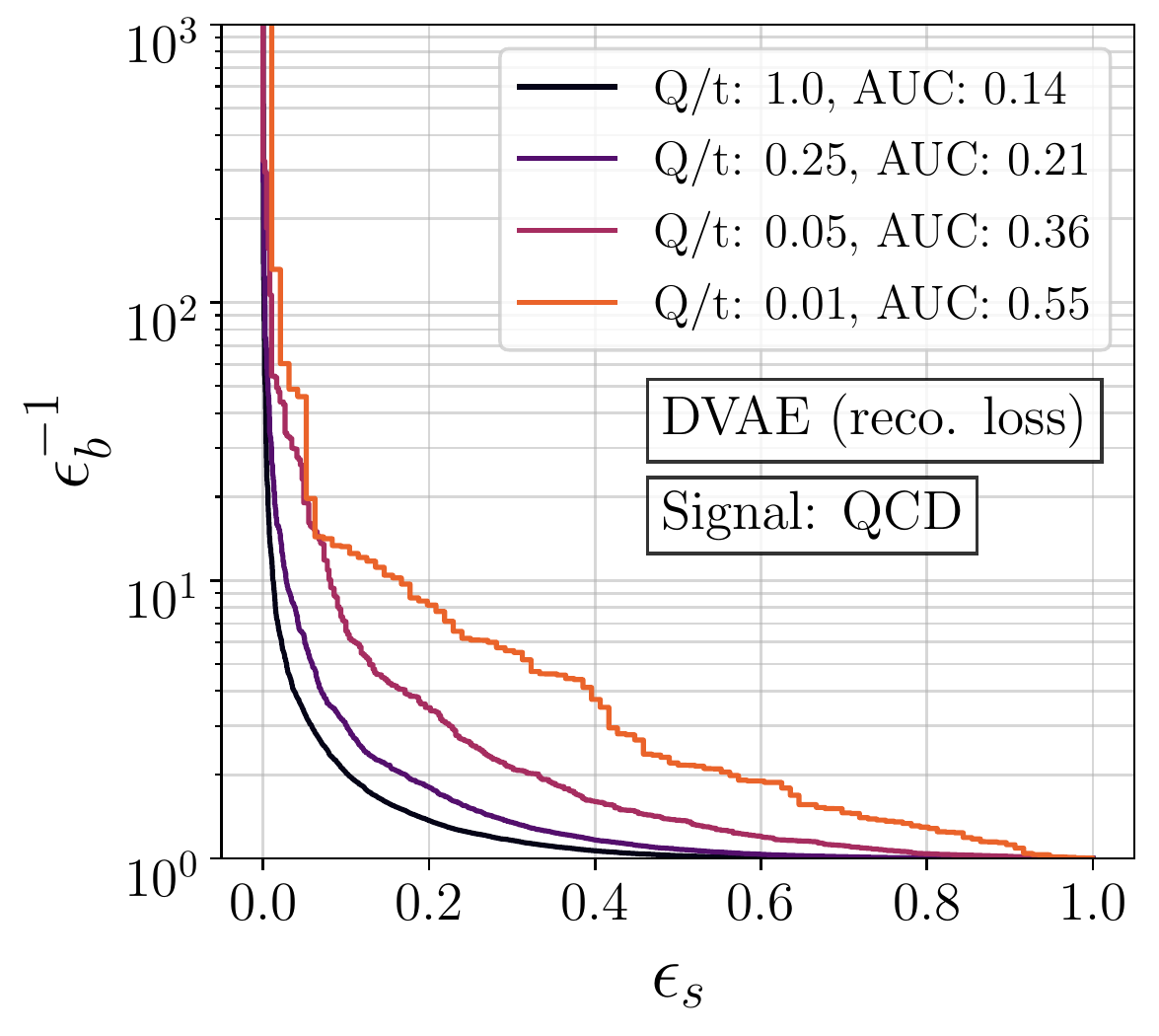}
\includegraphics[width=0.32\textwidth]{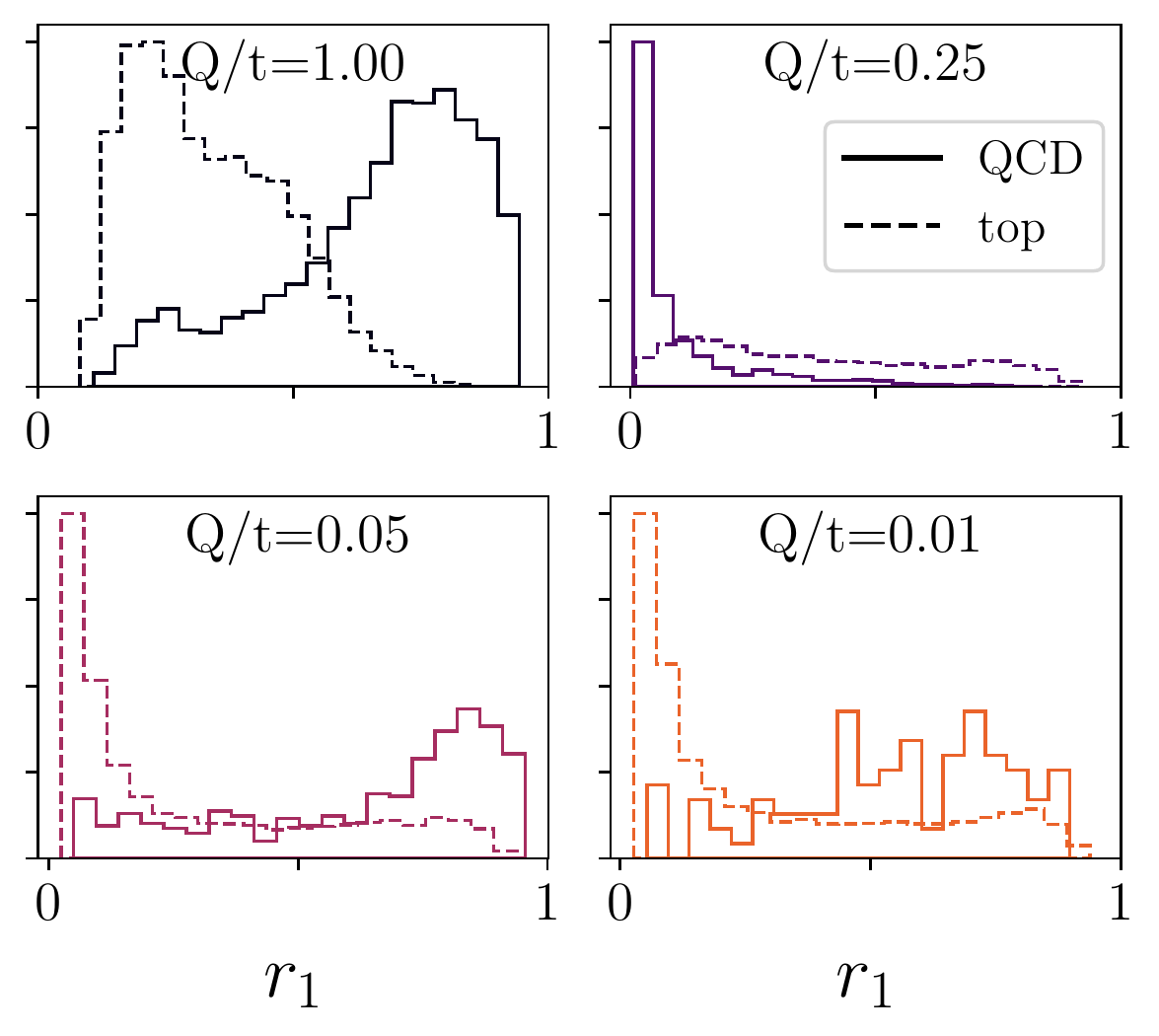}
\caption{DVAE results for various amounts of anomalous top (upper) and
  QCD (lower) jets in the sample, where the mixture weights (left) and
  the reconstruction loss (middle) are used for classification.  In
  the small panels we show the distributions of the top and QCD jets
  in the latent space.}
\label{fig:dvae_sbplots}
\end{figure}
%----------------------------------------------------------

Moving towards a more realistic scenario of unsupervised
classification, we need to study scenarios with a class imbalance,
where one of these classes is viewed as anomalous. We slightly adapt
our hyper-parameters to more hierarchic sub-class
proportions~\cite{Dillon_2020} and choose $\alpha_{1,2}=(1.0,0.25)$
when we have a class imbalance in the data.  The performance of the
network is not overly sensitive to the actual values.  The benefit is
that the DVAE will assign the dominant class of jets to $r_1\!=\!0$,
so we should find that anomalous jets are pushed towards $r_1\!=\!1$.

We consider four different signal-to-background ratios, $1.00$,
$0.25$, $0.05$, and $0.01$, where the latter two can be considered
anomalous.  In the top row of Fig.~\ref{fig:dvae_sbplots} we show
results for the case where the signal is top jets, and in the bottom
row for QCD jets.  We plot the ROC curves using the latent space and using
the reconstruction error, and then plot the latent space distributions
for the different signal-to-background ratios.

%----------------------------------------------------------
\begin{figure}[t]
\centering
\includegraphics[width=0.32\textwidth]{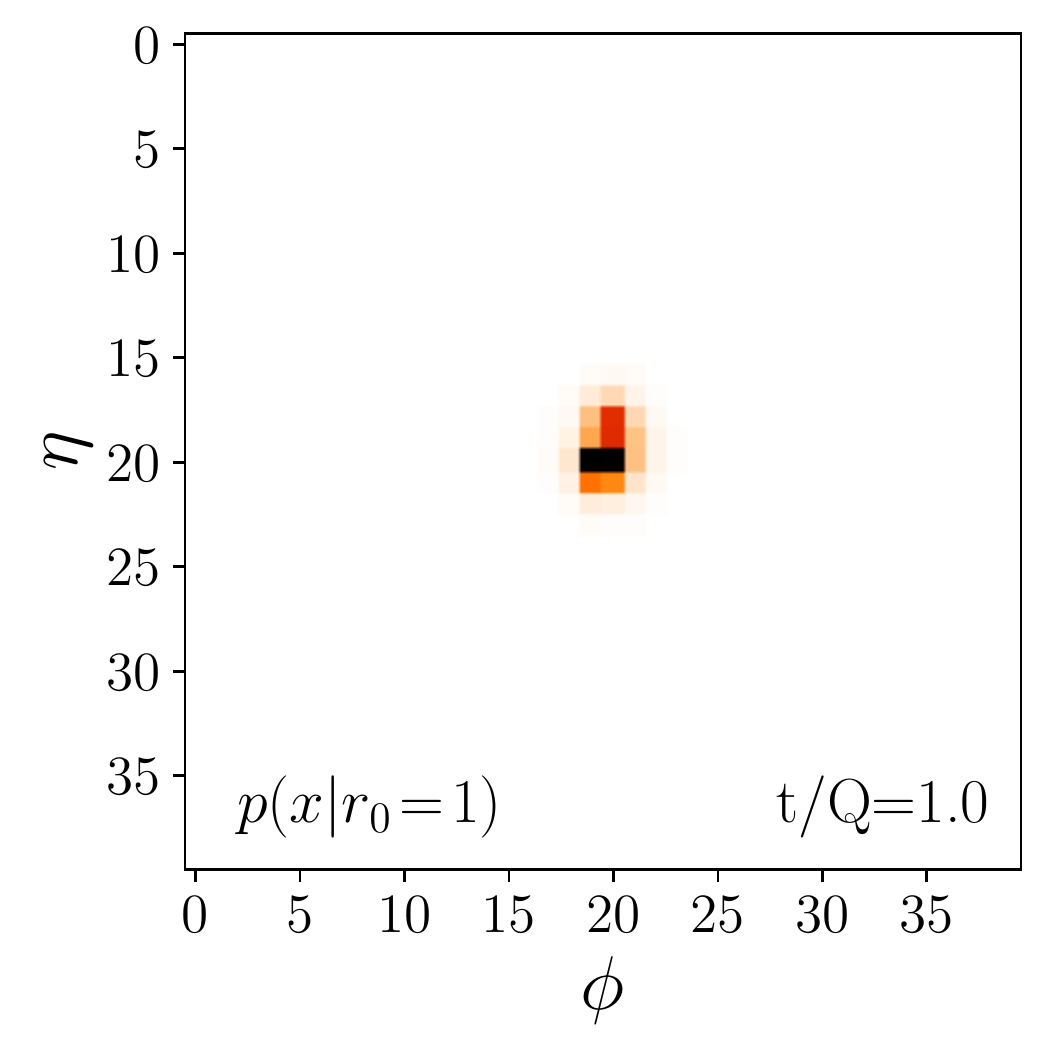}
\includegraphics[width=0.32\textwidth]{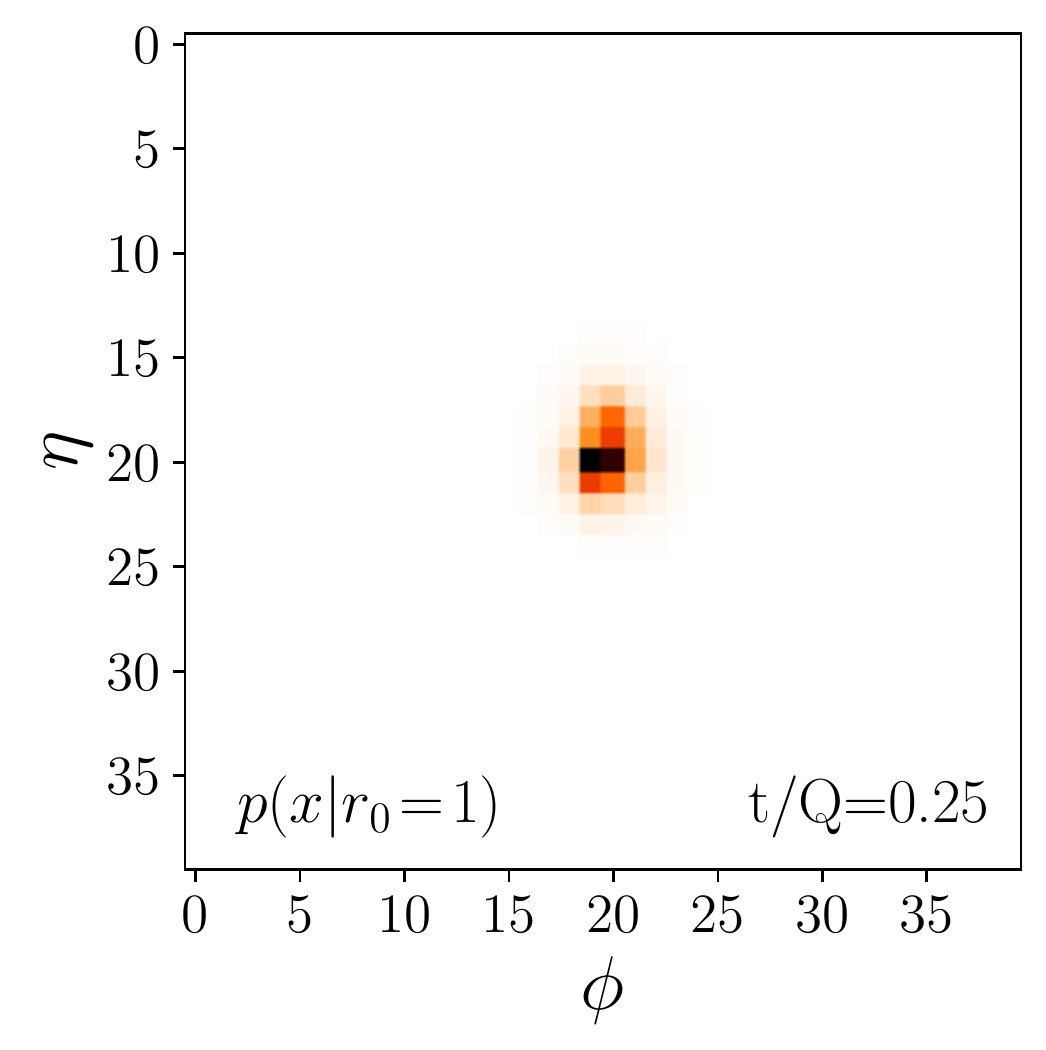}
\includegraphics[width=0.32\textwidth]{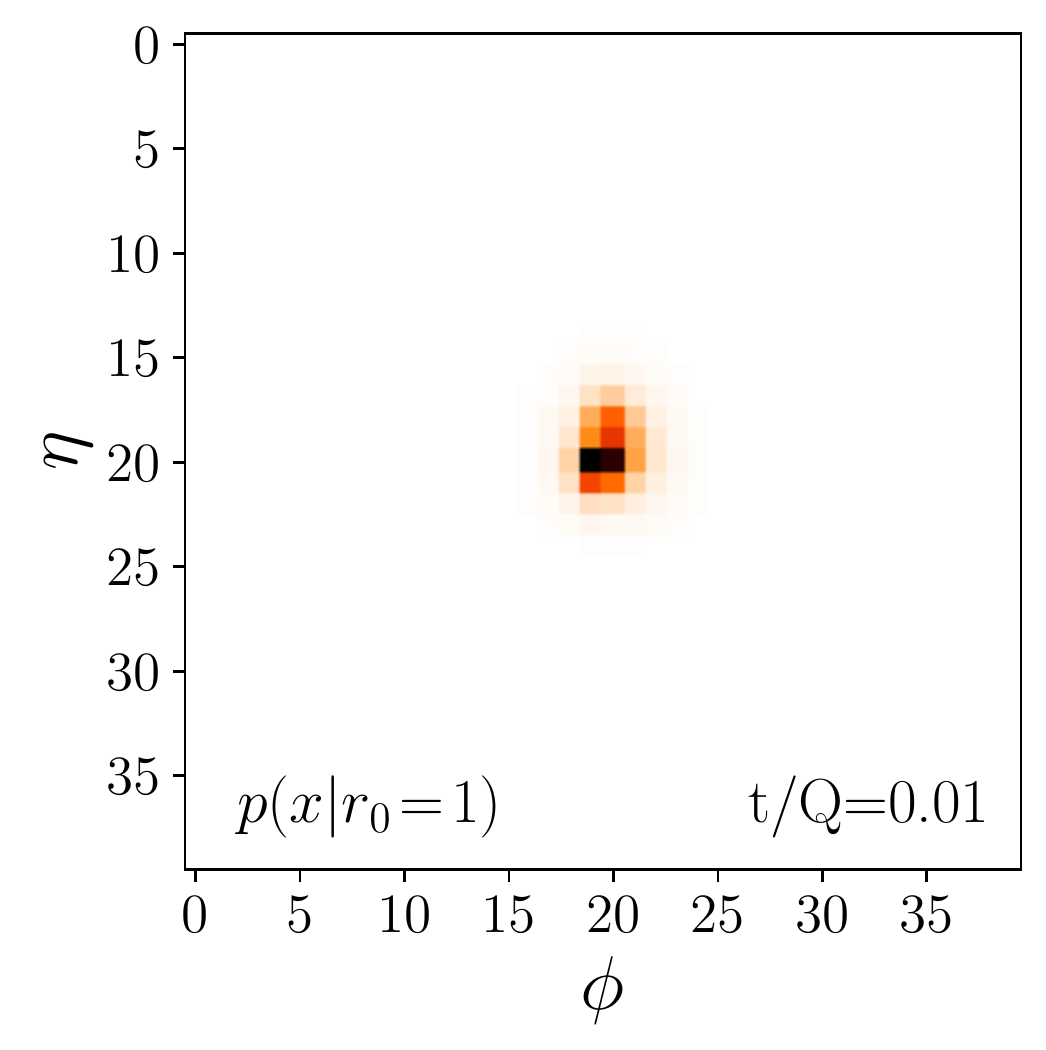} \\
\includegraphics[width=0.32\textwidth]{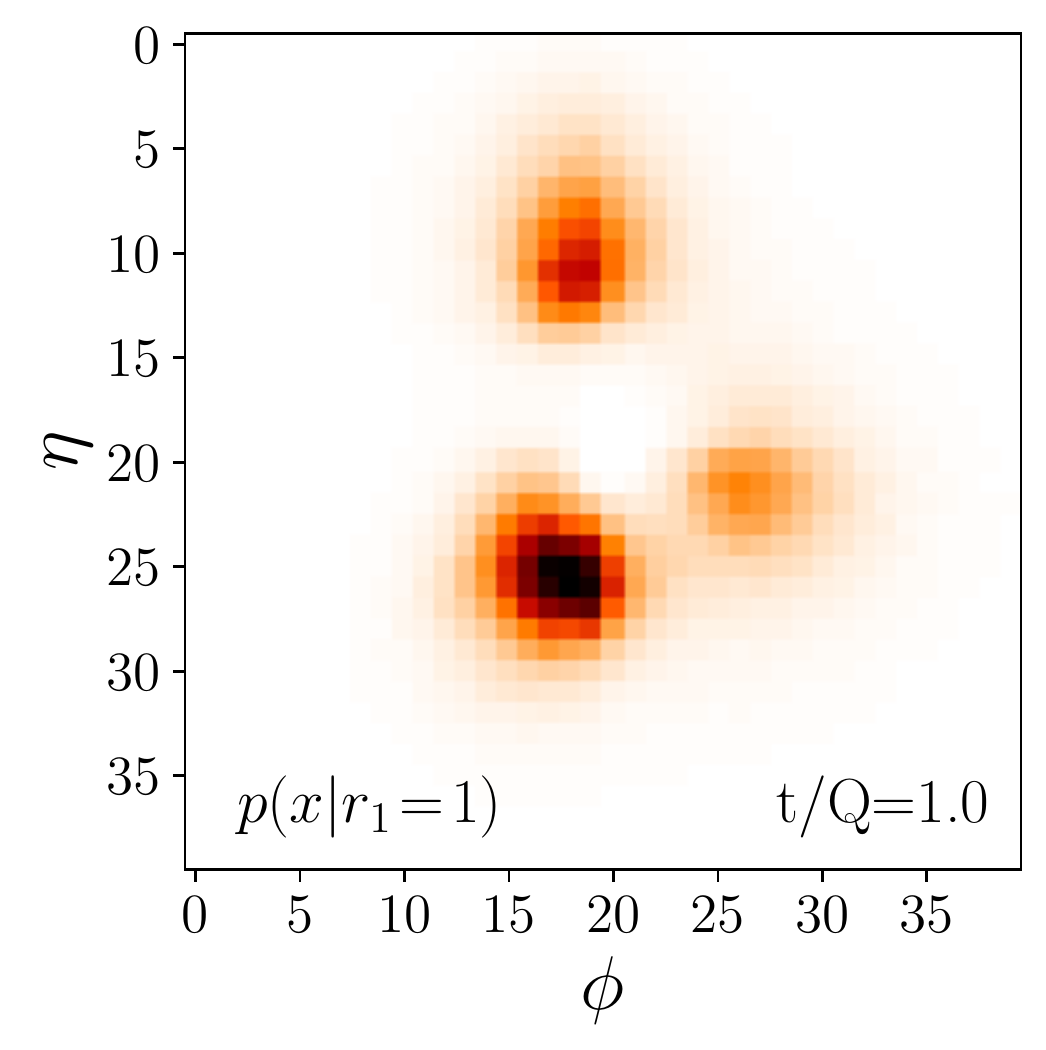}
\includegraphics[width=0.32\textwidth]{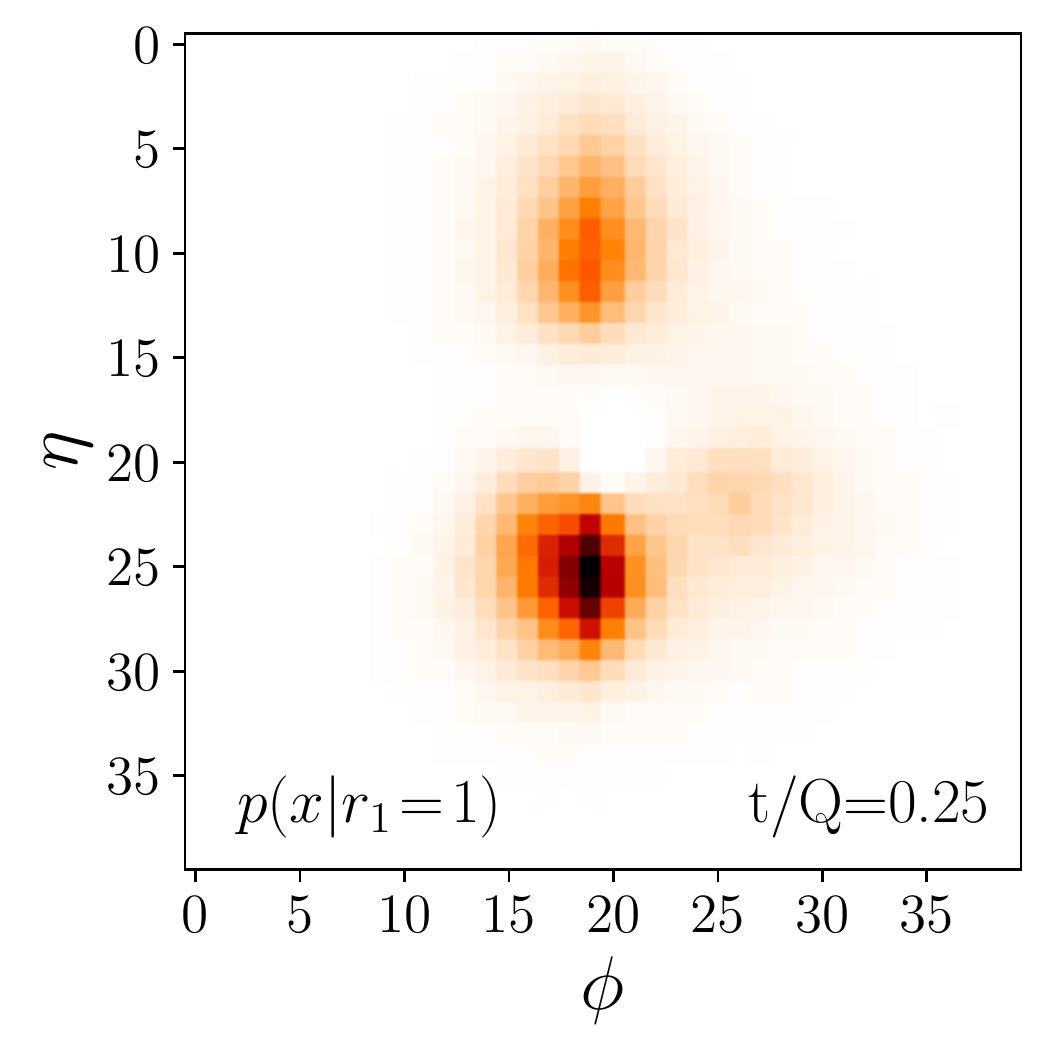}
\includegraphics[width=0.32\textwidth]{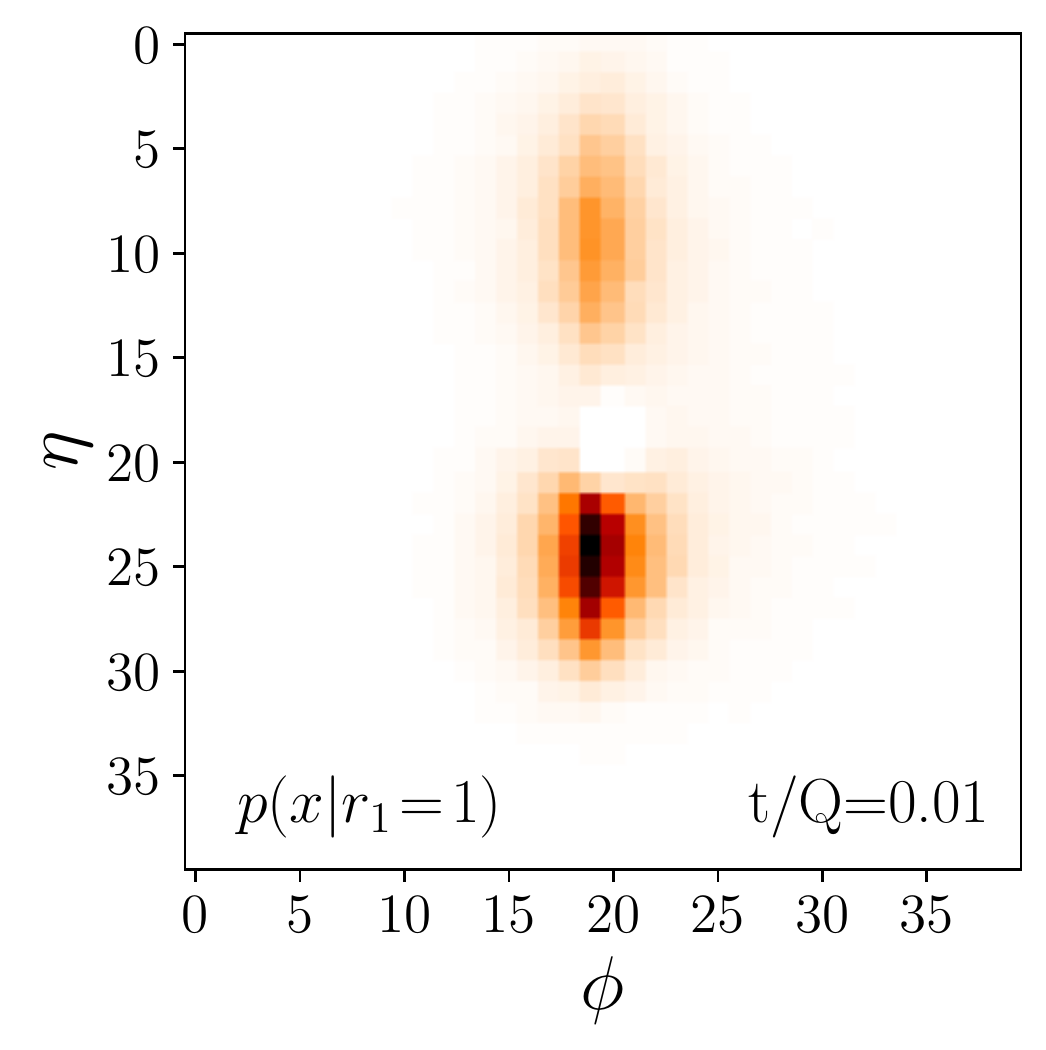} 
\caption{Learned mixture distributions extracted from the decoder
  weights with various amounts of anomalous top jets in the sample.
\label{fig:dvae_sbmixtures}}
\end{figure}
%----------------------------------------------------------

When the top jets are the signal, the classification based on mixture
weights works best for $t/Q=1.0$ and degrades towards smaller ratios
$t/Q$.  This is expected since the latent space requires information
about many jets of a class to accurately construct a good latent
representation for it.  In contrast, the performance of the
reconstruction loss improves as we decrease $t/Q$, since the network
will typically reconstruct an under-represented class poorly.  This is
why we find a reasonably good performance for anomalous top-tagging at
$t/Q=0.01$, as reported in
Ref.~\cite{Heimel:2018mkt,Farina:2018fyg,ivanthesis,thorbenthesis}. The picture
changes when the QCD jets are the signal. While the latent space
tagging with appropriate latent spaces is stable, the reconstruction
error fails as a classifier. This reflects the motivation of this
study, discussed in Sec.~\ref{sec:problem}, as well as the power of
our new approach.

The latent space distributions in Fig.~\ref{fig:dvae_sbplots} confirm
that when one class is anomalous, the Dirichlet prior helps in
assigning the dominant class to the mixtures $r_1\!=\!0$.  One outlier
here is the case $Q/t\!=\!0.25$, where QCD jets are, accidentally,
assigned to $r_1\!=\!0$. This can happen because also top jets have a
strong central prong and copy the typical QCD jet feature.  Because the
unsupervised DVAE does not know the truth label and only assigns
features to classes, the dominant feature even when $Q/t = 0.25$ turns out
QCD-like.

As before, we can use the visualisation of the decoder weights to
study what the DVAE has learned.  In Fig.~\ref{fig:dvae_sbmixtures} we
show this visualisation for the top-tagging runs in
Fig.~\ref{fig:dvae_sbplots}.  As $t/Q$ in the training data is
decreased, the $p_\theta(x|r_1\!=\!1)$ mixture transforms
from the 3-prong top-like structure to a 2-prong structure that is
quite prevalent already in the QCD jet sample.

%%%%%%%%%%%%%%%%%%%%%%%%%%%%%%%%%%%%%%%%%%%%%%%%%%%%%%%%%%%%%%%%%%%%%%%%
\subsubsection*{Enlarging the latent space}

%----------------------------------------------------------
\begin{figure}[t]
\centering
\includegraphics[width=0.32\textwidth]{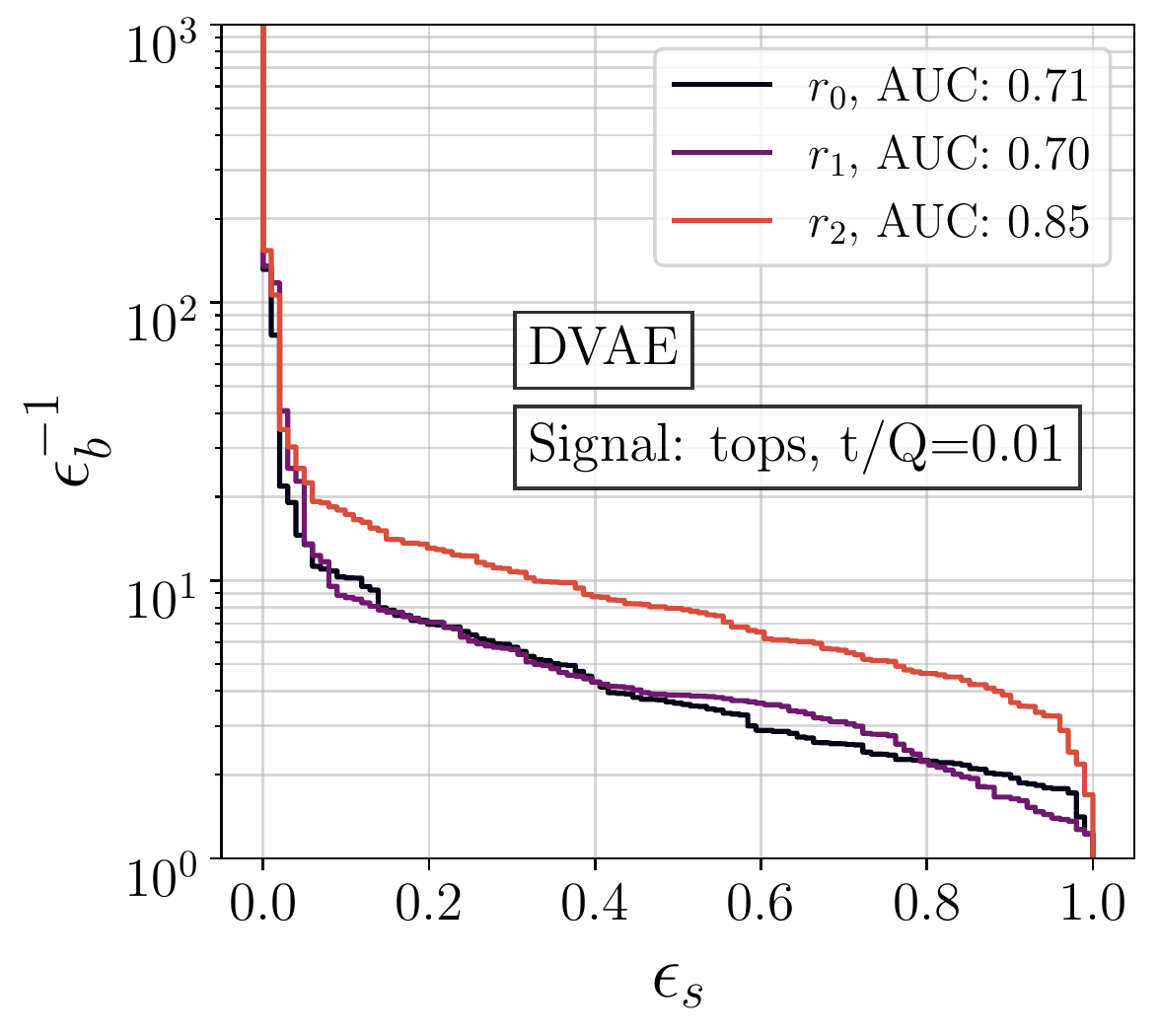}
\includegraphics[width=0.32\textwidth]{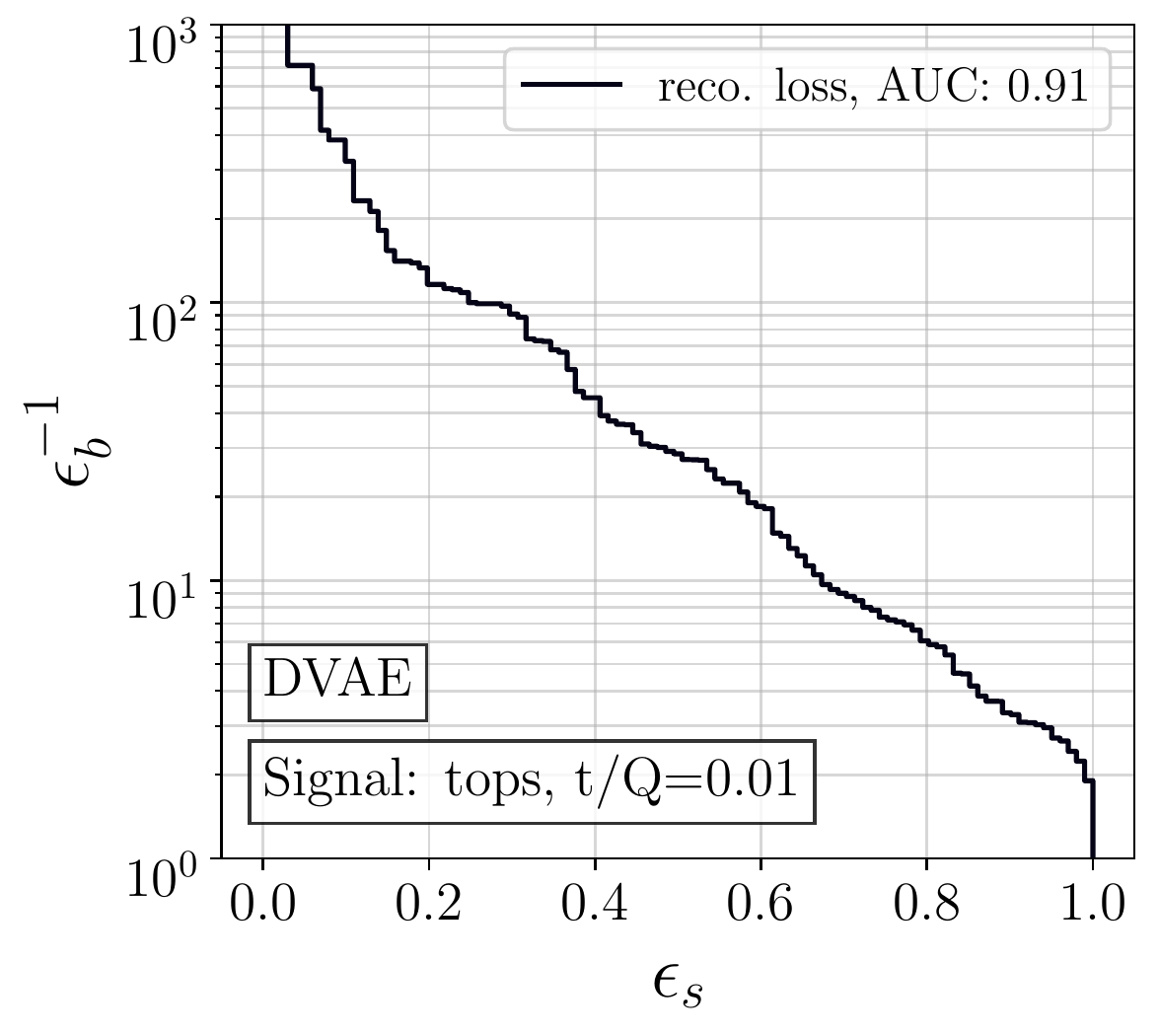}
\includegraphics[width=0.32\textwidth]{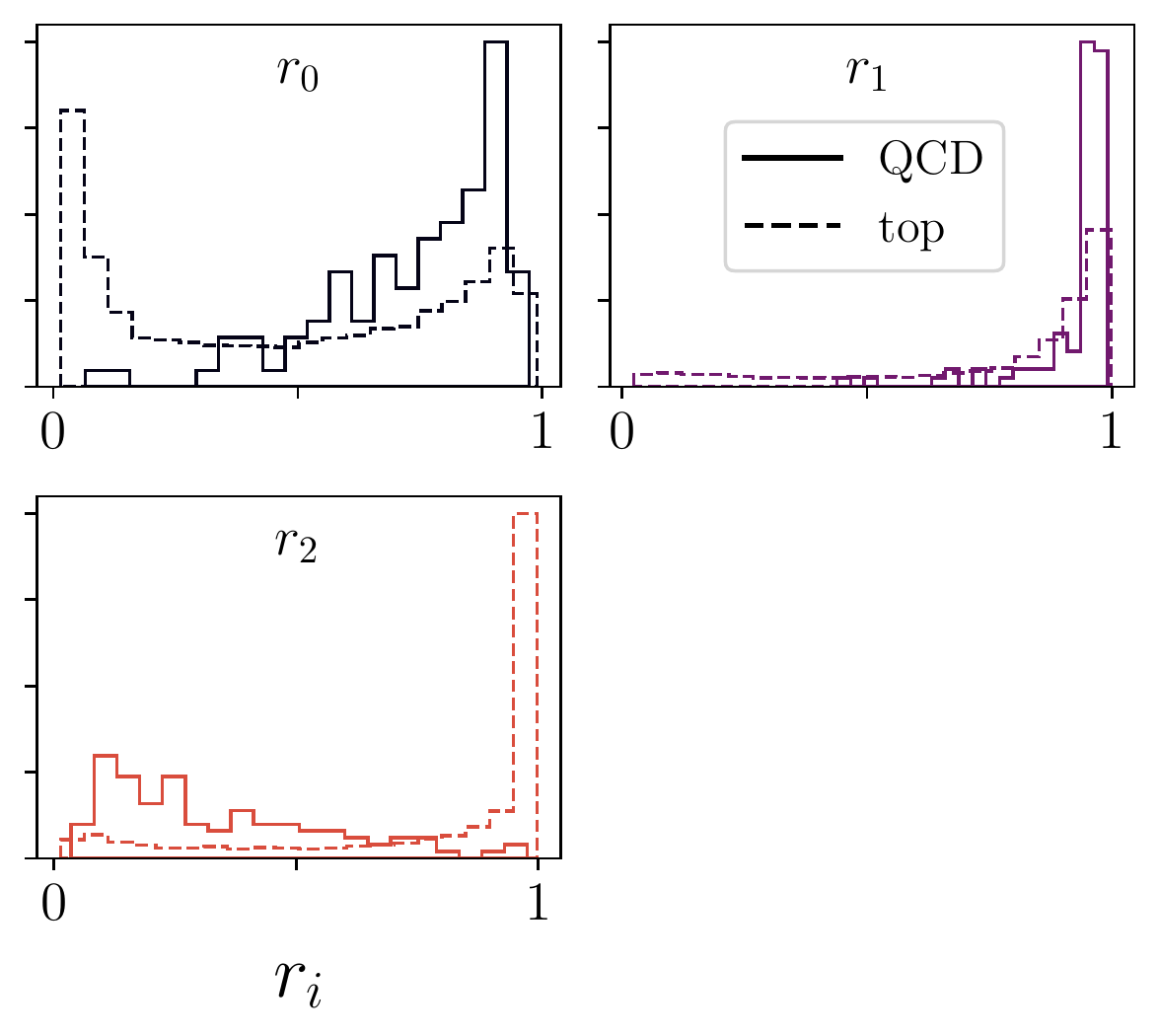}  \\
\includegraphics[width=0.32\textwidth]{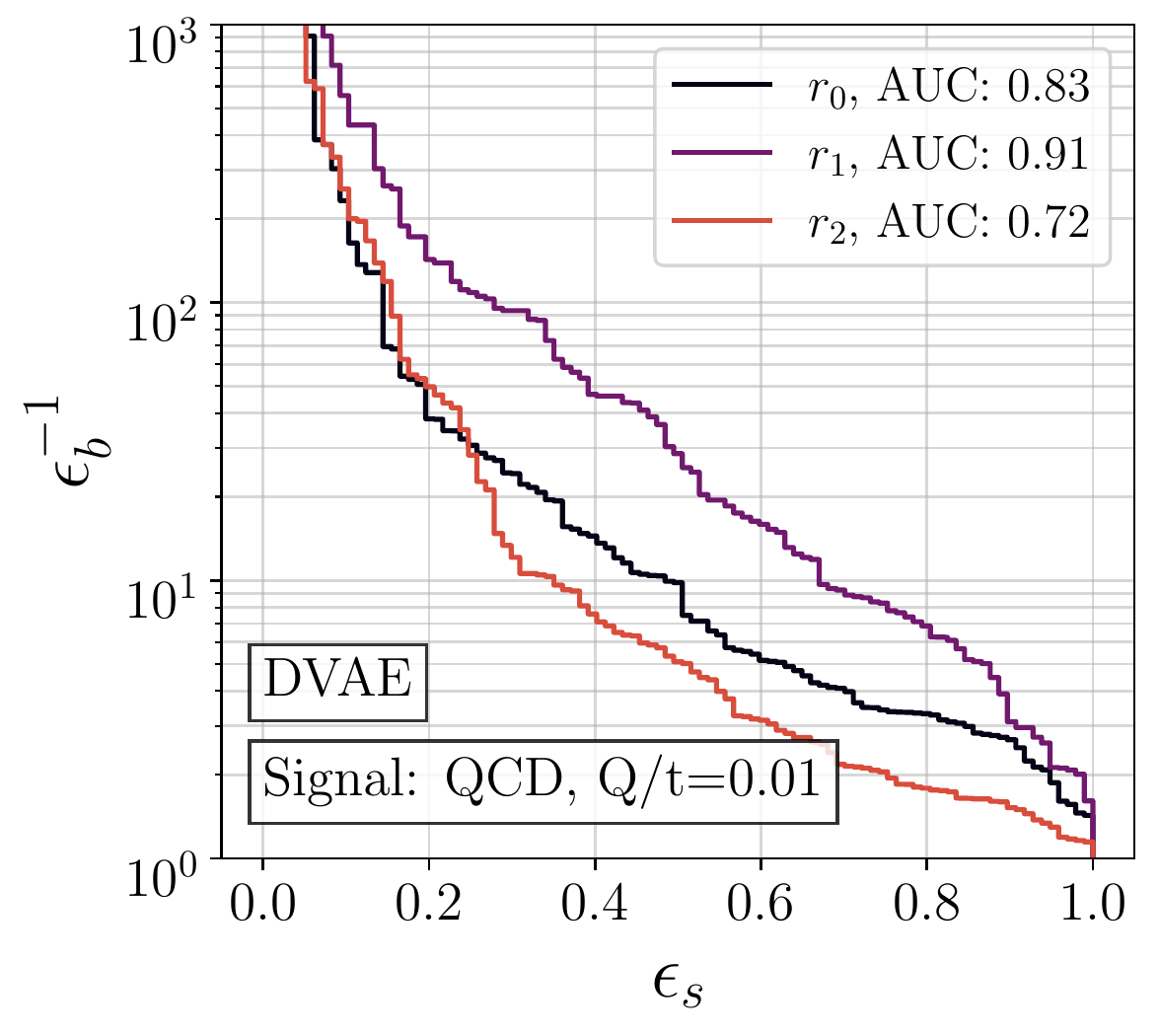}  
\includegraphics[width=0.32\textwidth]{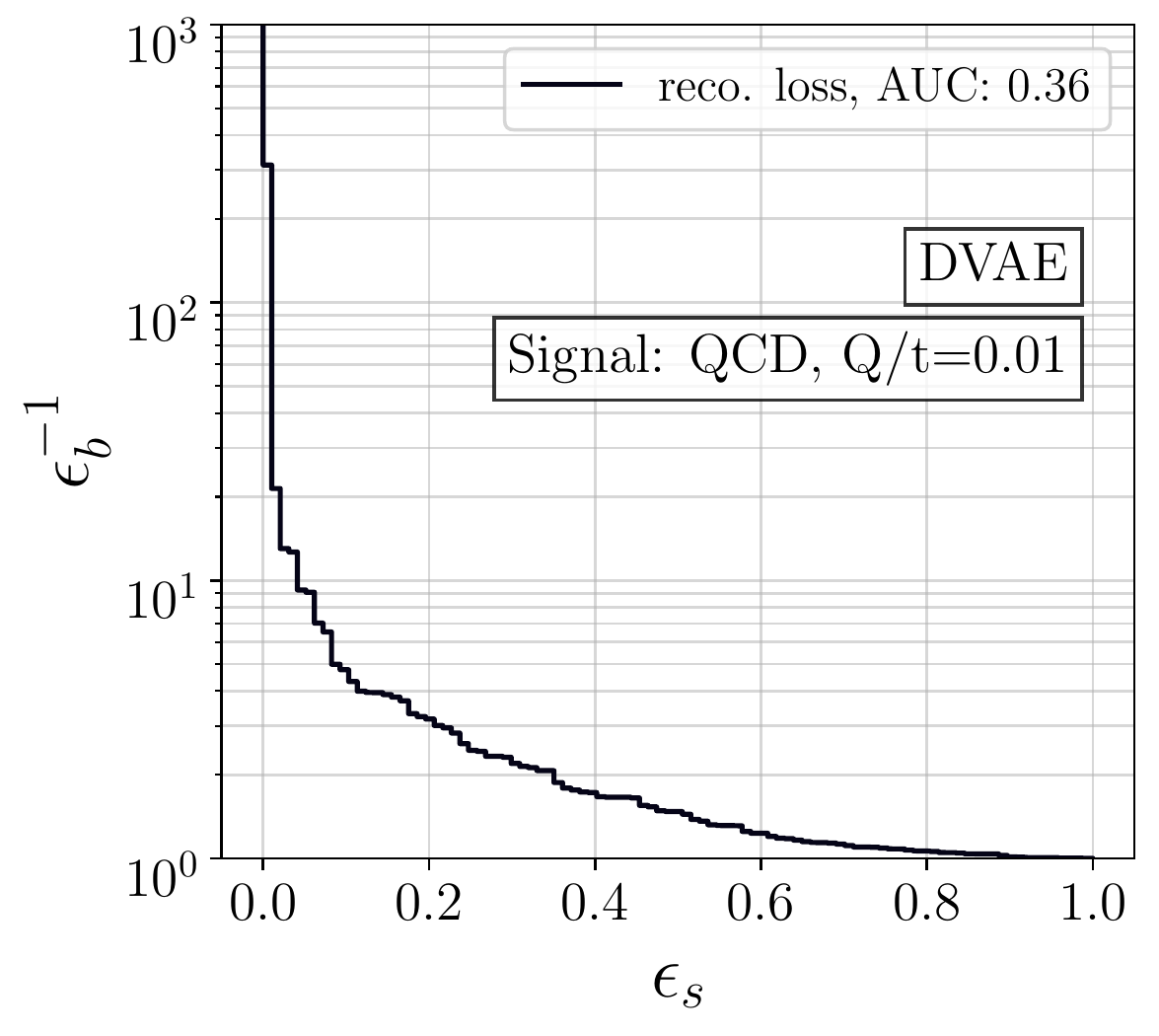}  
\includegraphics[width=0.32\textwidth]{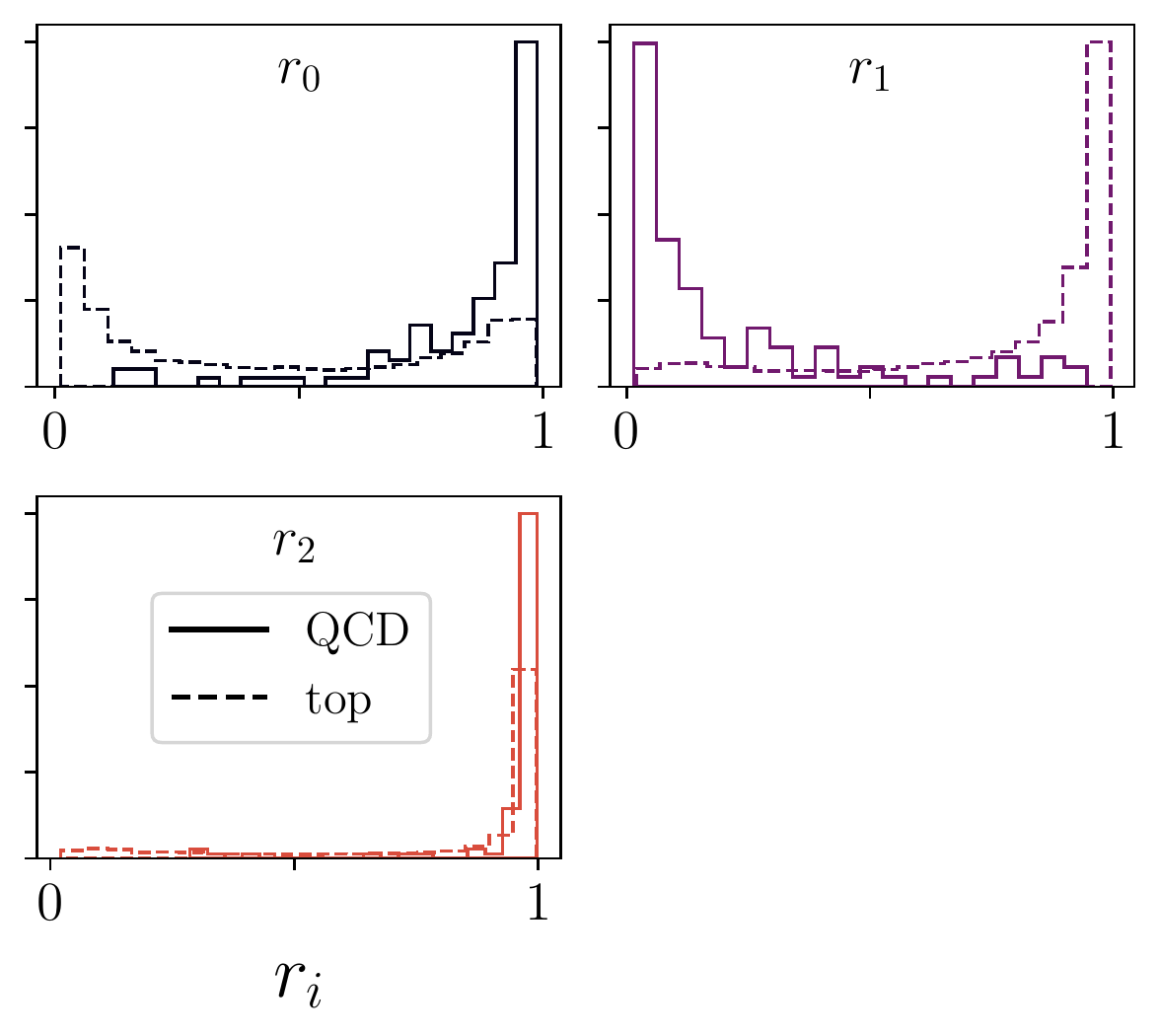} 
\caption{DVAE results using 3 mixture components ($R=3$) for various
  amounts of anomalous top (upper) and QCD (lower) jets in the sample,
  where the mixture weights (left) and the reconstruction loss
  (middle) are used for classification.  In the small panels we show
  the distributions of QCD and tops along each direction in latent
  space.}
\label{fig:dvae_3mix}
\end{figure}
%----------------------------------------------------------

Clearly, there will be applications where a 1D latent space or
$R\!=\!2$ is not enough to construct a sufficient representation of
the data for anomaly detection.  With this in mind we enlarge the
latent space to $R\!=\!3$ and again study the anomalous top ($t/Q\!=\!0.01$) and
anomalous QCD ($Q/t\!=\!0.01$) scenarios.  We choose a hierarchical prior through
$\alpha_i=(1.0,0.25,0.1)$, such that the mixtures are ordered by their
prevalence in the sample.  This forces the network to focus on
different scales in the data-set, instead of describing just the
most common patterns in the data.

In Fig.~\ref{fig:dvae_3mix} we show the DVAE results for $R\!=\!3$
using each of the latent mixture weights separately as the classifier.
With anomalous top jets, the $r_2$-mixture weight corresponding to the
weakest class always produces the best classifier. In contrast, for
anomalous QCD jets, the best classifier is the $r_1$-mixture, with a
larger weight than $r_2$ in the prior. Compared to the $R\!=\!2$ case,
the performance in anomalous QCD tagging here is strikingly better,
with an AUC of $0.91$.  A closer look at Fig.~\ref{fig:dvae_sbplots}
already indicates that the dominant, feature-rich top jets require
both mixtures to describe their range of features. This is enhanced by
the fact that there is a lot of variability within the top jet sample,
and is to some degree dependent on the pre-processing.

%----------------------------------------------------------
\begin{figure}[t]
\includegraphics[width=0.32\textwidth]{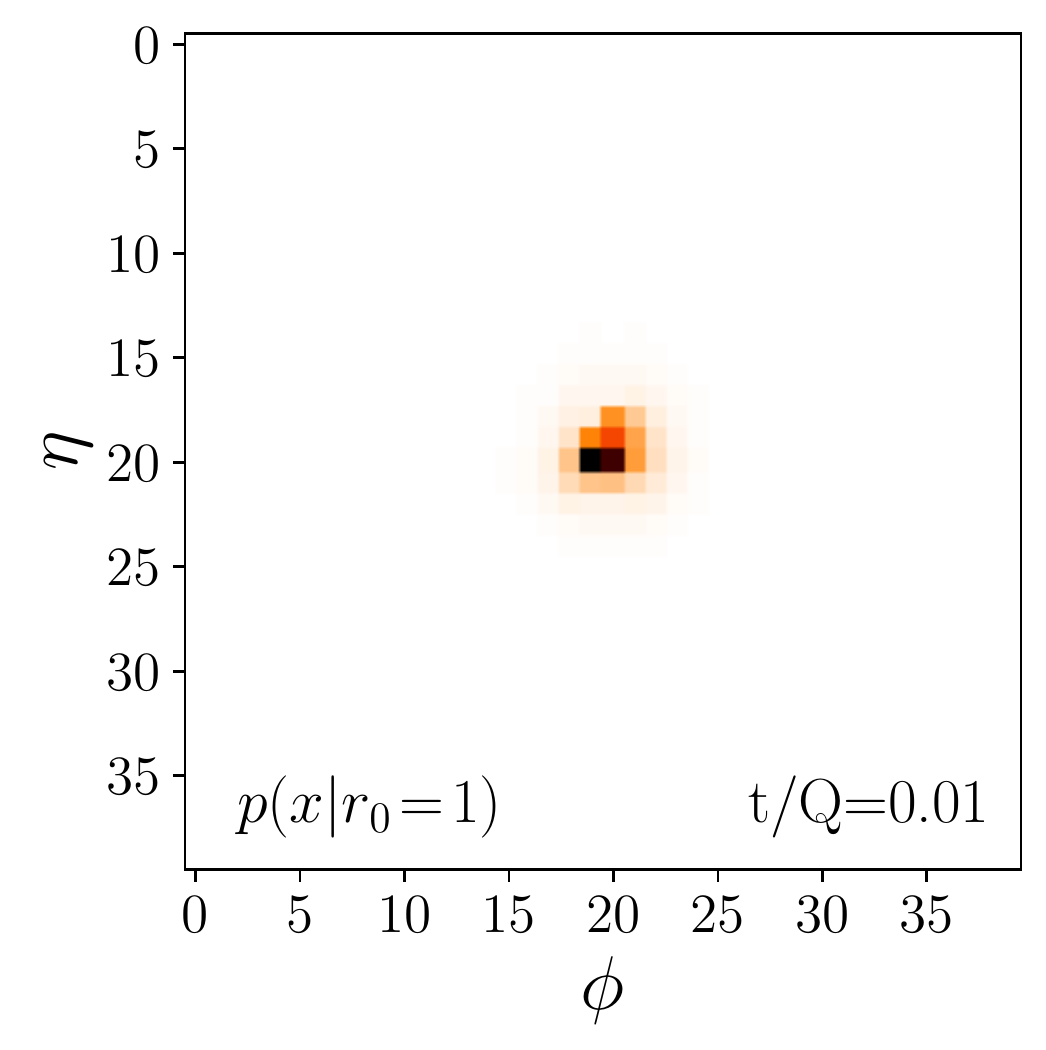}
\includegraphics[width=0.32\textwidth]{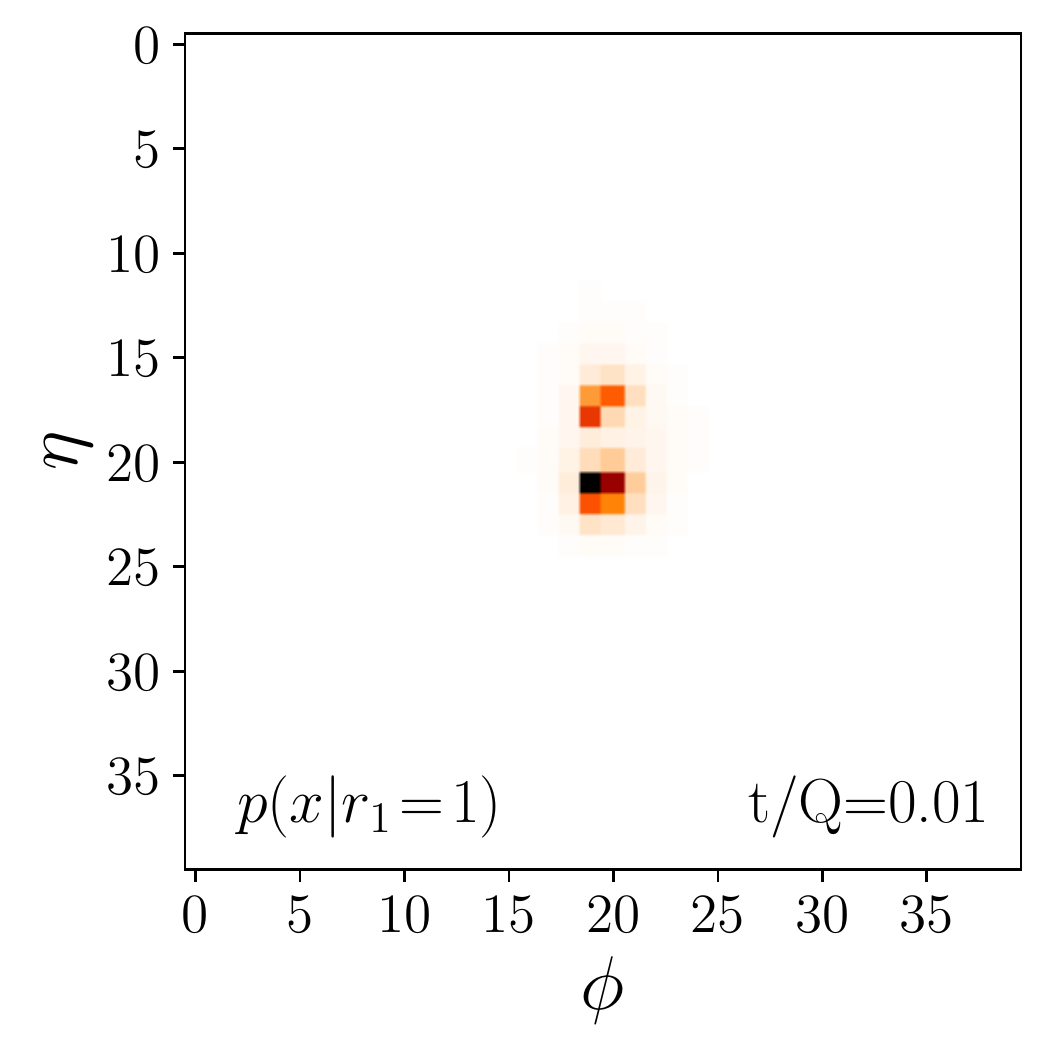} 
\includegraphics[width=0.32\textwidth]{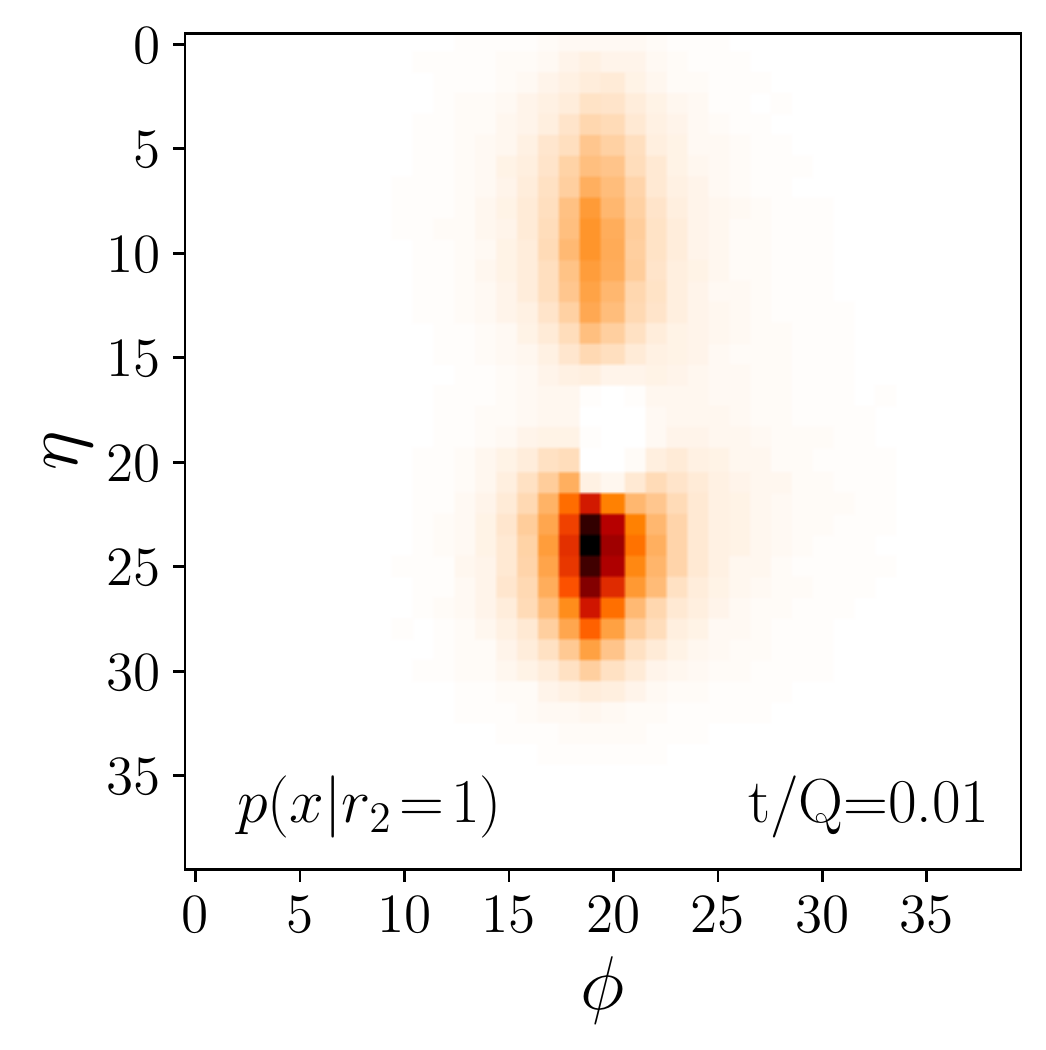} \\
\includegraphics[width=0.32\textwidth]{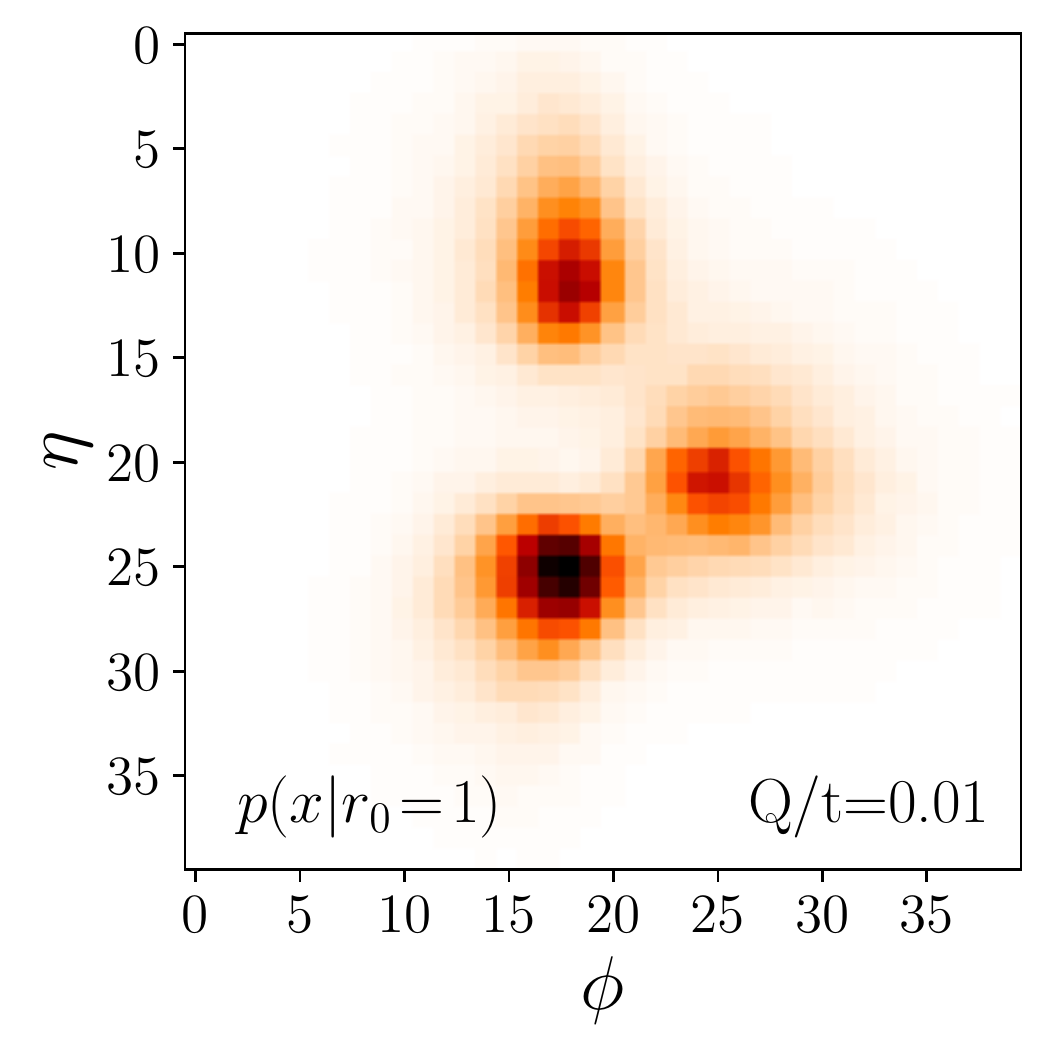}
\includegraphics[width=0.32\textwidth]{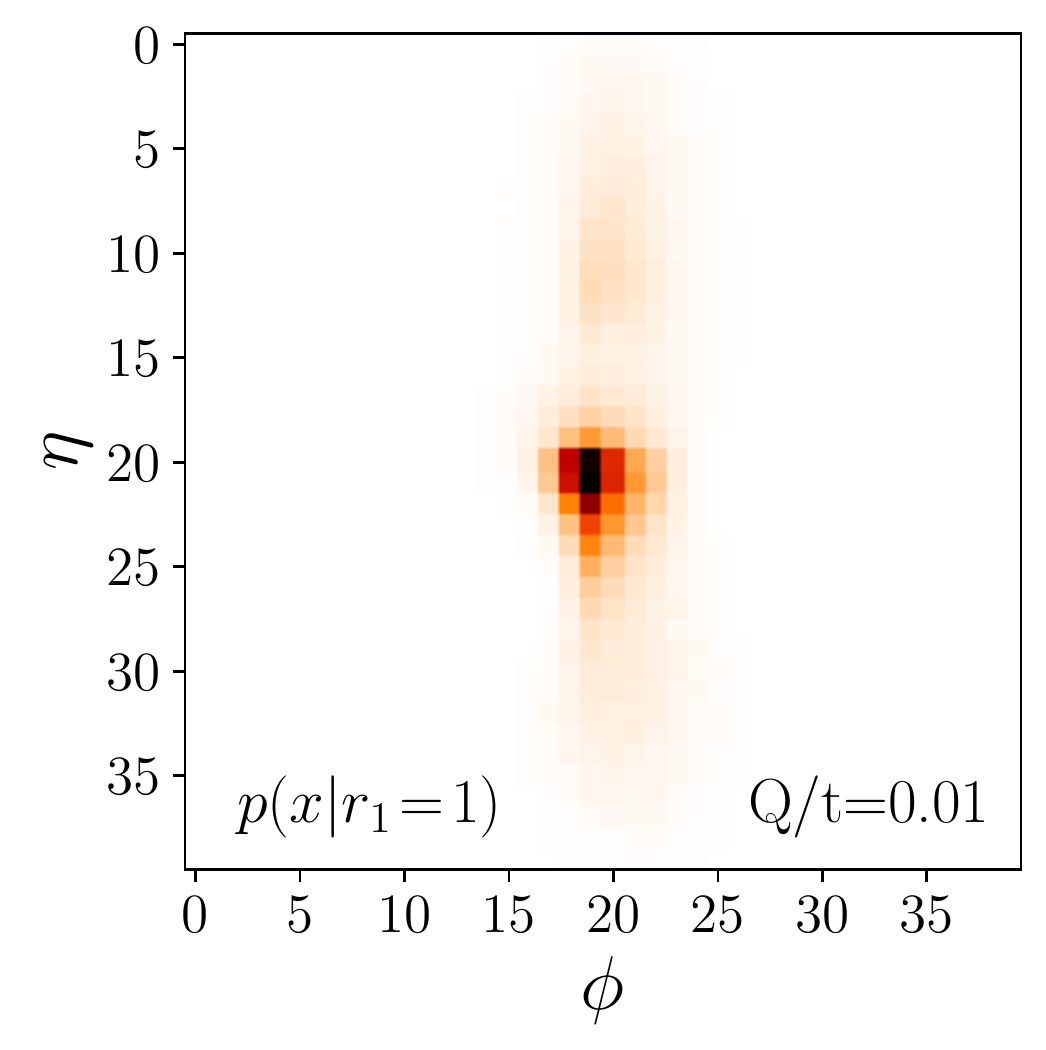} 
\includegraphics[width=0.32\textwidth]{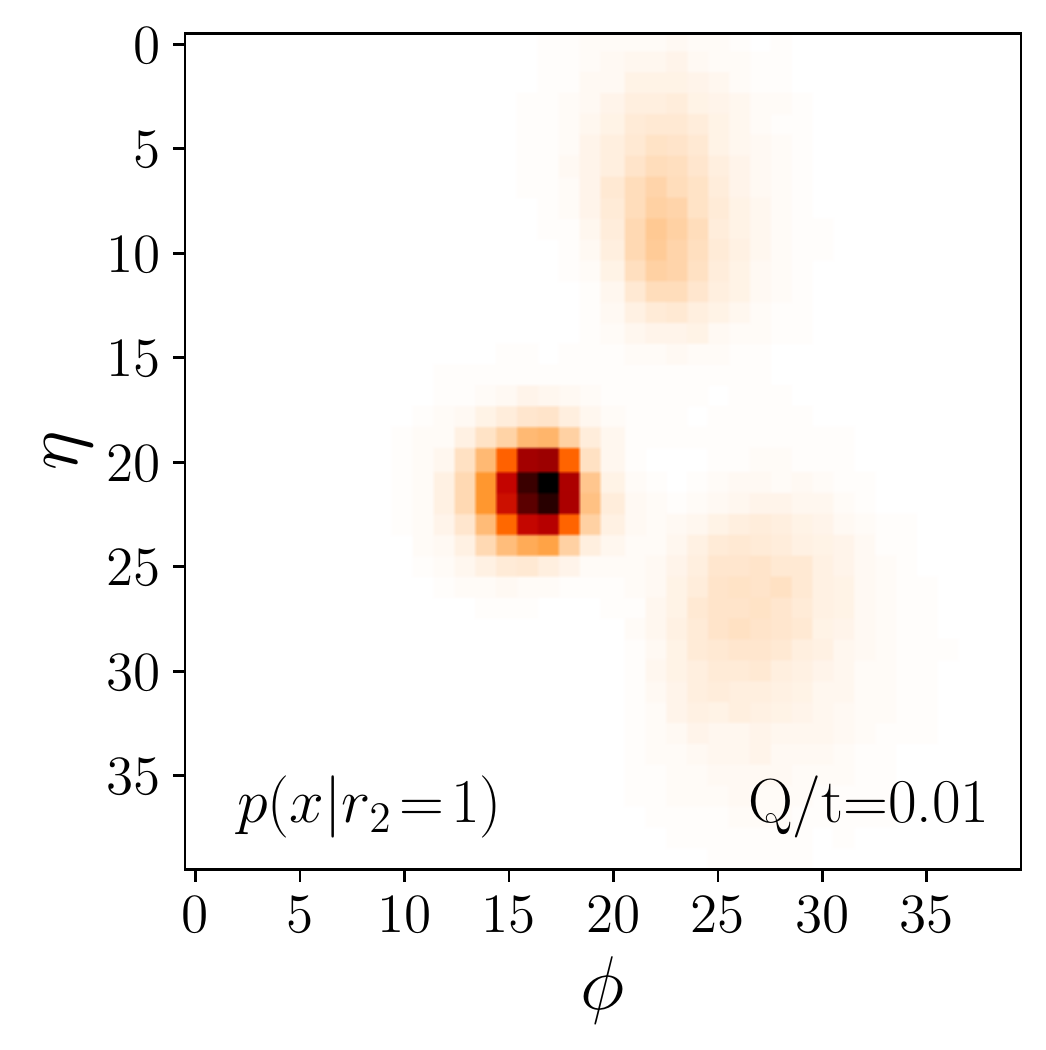} 
\caption{Three learned mixture distributions extracted from the
  decoder weights for $t/Q=0.01$ (upper) and $Q/t=0.01$ (lower). The
  mixtures are shown in order of their prevalence in the jet sample.}
\label{fig:dvae_3mix_dists}
\end{figure}
%----------------------------------------------------------

To understand the symmetric performance of the expanded DVAE, we look
at the three learned mixture distributions for $t/Q=0.01$ and for
$Q/t=0.01$ in Fig.~\ref{fig:dvae_3mix_dists}.  In the anomalous top
case (top row) the sample is mostly QCD jets, as reflected in the
mixture distributions. Because of the hierarchical prior, we see that
the less prevalent the mixture component the wider the angular
splitting in the mixture distribution gets.  The top jets typically
contain wide angle splittings, which explains why they are assigned
larger weights in the $r_2$ mixture than in the others.  Due to the
jet pre-processing and the fact that QCD are jets are mostly one-prong
or two-prong, the images are expected to be almost symmetric in the
$\eta$ and $\phi$ plane.  Interestingly, from the mixture
distributions we can see that the DVAE has learned these approximate
separate symmetries.  Because we use fully-connected networks with
pixelized input images and without any convolutions, the only
information about the locality of the pixels or the symmetries must be
learned from the jet images.

In the learned mixture distributions for the anomalous QCD case
(bottom row), the sample is mostly top jets, and the large differences
in the learned distributions highlights the amount of variance within
the top jets alone.  This is a strong indication that we do need more
latent dimensions to describe the top jets. We also see that 
the $r_1$ mixture is very QCD-like, even though it is not the
least prevalent mixture. This implies that QCD-like features are
prevalent in the top jets.  This is different from the anomalous top
jets, where key top-features are missing in the QCD jet
sample.  This explains why QCD jets do not generate a large
reconstruction error in a VAE trained on predominantly top jets.  It
also explains why the ROC curves for QCD-tagging have much lower
mis-tagging rates at low signal efficiencies than the ROC curves for
top-tagging.

\begin{figure}[t]
\begin{center}
\includegraphics[width=0.32\textwidth]{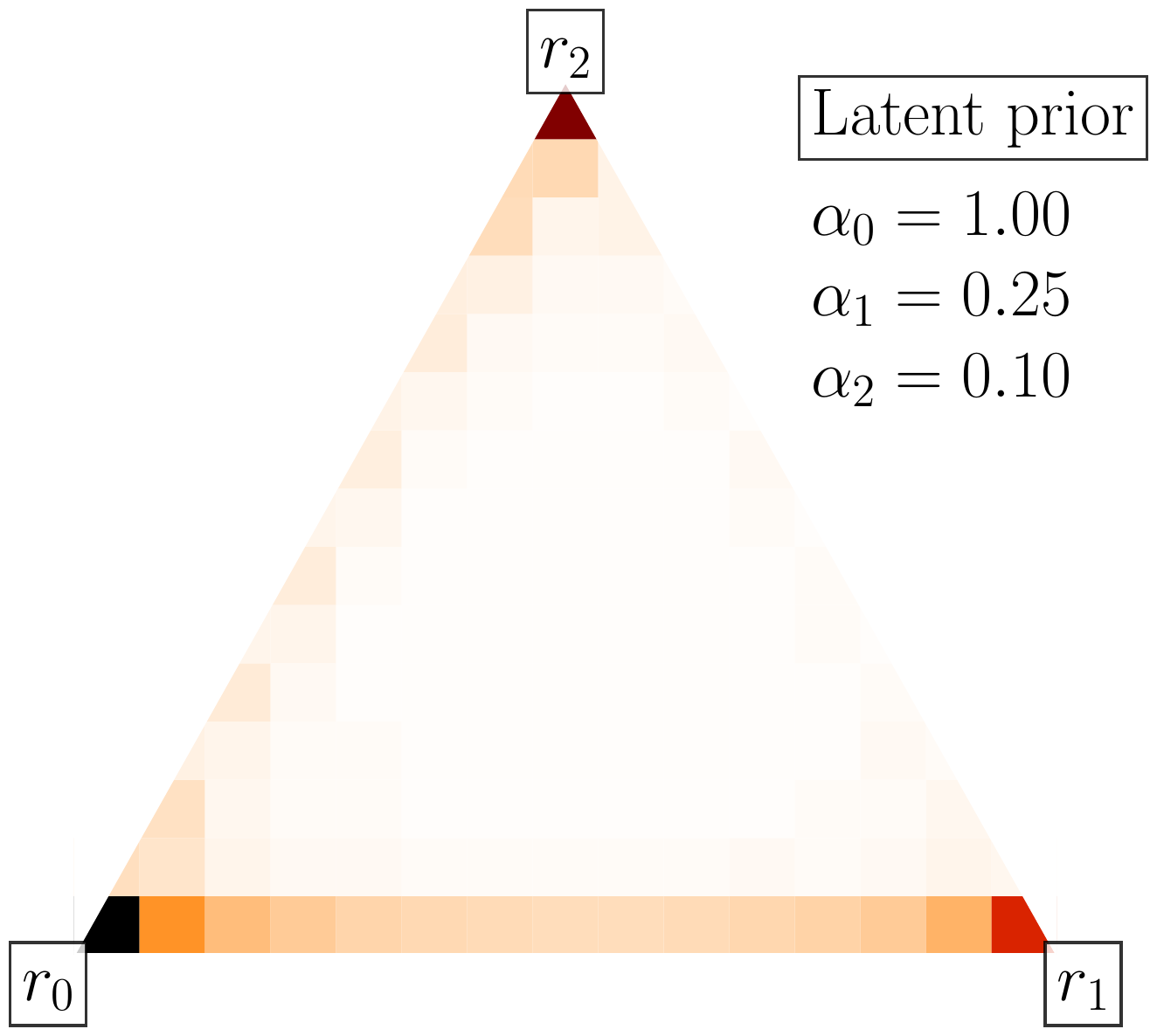}
\includegraphics[width=0.32\textwidth]{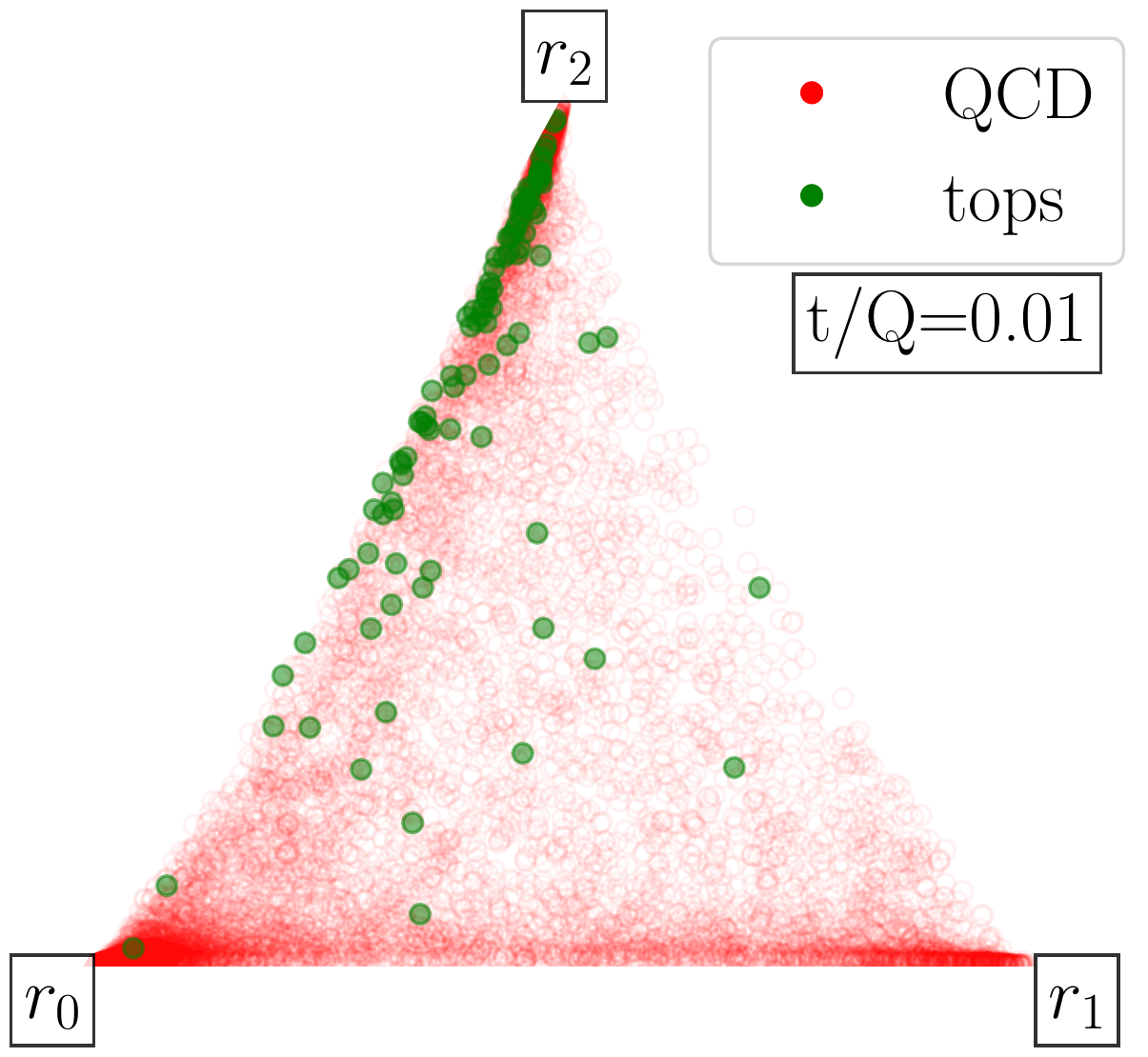} 
\includegraphics[width=0.32\textwidth]{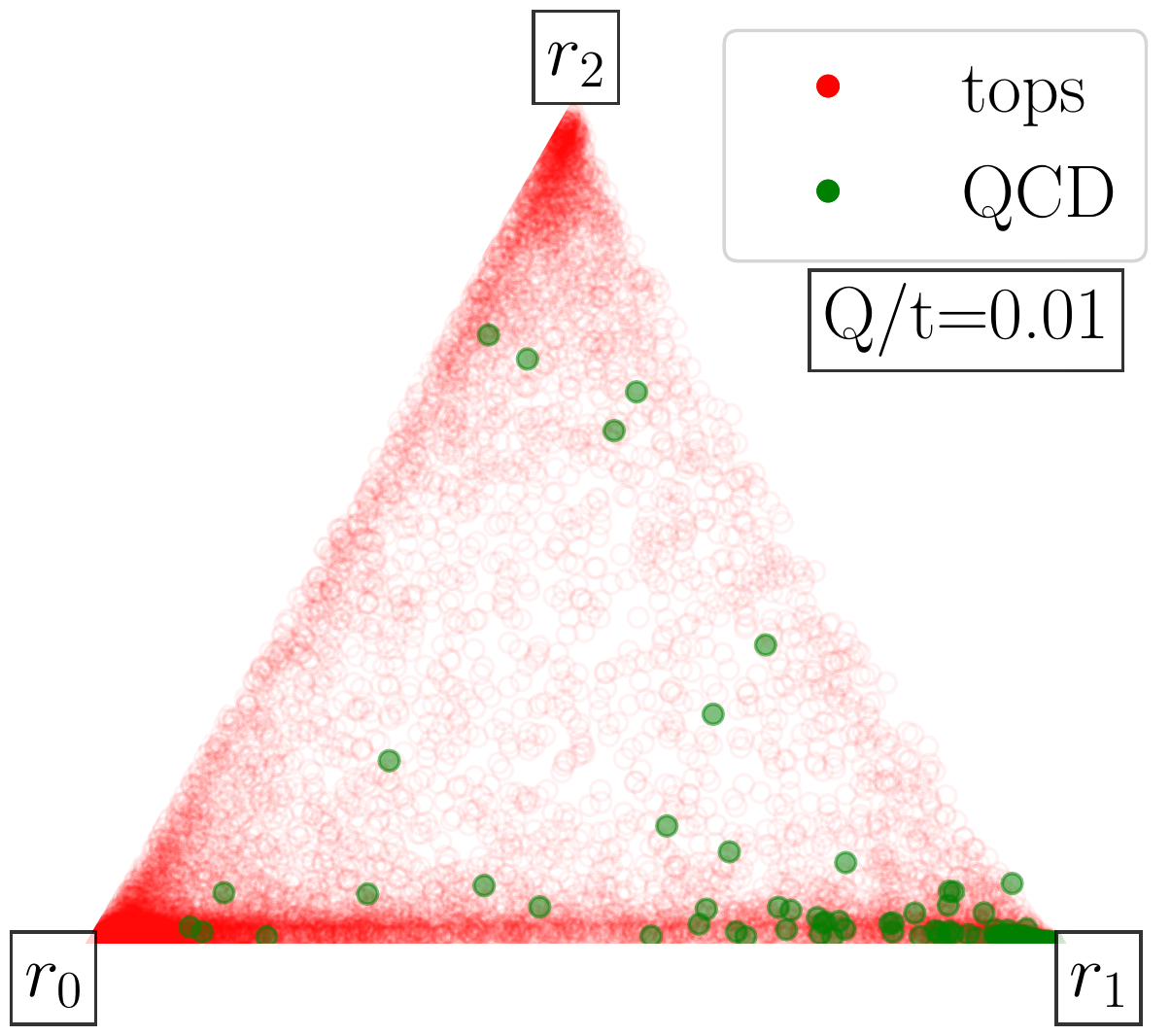}  \\
\hspace{0.32\textwidth}
\includegraphics[width=0.32\textwidth]{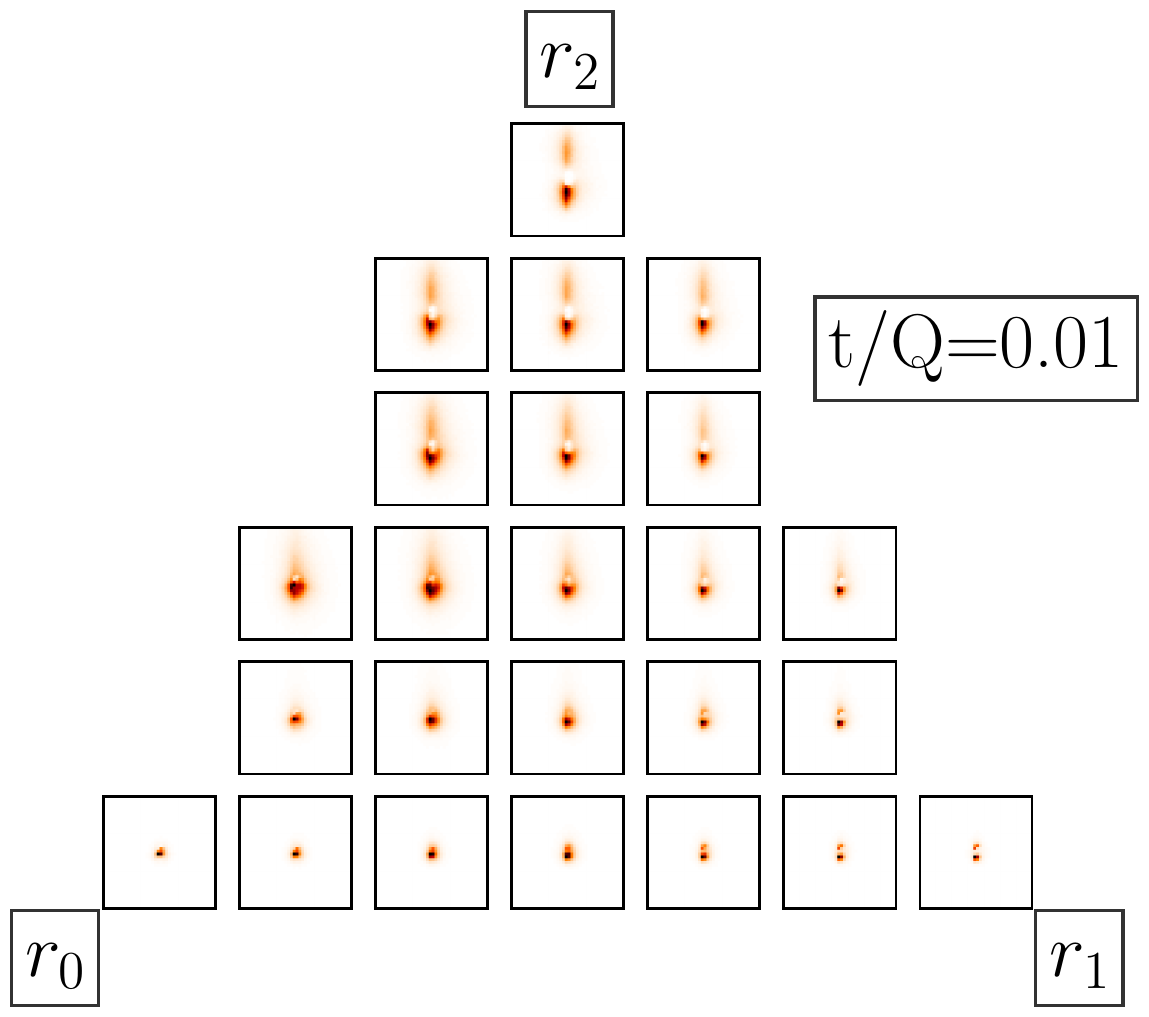} 
\includegraphics[width=0.32\textwidth]{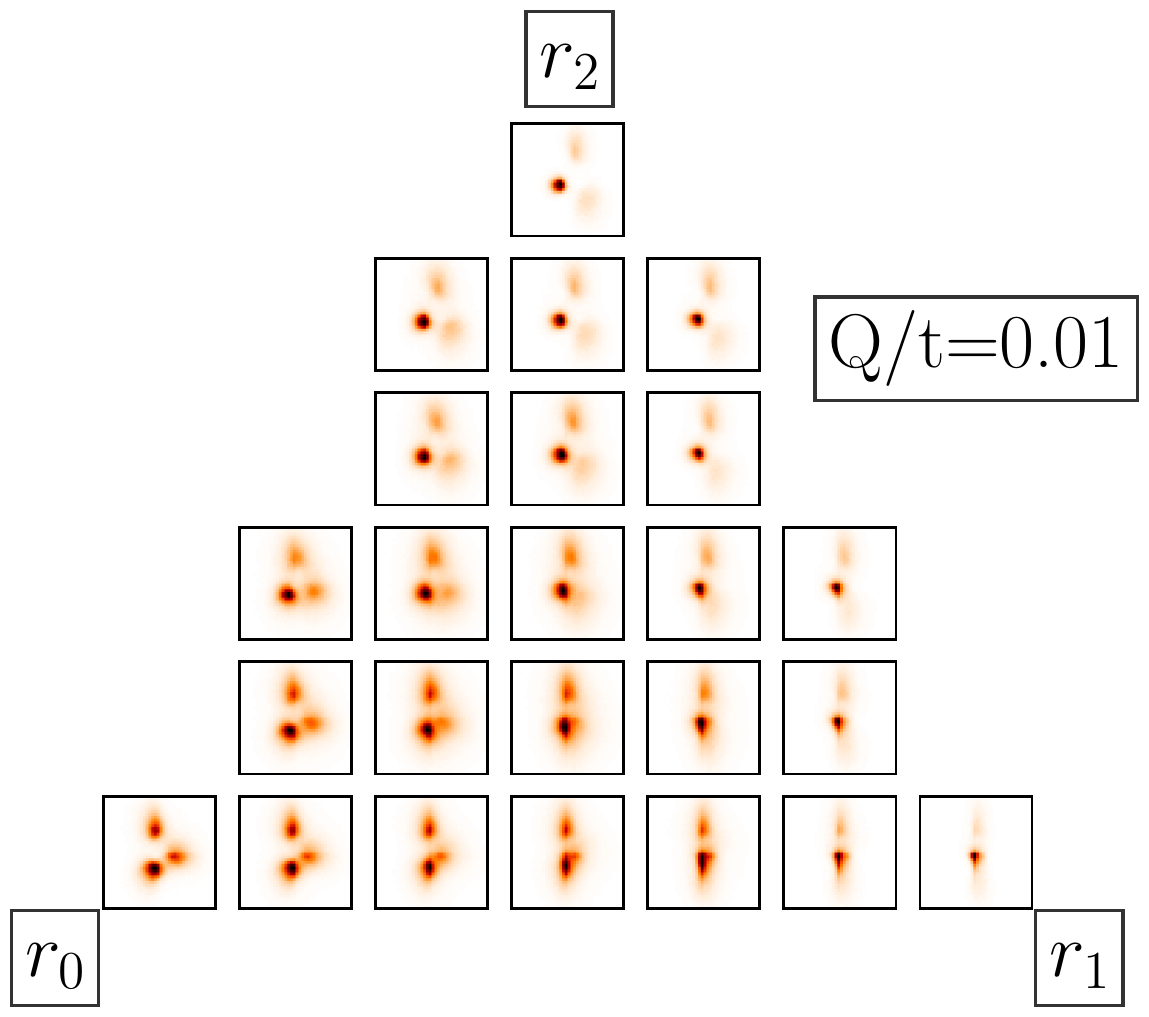} 
\end{center}
\caption{Gibbs triangle plots, also known as the Ternary plots or de
  Finetti diagrams, for the $R\!=\!3$ Dirichlet latent space.  Top
  row: the prior distribution (left) on the latent space, and the
  distributions of jets for anomalous top (centre) and QCD (right).
  Bottom row: visualisations from the latent space of the models
  trained in the anomalous tops and anomalous QCD scenarios.
\label{fig:dvae_gibbs_triangle}}
\end{figure}

The Gibbs triangles in Fig.~\ref{fig:dvae_gibbs_triangle} provide a
useful visualisation of how the jets are distributed in the $R\!=\!3$
Dirichlet latent space.  This visualisation is possible because the
latent space is defined on a simplex with $\sum_i r_i=1$.  Only two of
the variables are free due to the constraint, but rather than just
selecting two arbitrary directions we can use the constraint to
project the three variables onto a 2D plane that captures all three
directions equally.  The points on the triangle represent the latent
space coordinates $(1,0,0)$, $(0,1,0)$, and $(0,0,1)$, allowing us to
see exactly where the jets are encoded.  In the left panel we see the
prior distribution and the hierarchy imposed on the three directions
by the choice of hyper-parameters. On the top row we have the latent
representations for the jets in the anomalous top jet (center) and
anomalous QCD jet (right) cases, while in the second row we show
examples of jets generated from these specific points in latent space.
We see that the DVAE is learning a structured and hierarchic latent
space mapping for the jets, allowing the rarer features within the
data-set to occupy significant regions of latent space.  Importantly,
the anomalous jets in both, the anomalous top and the anomalous QCD
cases, are isolated in latent space. Going back to the start of this
paper, this is exactly what is not possible using the reconstruction
error.

%----------------------------------------------------------
%----------------------------------------------------------

%%%%%%%%%%%%%%%%%%%%%%%%%%%%%%%%%%%%%%%%%%%%%%%%%%%%%%%%%%%%%%%%%%%%%%%%
\section{Outlook}
\label{sec:outlook}

In this paper we studied how different latent spaces in VAEs
can facilitate improved unsupervised jet classification and can allow
for the direct interpretation of what is being learned by the neural
networks.  We started with a brief review of the AE and VAE and
discussed the drawbacks of both reconstruction error classification
and latent space classification.  In particular we discussed the
problem of classifying anomalous jets which have less structure than
the dominant class in the sample, with the prime example being tagging
anomalous QCD jets in a sample of predominantly top jets.

The work is guided by the idea that autoencoders should make use of
the latent space geometry to develop modes that represent classes of
features in the data-set, and that background and signal should
somehow separate into these modes.  We started by studying the
Gaussian-Mixture-VAE where we observe that the multi-modal latent
space did indeed provide a multi-modal encoding of the jets.  The
issue was that the modes did not represent all relevant classes of
features within the data-set.  Then we studied the Dirichlet-VAE,
which accurately determined hierarchic classes of features in the
data-set, and organised the jets in a way which separates the signal
and background.  With a simple decoder architecture we were able to
interpret the decoder weights as mixture distributions corresponding
to each of the latent space directions. They could be visualised for a
direct interpretation of what the network learns.

While the $R\!=\!2$ Dirichlet latent space worked well for
top-tagging, the network still did not tag anomalous QCD jets in a
sample of predominantly top jets symmetrically.  In going to an
$R\!=\!3$ latent space the network isolated the QCD-like features in
latent space leading to a massive improvement in the classification.
The larger latent space combined with a hierarchical prior provided a
latent space capable of extracting features of varying prevalences in
the data-set.  With these structures, we have shown that problems of
finding anomalous jets with less structure than the dominant class in
the data-set can be solved through choosing better latent space
structures.  This opens the door to more effective jet anomaly
detection methods for use in collider analyses.\bigskip

\textbf{Note added:} the submission of this paper is coordinate with
Ref.~\cite{aachen} and Ref.~\cite{rutgers} on related approaches to
the same problem with unsupervised learning and autoencoders.

%%%%%%%%%%%%%%%%%%%%%%%%%%%%%%%%%%%%%%%%%%%%%%%%%%%%%%%%%%%%%%%%%%%%%%
\begin{center} \textbf{Acknowledgments} \end{center}

This project was inspired by discussions and presentations during the
\textsl{Anomaly Detection Mini-Workshop --- LHC Summer Olympics 2020}
at Hamburg University~\cite{talk1,talk2} and during the TRR~257
mini-workshop on Machine Learning. As always, we would like to thank
Gregor Kasieczka for great discussion on all aspects of machine
learning at the LHC.  In addition, we are very grateful for many
discussions with the Aachen and Rutgers groups, leading to a
coordinated preprint submission.
BMD would like to thank the Ljubljana group for early discussions on Dirichlet latent spaces, and the general application of VAEs in anomaly detection.
The research of TP is supported by
the Deutsche Forschungsgemeinschaft (DFG, German Research Foundation)
under grant 396021762 -- TRR~257 \textsl{Particle Physics
  Phenomenology after the Higgs Discovery}.  CS is supported by the
International Max Planck School \textsl{Precision Tests of Fundamental
  Symmetries}.  PS is partly supported by the DFG Research Training
Group GK-1940, \textsl{Particle Physics Beyond the Standard Model}.
BMD is supported by funding from BMBF.

\appendix
%%%%%%%%%%%%%%%%%%%%%%%%%%%%%%%%%%%%%%%%%%%%%%%%%%%%%%%%%%%%%%%%%%%%%%%%
\section{Unsupervised Lund plane}
\label{app:lundjets}

%%%%%%%%%%%%%%%%%%%%%%%%%%%%%%%%%%%%%%%%%%%%%%%%%%%%%%%%%%%%%%%%%%%%%%%%
\subsubsection*{Lund-plane images}
\label{app:lund}

%----------------------------------------------------------
\begin{figure}[b!]
\centering
\includegraphics[width=0.32\textwidth]{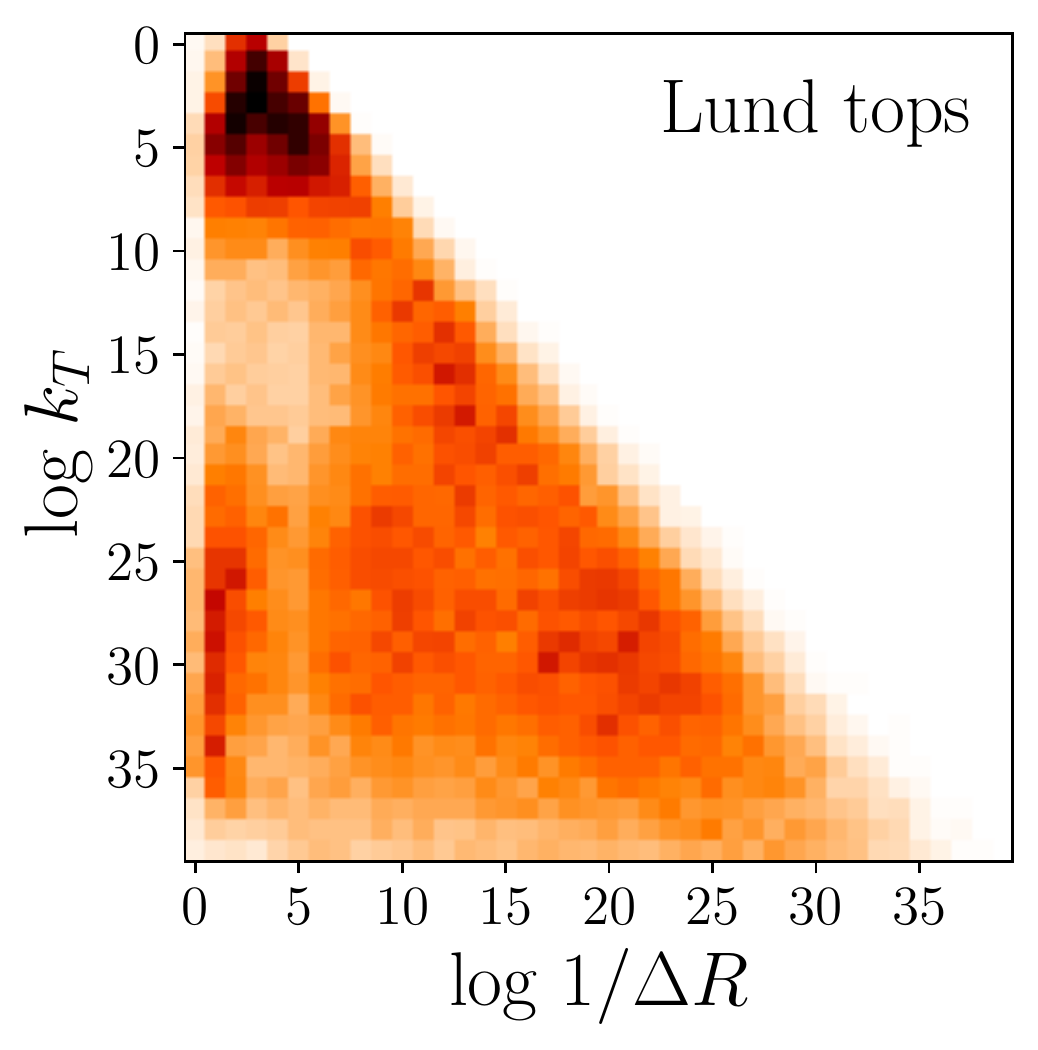}
\hspace*{0.1\textwidth}
\includegraphics[width=0.32\textwidth]{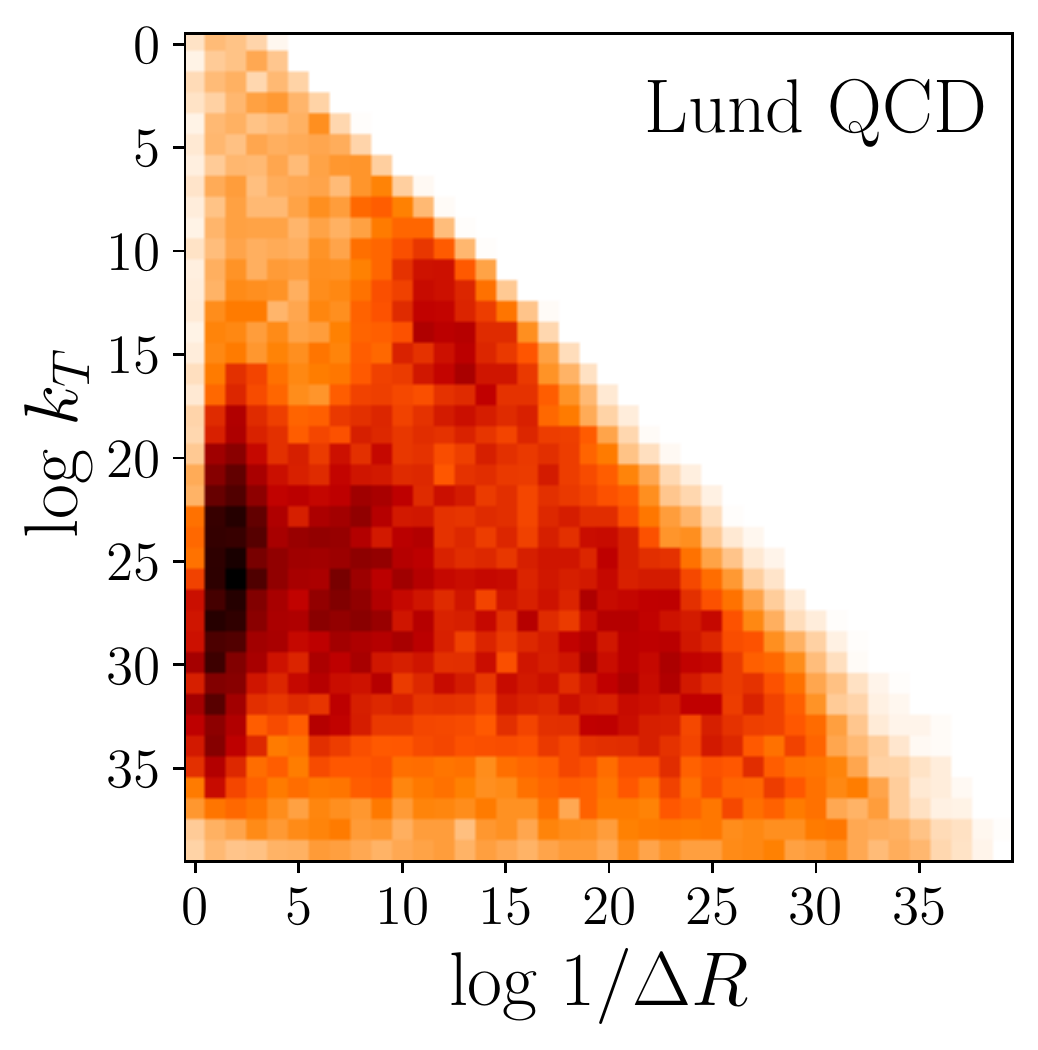} 
\caption{Average of 100k QCD and top jet images in the primary Lund
  plane representation.  The $x$ and $y$ axis show the pixel numbers,
  but these axis go from $0$ to $6$ and $-2$ to $5$, respectively.}
\label{fig:lundimages}
\end{figure}
%----------------------------------------------------------

To analyse jets in the Lund plane~\cite{Dreyer_2018,Carrazza:2019cnt},
we start by re-clustering them using the C/A algorithm.  The algorithm
finds the two constituents closest to each other in the $(\eta,\phi)$
plane and combines them to a single constituent or subjet.  This
$2\!\rightarrow\!1$ re-clustering continues until all constituents
(and subjets) are clustered into one jet.  The Lund plane
representation follows from undoing this clustering piece by piece.
At each step we split a subjet into two further subjets,
$j_a\rightarrow j_b j_c$, where $j_{b,c}$ are referred to as the
daughter subjets and $j_a$ as the parent subjet. From them we
calculate the observables
\begin{alignat}{7}
\Delta R_{b,c} &= \sqrt{ (\eta_b-\eta_c)^2 + (\phi_b-\phi_c)^2 } \qqquad
&k_T &= p_{T,b} \Delta R_{b,c} \quad \text{where} \; p_{T,b}>p_{T,c} \notag \\
m_a^2 &= p_a^2 = (p_b+p_c)^2 
&d & = \text{max}(m_b/m_a, m_c/m_a) \notag \\ 
z &= p_{T,b}/(p_{T,b}+p_{T,c}) 
&\kappa &=z\Delta R_{b,c}.
\end{alignat}
We then assign each splitting to a primary, secondary, etc Lund plane
using the simple algorithm 
\begin{enumerate}
\itemsep0pt
\item uncluster $j_a\rightarrow j_b j_c$
\item assign $l\!=\!0 (1)$ to the daughter with the larger (smaller) $p_T$
\item perform the next step in the unclustering, e.g. $j_c\rightarrow j_d,j_e$
\item assign $l\!=\!0 (1)$ to the daughter with the larger (smaller) $p_T$
\item add the $l$ from the parent $j_c$ to the $l$'s assigned to the daughters $j_d$ and $j_e$
\item repeat steps 3-5 until the jet is completely unclustered.
\end{enumerate}
At the end, each splitting comes with a set of observables and an $l$-value. 
We identify $l=0$ as the primary Lund plane, $l=1$ as the secondary Lund plane,
and so on.  This way the primary Lund plane contains all splittings
from the hardest $p_T$-core of the jet, while the secondary Lund plane
contains splittings once removed from this hardest $p_T$-core, and so
on.

%----------------------------------------------------------
\begin{figure}[t]
\includegraphics[width=0.32\textwidth]{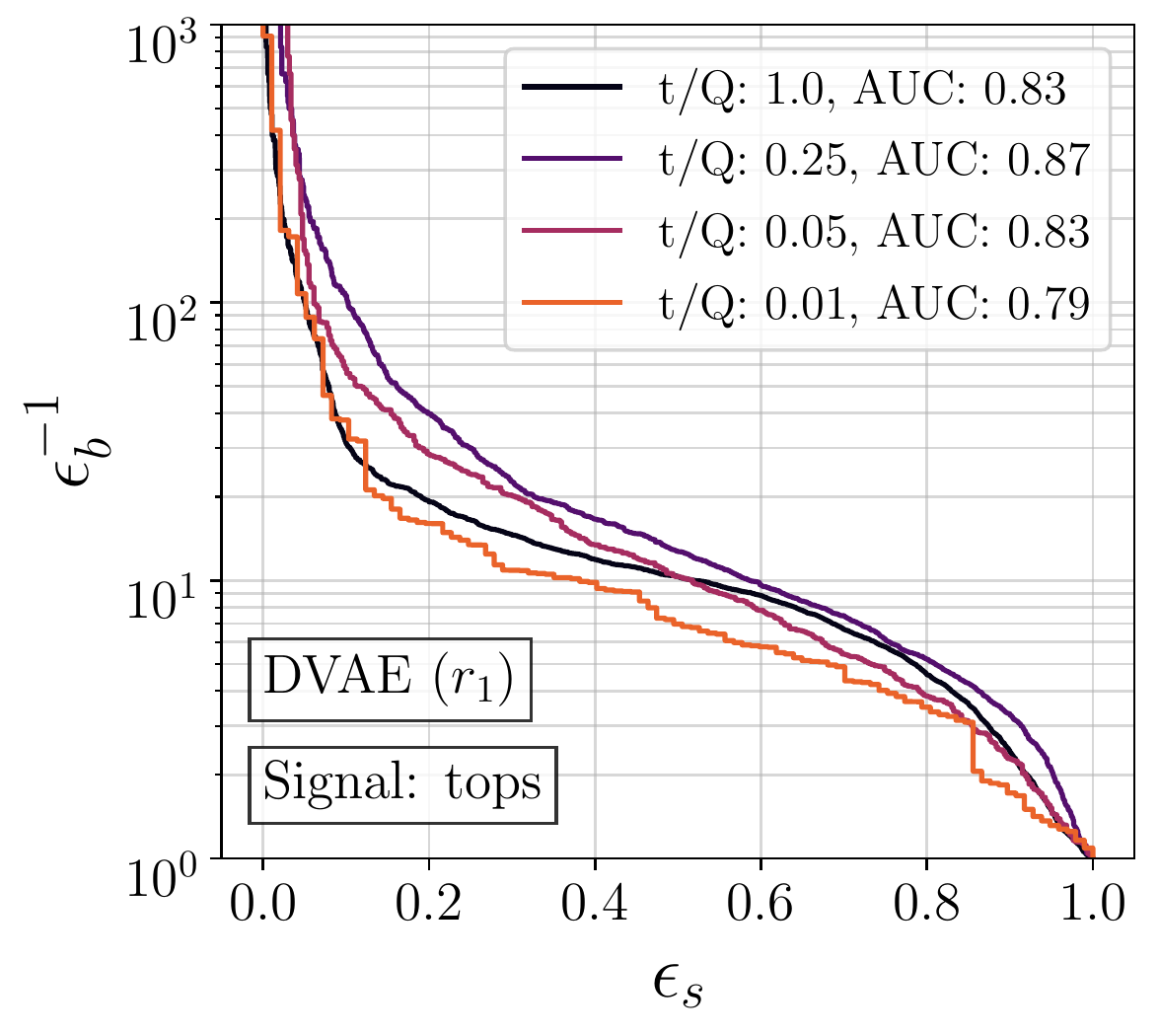}
\includegraphics[width=0.32\textwidth]{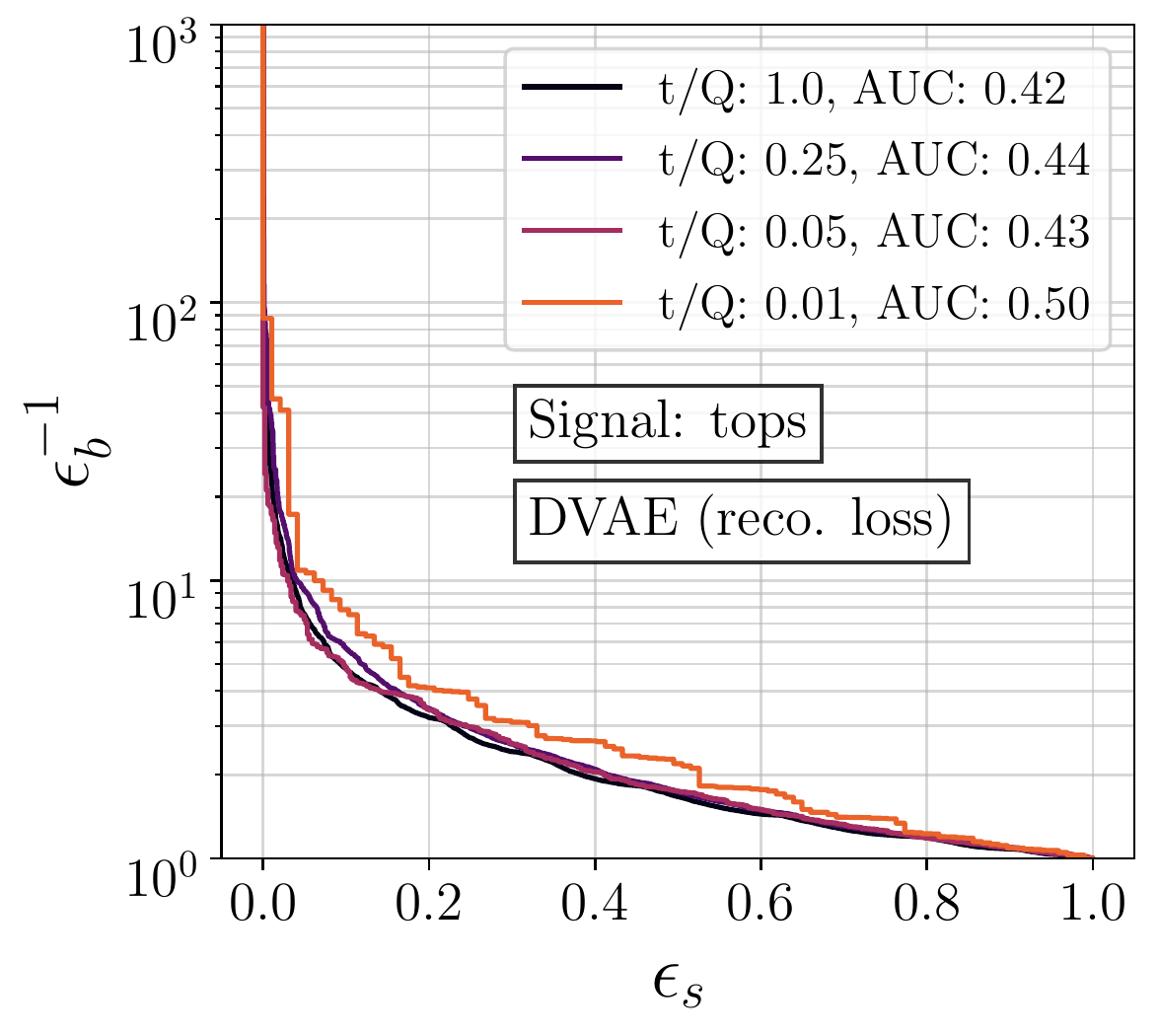} 
\includegraphics[width=0.32\textwidth]{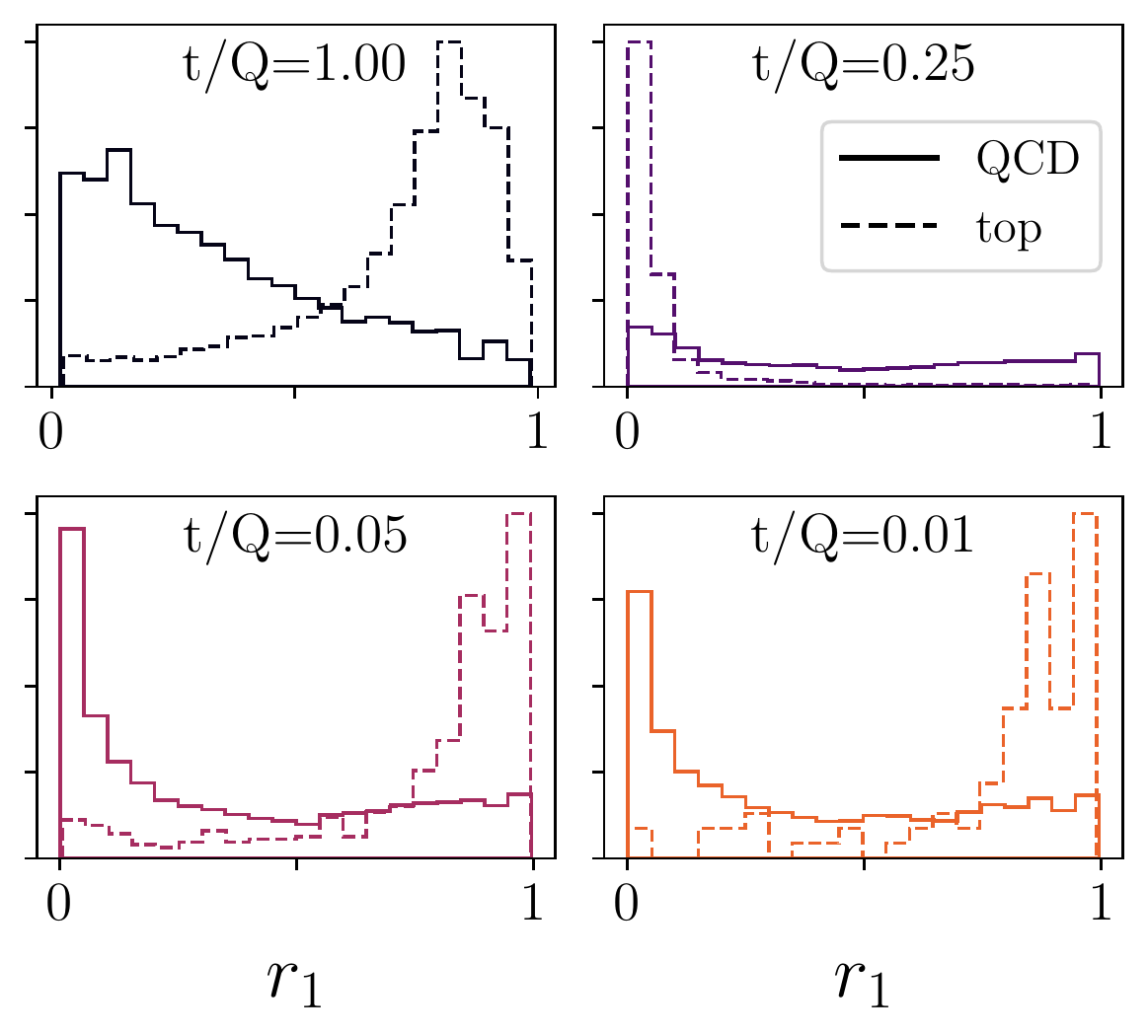} \\
\includegraphics[width=0.32\textwidth]{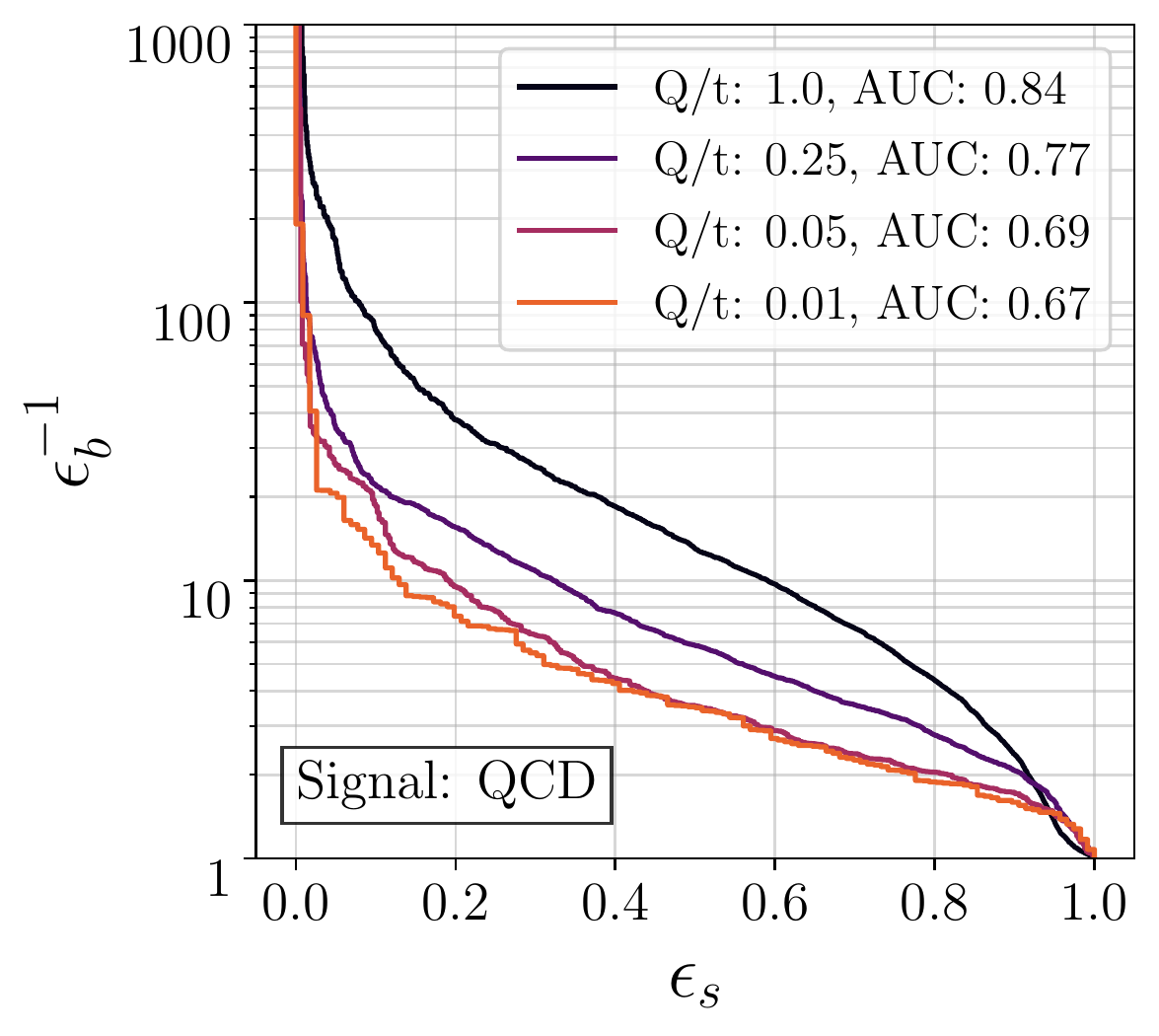}
\includegraphics[width=0.32\textwidth]{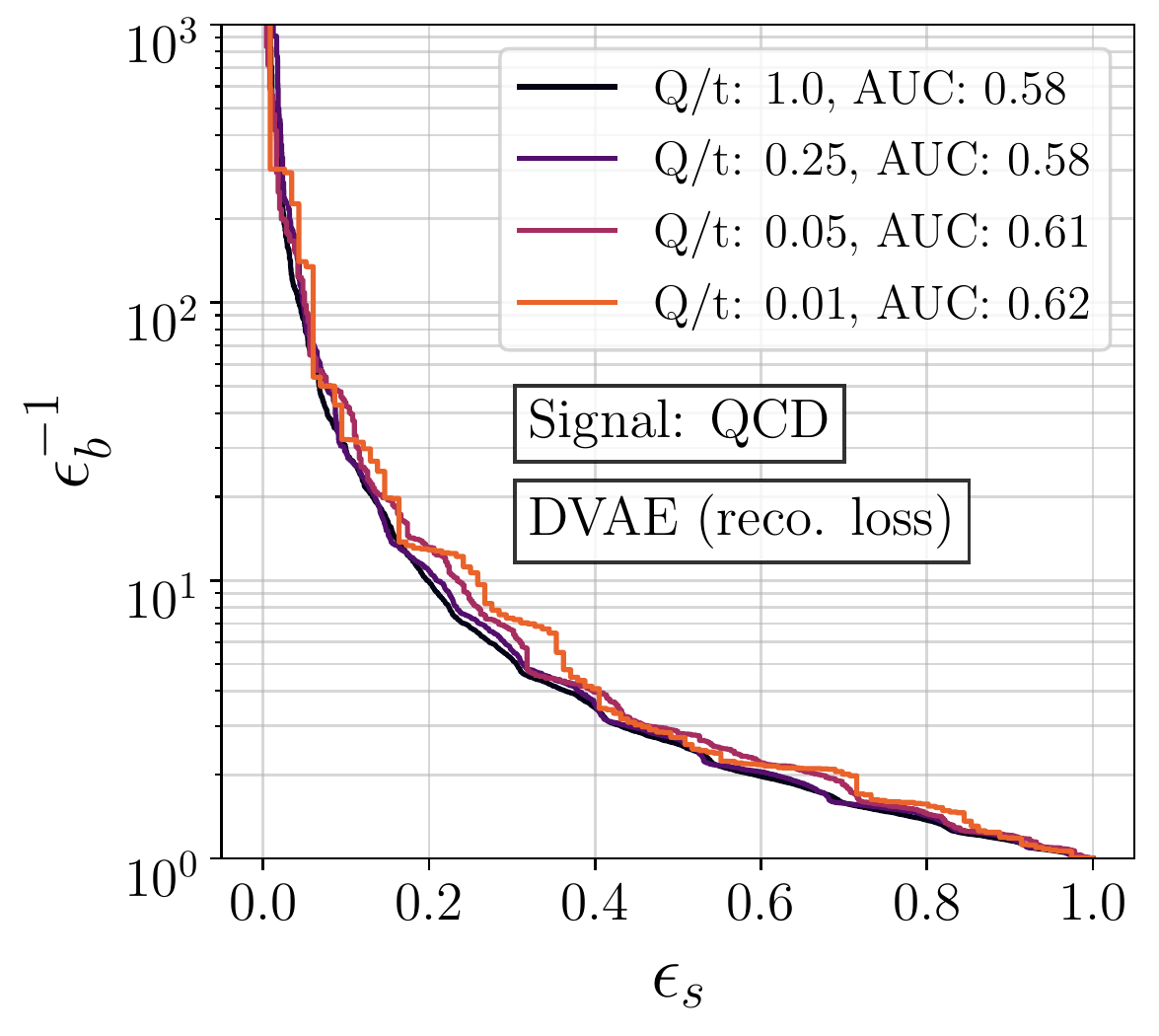} 
\includegraphics[width=0.32\textwidth]{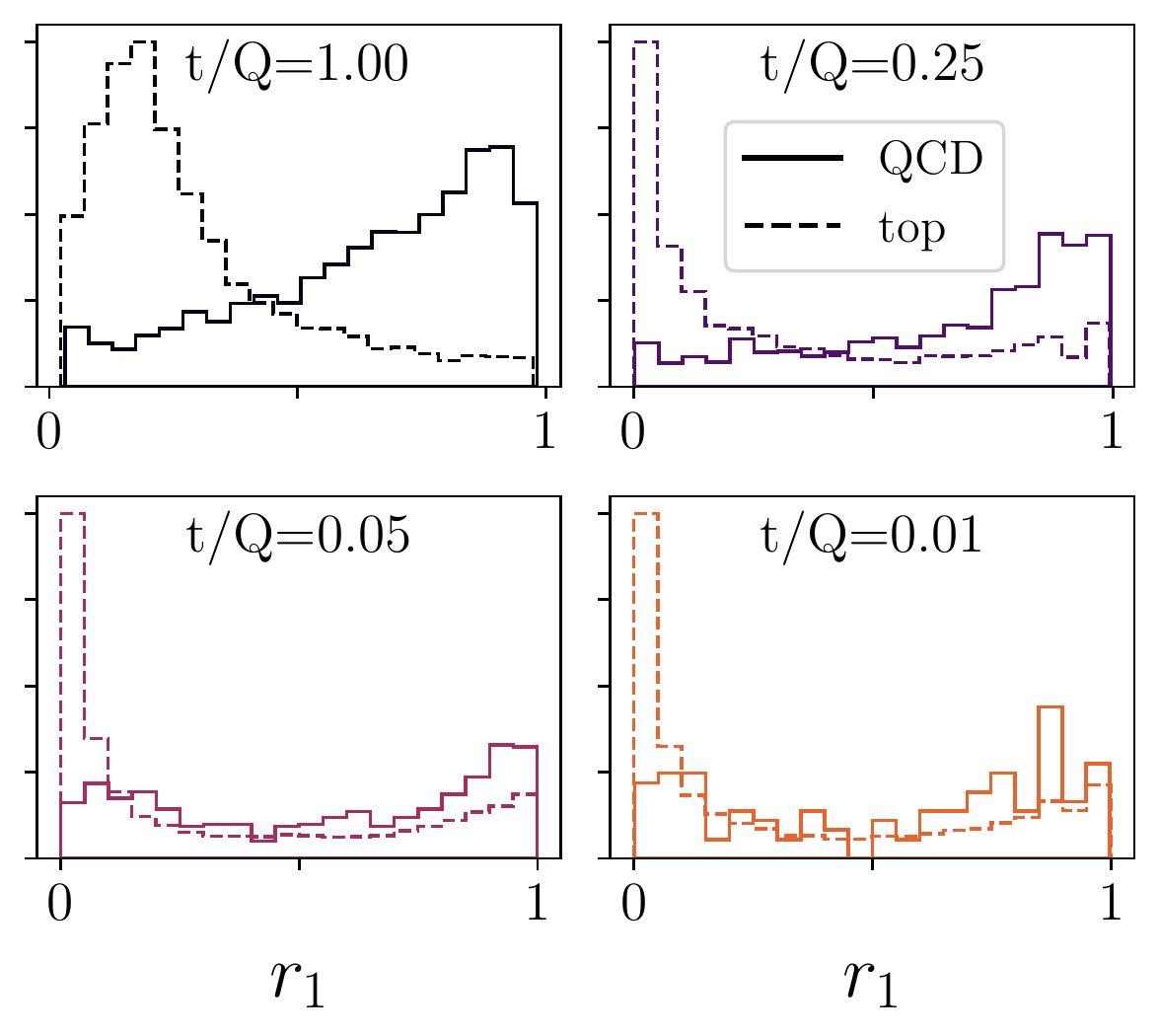} 
\caption{DVAE results using the Lund plane for various amounts of top
  (upper) and QCD (lower) jets in the sample, where the mixture
  weights (left) and the reconstruction loss (middle) of each jet are
  used for classification.  In the small panels we show the
  distributions of the top and QCD jets in the latent space.}
\label{fig:dvae_lund}
\end{figure}
%----------------------------------------------------------

For our DVAE analysis we only use only the observables $\log k_T \in
[-2,5]$ and $\log 1/\Delta R \in [0,6]$ and represent the Lund plane
as $40 \times 40$ images, just as before.  In
Fig.~\ref{fig:lundimages} we plot the average of 100k jet images for
top and QCD, in the primary Lund plane.  Because the main differences
between the QCD and top jets should appear in the primary Lund plane,
we will only use it for now. The DVAE training and architecture used here is identical to what we used for the
jet images (Fig. \ref{fig:dvae_arch}).

In Fig.~\ref{fig:dvae_lund} we show results in scenarios with varying
numbers of top jets and varying numbers of QCD jets.  The unique
aspect of the Lund plane results in comparison to the jet images is
that the reconstruction error does not work here for tagging anomalous
QCD jets or anomalous top jets.  It seems that this is due to the
truth level distributions for the QCD and top jets in the Lund plane
representation having a very large overlap, differing only in the
large angle and large $k_T$ region.  Despite this, the DVAE still
separates the two jet classes in the latent space.  In viewing these
results we note that there may be several routes to improving them; (i) weight the pixels by $k_T$ or another Lund observable, (ii) cut off
non-perturbative regions of the Lund plane during the pre-processing.
We did perform some tests with the pixel weights but the results did not show an improvement, alternatives were not fully explored in this analysis.

\clearpage

\bibliography{literature}
\end{document}